\documentstyle[12pt]{article}

\newcommand{\supscrpt}[2]{{#1}^{{#2}}}
\newcommand{\subscrpt}[2]{{#1}_{{#2}}}
\newcommand{\mixten}[3]{{#1}^{#2}_{\phantom{{#2}} #3}}
\newcommand{\supsub}[3]{{#1}^{#2}_{#3}}

\newcommand{\bfid}{{\bf 1}}
\newcommand{\bfc}{{\bf c}}
\newcommand{\bfcsup}[1]{\supscrpt{\bfc}{#1}}
\newcommand{\bfci}{\bfcsup{i}}
\newcommand{\bfe}{{\bf e}}
\newcommand{\bfesup}[1]{\supscrpt{\bfe}{#1}}
\newcommand{\bfei}{\bfesup{i}}
\newcommand{\bfk}{{\bf k}}
\newcommand{\bfu}{{\bf u}}
\newcommand{\bfw}{{\bf w}}
\newcommand{\bfx}{{\bf x}}
\newcommand{\bfxci}{\bfx + \bfci}
\newcommand{\bfxt}{(\bfx, t)}
\newcommand{\bfy}{{\bf y}}
\newcommand{\bfz}{{\bf z}}
\def\bfgamma{\mbox{\boldmath $\Gamma$}}
\def\bfdelta{\mbox{\boldmath $\delta$}}
\def\bfzeta{\mbox{\boldmath $\zeta$}}
\def\bfomega{\mbox{\boldmath $\Omega$}}

\newcommand{\bge}{\begin{equation}}
\newcommand{\ee}{\end{equation}}
\newcommand{\bgc}{\begin{center}}
\newcommand{\ec}{\end{center}}
\newcommand{\bgea}{\begin{eqnarray}}
\newcommand{\eea}{\end{eqnarray}}
\newcommand{\bgeas}{\begin{eqnarray*}}
\newcommand{\eeas}{\end{eqnarray*}}
\newtheorem{thm}{Theorem}
\newtheorem{corr}{Corollary}
\newtheorem{lem}{Lemma}
\newtheorem{definition}{Definition}

\newcommand{\advop}[2]{\mixten{{\cal A}}{#1}{#2}}
\newcommand{\advco}[1]{\supscrpt{\mbox{\boldmath ${\cal A}$}}{#1}}
\newcommand{\ctij}{\mixten{C}{(\tau)i}{j}}

\newcommand{\cdop}[2]{\mixten{\mbox{\boldmath ${\cal D}$}}{#1}{#2}}

\newcommand{\jop}[2]{\mixten{J}{#1}{#2}}
\newcommand{\jij}{\jop{i}{j}}

\newcommand{\tjop}[2]{\mixten{\tilde{J}}{#1}{#2}}
\newcommand{\tjij}{\tjop{i}{j}}
\newcommand{\tjt}{\supscrpt{\tilde{J}}{(\tau)}}
\newcommand{\tjtop}[2]{\tjop{(\tau) {#1}}{#2}}
\newcommand{\tjtij}{\tjtop{i}{j}}
\newcommand{\cjop}[2]{\mixten{{\cal J}}{#1}{#2}}
\newcommand{\kop}[2]{\mixten{K}{#1}{#2}}
\newcommand{\ckop}[2]{\mixten{{\cal K}}{#1}{#2}}
\newcommand{\sop}[2]{\mixten{\Sigma}{#1}{#2}}
\newcommand{\bsop}[2]{\mixten{\bar{\Sigma}}{#1}{#2}}
\newcommand{\dsop}[2]{\mixten{\dot{\Sigma}}{#1}{#2}}
\newcommand{\sinvop}[2]{\mixten{(\Sigma^{-1})}{#1}{#2}}
\newcommand{\bsinvop}[2]{\mixten{(\bar{\Sigma}^{-1})}{#1}{#2}}
\newcommand{\dsinvop}[2]{\mixten{(\dot{\Sigma}^{-1})}{#1}{#2}}
\newcommand{\vop}[2]{\mixten{V}{#1}{#2}}
\newcommand{\ctop}[2]{\mixten{{\cal T}}{#1}{#2}}
\newcommand{\cvop}[2]{\mixten{{\cal V}}{#1}{#2}}
\newcommand{\xop}[2]{\mixten{X}{#1}{#2}}
\newcommand{\yop}[2]{\mixten{Y}{#1}{#2}}

\newcommand{\csup}[1]{\supscrpt{c}{#1}}
\newcommand{\ccsup}[1]{\supscrpt{C}{#1}}
\newcommand{\ccsupo}[2]{\supsub{C}{#1}{#2}}
\newcommand{\fsup}[1]{\supscrpt{f}{#1}}
\newcommand{\gsup}[1]{\supscrpt{g}{#1}}
\newcommand{\nsup}[1]{\supscrpt{n}{#1}}
\newcommand{\nixt}{\nsup{i}\bfxt}
\newcommand{\nnsup}[1]{\supscrpt{N}{#1}}
\newcommand{\nnsupo}[2]{\supsub{N}{#1}{#2}}
\newcommand{\nnixt}{\nnsup{i}\bfxt}
\newcommand{\ppsup}[1]{\supscrpt{P}{#1}}
 \newcommand{\qrow}[2]{\supsub{q}{#1}{#2}}
 \newcommand{\qcol}[2]{\supsub{q}{#2}{#1}}
\newcommand{\qrowd}{\qrow{\mu}{i}}
\newcommand{\qcold}{\qcol{\mu}{j}}
\newcommand{\qsup}[1]{\supscrpt{q}{#1}}
\newcommand{\qsub}[1]{\subscrpt{q}{#1}}
 \newcommand{\qsupp}[1]{\supscrpt{q}{#1}}
 \newcommand{\qsubp}[1]{\subscrpt{q}{#1}}
 \newcommand{\qqsupp}[1]{\supscrpt{Q}{#1}}
 \newcommand{\qqsuppo}[2]{\supsub{Q}{#1}{#2}}
\newcommand{\ggamsup}[1]{\supscrpt{\Gamma}{#1}}
 \newcommand{\lamsupp}[1]{\supscrpt{\lambda}{#1}}

 \newcommand{\alpsupp}[1]{\supscrpt{\alpha}{#1}}
 \newcommand{\alpsubp}[1]{\subscrpt{\alpha}{#1}}
 \newcommand{\betsupp}[1]{\supscrpt{\beta}{#1}}
 \newcommand{\etasupp}[1]{\supscrpt{\eta}{#1}}
 \newcommand{\thetasupp}[1]{\supscrpt{\theta}{#1}}
\newcommand{\phisup}[1]{\supscrpt{\Phi}{#1}}
\newcommand{\chisup}[1]{\supscrpt{\chi}{#1}}

\newcommand{\pdv}{\partial}
\newcommand{\dt}{\Delta t}
\newcommand{\tpdt}{t+\dt}

\def\glossitem#1!#2!#3{$#1$ \> \if !#2! \else \pageref{#2} \fi\>
    \parbox[t]{3.5truein}{#3} \\}

\def\registeredsymbol{
  {\ooalign{\hfil\raise.05ex\hbox{\hskip.02ex$\rm
   \scriptscriptstyle R$}\hfil\crcr
   \hbox{$\scriptstyle\mathchar"20D$}}}}
\def\registered{
  \ifmmode^\registeredsymbol\else$^\registeredsymbol$\fi}

\def\vertex#1#2#3#4{\begin{picture}(8,8)(- 4,- 4)
\put(0,0){\circle*{3}}
\ifnum#1=1 \put(0,0){\line(1,0){4}}\fi
\ifnum#2=1 \put(0,0){\line(0,1){4}}\fi
\ifnum#3=1 \put(0,0){\line(-1,0){4}}\fi
\ifnum#4=1 \put(0,0){\line(0,-1){4}}\fi
\end{picture}}

\begin{document}

\title{{\small comp-gas/9403003} \hfill {\small MIT-CTP-2265}\\[0.5in]
       Correlations and Renormalization \\
       in Lattice Gases\footnote{
\small \baselineskip=11pt This work was supported in part by
Thinking Machines Corporation, and in part by the divisions of Applied
Mathematics of the U.S. Department of Energy under contracts
DE-FG02-88ER25065 and DE-FG02-88ER25066.}
}
\author{
Bruce M. Boghosian\\
{\small \sl Thinking Machines Corporation,}\\
{\small \sl 245 First Street, Cambridge, Massachusetts 02142-1264, U.S.A.} \\
{\small \tt bmb@think.com} \\[0.3cm]
Washington Taylor\\
{\small \sl Center for Theoretical Physics,}\\
{\small \sl Laboratory for Nuclear
Science and Department of Physics,}\\
{\small \sl Massachusetts Institute of
Technology; Cambridge, Massachusetts 02139, U.S.A.} \\
{\small \tt wati@mit.edu}}
\date{(\today)}
\maketitle

\begin{abstract}
A complete formulation is given of an exact kinetic theory for lattice
gases.  This kinetic theory makes possible the calculation of
corrections to the usual Boltz\-mann / Chap\-man-Ens\-kog analysis of
lattice gases due to the buildup of correlations.  It is shown that
renormalized transport coefficients can be calculated perturbatively by
summing terms in an infinite series.  A diagrammatic notation for the
terms in this series is given, in analogy with the diagrammatic
expansions of continuum kinetic theory and quantum field theory.  A
closed-form expression for the coefficients associated with the vertices
of these diagrams is given.  This method is applied to several standard
lattice gases, and the results are shown to correctly predict
experimentally observed deviations from the Boltzmann analysis.
\end{abstract}

\newpage
\section{Introduction}
\setcounter{equation}{0}
\label{sec:intro}

Lattice gases are a class of dynamical systems in which particles move
on a lattice in discrete time steps.  In much the same way that the
Ising model and other similar lattice spin models provide simple
examples of equilibrium statistical mechanical systems, lattice gases
capture many features of nonequilibrium statistical mechanical systems
such as fluids.

Most work on lattice gases has been done in a framework where the
Boltzmann assumption is made, neglecting all correlations between the
particles moving on the lattice.  In this paper we give a complete
description of a theory of lattice gases in which the effects of all
correlations are included.  For a wide variety of lattice gases, the
effect of these correlations is only to correct or {\sl renormalize} the
hydrodynamic transport coefficients, and we show how to compute this
correction.  For some lattice gases, the resulting renormalization can
change parameters in the theory by substantial amounts, so that in any
situation where lattice gases are used to make precise quantitative
predictions, the effects of correlations should be considered, and the
magnitude of the resulting correction to the usual analysis should be
estimated.

\subsection{Background}

In 1986, it was shown~\cite{fhp} that lattice gases could be used to
simulate two-dimensional Navier-Stokes flow.  Since then, lattice gases
have been developed to describe a wide variety of physical systems.
Examples are three-dimensional Navier-Stokes flow~\cite{fchc},
magnetohydrodynamics~\cite{chma}, immiscible fluids with a surface
tension interface~\cite{rk,sr}, convection~\cite{bz}, two-phase
liquid-gas flow~\cite{az}, Burgers' equation~\cite{bmbcdl}, and
reaction-diffusion equations~\cite{lw}.  For a summary of recent works
on the subject, see the proceedings edited by
Doolen~\cite{doolena,doolenb}, by Monaco~\cite{monaco}, by Manneville
et. al.~\cite{manneville}, by Alves~\cite{alves}, and by Boon and
Lebowitz~\cite{nicemtng}.

Why should we expect the bulk behavior of particles moving and colliding
on a lattice to be that of a fluid?  In nature, we observe that many
different fluids, with drastically differing intermolecular force laws,
all satisfy the Navier-Stokes equations to a reasonable degree of
approximation.  In spite of all the differences between the
intermolecular collisions of, say, water and molasses, both types of
collisions conserve mass and momentum; ultimately the existence of these
conserved quantities is what gives rise to fluid-like behavior at the
macroscopic level\footnote{Conservation of mass and momentum are
sufficient to get the correct behavior in the incompressible regime.  To
get correct {\sl compressible} behavior, it is necessary to conserve
energy as well.}.  A lattice gas model can be thought of as an attempt
to find the simplest possible dynamical system with these conservation
laws.

The idea that macroscopic properties of a physical system should be
independent of the microscopic definition of the system is also a
familiar concept in equilibrium statistical mechanics and in quantum
field theory.  In such theories, the effect of looking at the physics of
the system at larger and larger scales is mathematically described by
the renormalization group flow of the system~\cite{Wilson}.  Generally,
as the scale of the physics of interest becomes extremely large compared
to the scale at which the system is defined, one finds that the
renormalization group flow takes the system towards certain fixed
points, which describe entire universality classes of theories with
identical macroscopic behavior.  The emergence of similar hydrodynamic
equations in a variety of systems with different microscopic dynamics is
an equivalent phenomenon in nonequilibrium statistical mechanics.

Of course, there are some properties of fluids in which the details of
the microscopic collisions manifest themselves at the macroscopic level.
The most obvious such properties are the transport coefficients.  While
water and molasses both satisfy the Navier-Stokes equations, they do so
with very different viscosities.  Long-time tails in velocity
autocorrelation functions are also well-known to be sensitive to the
microscopic dynamics.

The central problem in the theoretical analysis of a lattice gas is thus
the determination of the macroscopic hydrodynamic equations obeyed by
the conserved quantities of the system, with the transport coefficients
expressed as functions of those conserved quantities.  This is the
classical problem of kinetic theory.  The methods needed to perform this
analysis for continuum fluids have been well developed over the past
century~\cite{classical}.  Two key approximations are used in the course
of such analyses:
\begin{itemize}
\item The {\sl Boltzmann molecular chaos approximation} (or {\sl
Stosszahlansatz}) neglects correlations between molecules entering a
collision.  This makes it possible to obtain a single closed equation,
called the {\sl Boltzmann equation}, for the single-particle
distribution function.
\item The {\sl Chapman-Enskog analysis} is an asymptotic expansion in
Knudsen number\footnote{The Knudsen number is the ratio of mean free
path to the macroscopic scale length.} that yields closed hydrodynamic
equations for the conserved quantities.
\end{itemize}

During the 1960's and 1970's, much work centered on the removal of the
first of these two approximations for continuum fluids~\cite{kringfl}
and for Lorentz gases~\cite{kringlo}.
It was found that the buildup of correlations
between the molecules of a fluid could seriously alter, or {\sl
renormalize}, the transport coefficients predicted by the Boltzmann
theory.  Expressions for these corrections were derived in terms of
diagrammatic sums.  The propagators in these diagrams can be thought of
as representing correlated quantities, and the vertices as collisions
in which
correlated quantities can interact.  Thus, an event in which two
particles emerge from a collision (and thereby acquire a correlation),
move about in a background of uncorrelated particles, and later
recollide, can be thought of as a one-loop correction to the Boltzmann
approximation.  Further refinements can be obtained by including more
intricate diagrams -- with multiple loops, nested loops, etc. -- to
account for the interaction of the correlated quantities with the
background.  Standard field-theoretic techniques can then be used to
approximate these diagrammatic sums.

As part of the recent flurry of interest in lattice gases, much of
classical kinetic theory has been extended to lattice gases.  In
particular, the derivation of Boltzmann equations for lattice gases and
the application of the Chapman-Enskog theory has been well understood
for several years now thanks to the works of a number of authors (see,
e.g., \cite{swolf,fchc}).  While these works have generally capitalized
on the similarities between lattice gases and continuum fluids, they
have also pointed to some very important differences between the two.

The most important of these differences has to do with the assumptions
leading up to the derivation of the Boltzmann equation.  Though the
assumption of molecular chaos is made in either case, the analysis of
continuum fluids generally proceeds under
the additional assumption that the
fluid is {\sl dilute} -- that is, that collisions involving more than
two molecules can be ignored.  For lattice gases, on the other hand, it
is essential that tertiary and higher collision events be treated
properly.  For example, tertiary collisions are essential to the success
of the FHP-I lattice gas because they break an unphysical (spurious)
conserved quantity~\cite{fhp}.  In addition, the inclusion of
higher-order collision events has been shown to decrease the viscosity
of fluid lattice gases, thereby making possible higher Reynolds' number
simulations~\cite{swolf}.

Thus, collision operators for lattice gases routinely include terms for
tertiary and higher-order collisions.  While this is really only a minor
nuisance in the Boltzmann and Chapman-Enskog analyses, it introduces a
significant complication to the exact kinetic theory.  Continuum kinetic
theory needs to consider only those vertices for which two or fewer
correlated quantities enter and exit.  Lattice gas kinetic theory, on
the other hand, must additionally treat vertices involving three or more
entering and/or exiting correlated quantities.  The number of correlated
quantities that can enter and/or exit a vertex is limited only by the
number of lattice vectors at a site.  This gives rise to a much richer
diagrammatic series than is the case for a continuum fluid.

Another difference between continuum fluids and lattice gases, which
gives the kinetic theories for these types of systems very different
flavors, is the simple fact that the continuum fluids are defined on
continuous spaces while the lattice gases are defined on discrete
spaces.  The consequence of this difference is that while the
corrections due to correlations in continuum fluids are expressed in
terms of complicated integral expressions, the corrections in lattice
gases are given by combinatorial sums over countable sets of graphs.
Thus, the problem of calculating exact transport coefficients in a
lattice gas becomes essentially a combinatoric, rather than an analytic,
problem.

For some lattice gases the dominant corrections to the transport
coefficients arise from diagrams which extend spatially and temporally
over only a short distance on the lattice\footnote{i.e., a distance that
goes to zero in the scaling limit}; this dramatically simplifies the
numerical calculation of renormalized transport coefficients.  The
resulting simplicity in the formulae describing diagrammatic corrections
to transport coefficients in lattice gases make these systems extremely
attractive both as models with which to study complicated properties of
related physical systems in the continuum, and as pedagogical tools with
which to describe the essential features of a complete kinetic theory
without the complications of many continuum systems.

\subsection{Results Presented in This Paper}

In this paper, we give for the first time a complete formulation of an
exact kinetic theory of lattice gases.  We prove that for a wide class
of lattice gases, the effects of correlations can be completely
described by a renormalization of the transport coefficients.
Furthermore, we give explicit formulae for the renormalized transport
coefficients in terms of a sum over an infinite set of diagrams.  We
describe the renormalization calculation in detail for several specific
lattice gases, and discuss a variety of methods for approximating the
complete diagrammatic sum.

We begin with a review of the Boltzmann and Chapman-Enskog theories.
Though, as has been noted, these methods have been successfully extended
to lattice gases for several years now, our presentation contains
several novel features and results:
\begin{itemize}
\item Our final results for the hydrodynamic equations and the transport
coefficients have, to our knowledge, never before been written down in
this generality.
\item  Our results are presented in a mathematical format that is
interesting in its own right.  We introduce a family of metrics on the
space of the Boltzmann distribution function, and show that the
transport coefficients can be expressed quite naturally in terms of
these metrics and their associated Christoffel symbols.
\item For the special case of lattice fluids on a regular
lattice, we derive an expression for the viscosity that agrees with that
of H\'{e}non, and yields a new and compact form for the quantity that he
calls $\lambda$ in his paper on the subject~\cite{henonvis}.
\end{itemize}

Next we present the exact kinetic theory for lattice gases.  The main
features of this theory are as follows:
\begin{itemize}
\item We give a simple expression for the renormalized transport
coefficients in terms of a sum over an infinite set of terms, each
associated with a diagram describing the propagation and interaction
of a set of correlations between lattice sites.
\item We show that the correction associated with each diagram can be
decomposed into a product of elementary factors associated with the
vertices in the diagram; these factors describe the interactions of
correlated quantities at a single lattice site.  Furthermore, we give a
closed-form expression for these vertex coefficients that is valid for
any lattice gas obeying semi-detailed balance at lowest order in the
Chapman-Enskog expansion parameter.
\item We show that when a lattice gas  violates semi-detailed balance
at higher order in the expansion parameter, the effects of
correlations can give rise to renormalized source terms in the
hydrodynamic equations in addition to the usual renormalization of
transport coefficients.  We describe these renormalized source terms
in terms of the same types of diagrams used to describe the
renormalized transport coefficients.
\item We formulate the theory in terms of connected
correlation functions.  A similar expansion in terms of products of
fluctuations was previously used in~\cite{Ern,ernstd} and
in~\cite{vanvel,kringlob}, where the ring kinetic theory was derived
for lattice gases, and for lattice Lorentz gases, respectively.  The
use of the connected correlation functions simplifies the form of the
complete diagrammatic expansion for the kinetic theory significantly.
\item We describe a variety of simplifications and approximations
which can be made to simplify the numerical computation of the
renormalized transport coefficients.  The simplest of these
approximations is the ring approximation, in which correlations between
more than 2 particles are neglected.  Thus, we find that the results
of~\cite{Ern} are described in our theory by the restriction of the
diagrammatic sum to the simple set of ring diagrams.
\item Previous work on the kinetic ring theory of lattice gases has
used the Green-Kubo formalism to obtain the series for the transport
coefficients.  In this paper, we use the Chapman-Enskog theory instead.
In this way, we get the Boltzmann approximation at zeroth order.  All
higher-order terms in our series are thus corrections to the Boltzmann
approximation.  In the Green-Kubo theory, by contrast, it is necessary
to sum an infinite number of terms just to get the Boltzmann
approximation.
\item We work out the diagrammatic expansion in detail for four model
lattice gases, and we compare some of these results to lattice gas
simulations.
\item For one of these model problems -- a lattice gas for Burgers'
equation~\cite{bmbcdl} -- we show that the Boltzmann approximation is
exact.  This result has been previously obtained by other
means~\cite{lebopres}.  We show that the diagrammatic formalism
introduced here reduces the proof to essentially a triviality.
\end{itemize}

\subsection{Layout of This Paper}

Section~\ref{sec:reviewlgt} begins with a review of lattice gas
theory, and establishes notation that is used throughout the rest of
the paper.  Section~\ref{sec:chen} then reviews the Boltzmann and
Chapman-Enskog analyses, and introduces a novel mathematical formalism
with which the results for the transport coefficients can be expressed
in a simple form.  Section~\ref{sec:exboce} contains examples in which
we work out the details of this formalism for four model lattice
gases.

Section~\ref{sec:exact} formulates the complete lattice gas kinetic
theory, proves that the exact transport coefficients are expressible as
the sum of a diagrammatic series, and gives a closed-form expression for
the vertex coefficients.  Section~\ref{sec:examples} then works out
these vertex coefficients for the same four model lattice gases
described in Section~\ref{sec:exboce}, and presents the results for the
transport coefficients of these lattice gases as formal diagrammatic
series.  For the Burgers' equation lattice gas we show that the
diagrammatic series of corrections to the Boltzmann approximation
vanishes, and thereby recover the result of~\cite{lebopres}.

Since, with the exception of the Burgers' equation example, these
diagrammatic series are difficult to sum without resorting to numerical
methods, Section~\ref{sec:exact} includes a discussion of approximation
schemes for these diagrammatic sums, including summations for short
times, summations over one-loop diagrams (ring approximation), and
summations over subsets of diagrams which correspond to truncations of
the BBGKY hierarchy.  In Section~\ref{sec:approximations}, we apply
several of these approximation schemes to one of the model problems (the
1D3P lattice gas), and we present comparisons to experiment, which show
that our theory correctly predicts observed deviations from the
predictions of the Boltzmann theory to within measurable accuracy.  In
the course of doing this, we expand on the discussion of approximations
in Section~\ref{sec:exact}, further developing techniques valid for
general lattice gases.

At the end of the paper, a glossary of notation is included due to the
large number of symbols introduced in the text.

\newpage
\section{Review of Lattice Gas Theory}
\setcounter{equation}{0}
\label{sec:reviewlgt}

\subsection{Definitions and Notation}

A lattice gas is generally described by a state space and a
time-development rule.  The state space is defined by associating $n$
bits with each point on a lattice $L$.  (Bits are variables taking
values in $\{0,1\}$.)  We define the set of bits at a general lattice
site to be $B$, so that $|B|=n$.  We denote the total number of bits on
the lattice by $N = n|L|$.  For each value of the discrete time
parameter $t$, we write the values of the bits as $\nixt$, where
$\bfx\in L$ and $i\in B$.  As noted in Section~\ref{sec:intro}, these
bits can be thought of as a set of occupation numbers for individual
{\sl particle} states.

We define the set $S$ of possible states of the bits at a general
lattice point at a fixed value of $t$ to be
\[
   S=\{s: s\subseteq B\},
\]
where a state $s$ is associated with the set of bits taking the value
$1$.  We  write the value of bit $i$ in state $s$  as
\[
\supscrpt{s}{i} = \left\{\begin{array}{ll}
                         1 & \mbox{if $i\in s$} \\
                         0 & \mbox{otherwise}
                         \end{array}
                  \right.
\]
Note that $|S|=2^n$.  \label{pg:al}

We shall often want to refer to the $N=n|L|$ bits of the lattice in a
uniform fashion, so we introduce an enumeration of these $N$ bits, given
by a 1-1 correspondence between the sets ${\cal B}= \{1, 2, \ldots, N\}$ and
$B \times L = \{(i,\bfx): i\in B,\bfx\in L\}$.  In this notation, a single
bit of the lattice gas is written as $\nsup{a}$, $a\in{\cal B} $.
To relate this notation to the more explicit $(i,\bfx)$ notation, we
express the above 1-1 correspondence by writing $a$ and $(i,\bfx)$ as
functions of one another, so that
\[
   \nsup{a}(t) = \nsup{i(a)} (\bfx (a), t)
\]
and
\[
   \nixt = \nsup{a(i,\bfx)}(t).
\]
We shall use both notations interchangeably throughout this paper.

\subsection{Microscopic Dynamical Equation}

For each value of $i\in\{1,\ldots,n\}$ there is a lattice vector $\bfci$
such that $\bfxci\in L$ for every $\bfx\in L$.  \label{pg:aq} The
evolution of a lattice gas for one timestep can be divided into two
substeps: (1) A {\sl collision} substep in which the $n$ bits at each
site may alter their values to model a local {\sl interaction} between
the particles that they represent, and (2) an {\sl advection} substep in
which the new value of bit $i$ at site $\bfx$ moves to that of bit $i$
at site $\bfxci$.

We wish to express this evolution in the form of an equation for the
microscopic dynamics of the lattice gas; that is, we desire an equation
for $\nsup{i}(\bfx,\tpdt)$ in terms of $\nixt$, where $\dt$ denotes the
timestep.  \label{pg:bp} Suppose, for a moment, that the particles
simply advected without colliding.  Then the dynamics would be described
by
\[
   \nsup{i}(\bfxci,\tpdt) = \nixt.
\]
The addition of collisions introduces a {\sl collision operator} on the
right-hand side of the above equation.  That is, we have
\bge
   \nsup{i}(\bfxci,\tpdt) =
   \nixt + \csup{i}\left(\nsup{\ast}\bfxt\right),
   \label{eq:microdyn}
\ee
where the collision operator, $\csup{i}$, describes the change in bit
$i$ due to collisions.  Note that the form \label{pg:aa}
$f(\supscrpt{z}{\ast})$ is used to indicate that a function $f$ depends
on a quantity $z$ for all possible values of the index that has been
replaced by the asterisk.

\subsection{The Collision Operator}

The collision process at a fixed lattice site and timestep can be fully
specified by a $2^n$ by $2^n$ Boolean {\sl transition matrix}, $a$,
whose element $a(s\rightarrow s')$ is unity if and only if the particles
in state $s$ collide to yield particles in state $s'$.  \label{pg:am}
Since each incoming state gives rise to exactly one outgoing state,
\bge
   \sum_{s'} a(s\rightarrow s') = 1.
   \label{eq:consprob}
\ee

Stochastic lattice gases deserve special mention at this point.  They
also have $a(s\rightarrow s')$ equal to unity for exactly one value of
$s'$ for each value of $s$, but that value of $s'$ may differ from site
to site, and from timestep to timestep for a fixed value of $s$.  For
example, we might have $a(s\rightarrow s_1)=r$, $a(s\rightarrow
s_2)=1-r$, and $a(s\rightarrow s')=0$ for all other $s'$, where $r$ is a
random bit that is sampled at each site at each timestep with some
specified mean, $R=\langle r\rangle$.  This would effectively mean that
the outcome of a collision for incoming state $s$ is state $s_1$ with
probability $R$, and state $s_2$ with probability $1-R$.  Note that
Eq.~(\ref{eq:consprob}) is still always satisfied.

If we define the Kronecker delta function of two bits \label{pg:ah},
\bgea
   \delta (x,y) &=&
   \left\{
   \begin{array}{ll}
   1 & \mbox{if $x=y$} \\
   0 & \mbox{otherwise}
   \end{array}
   \right.\nonumber \\
   &=& 1-x-y+2xy,
   \label{eq:kronalg}
\eea
then the product
\[
   \prod_{j=1}^n \delta(\nsup{j},\supscrpt{s}{j})
\]
is unity at a given site if and only if that site is in state $s$.
Since the collision operator $\csup{i}$ is nothing more than the total
change in bit $i$ due to collision, it can be expressed in terms of the
transition matrix as follows:
\bge
   \csup{i}(\nsup{\ast}) =
   \sum_{s,s'} a(s\rightarrow s')
   (\supscrpt{s}{\prime i} - \supscrpt{s}{i})
   \prod_{j=1}^n \delta(\nsup{j},\supscrpt{s}{j}).
   \label{eq:collop}
\ee

\subsection{Conserved Quantities}

Another distinguishing feature of lattice gases is the presence of some
number of additive conserved quantities that are linear in the bit
values.  For example, in some lattice gases, the total number of
particles is a conserved quantity; it is clearly conserved by the
advection phase of the timestep, and we can choose collision rules that
conserve particle number as well.

Let us assume that we have a lattice gas with some number, $n_c$, of
conserved quantities.  \label{pg:bh} We assume that all conserved
quantities are linear in the configuration bit values, so we can write
the value of the $\mu$th conserved quantity at site $\bfx$ and at time
$t$ in the form\footnote{Throughout this paper we adopt the modified
Einstein summation convention that if the same index appears in at least
one contravariant position and at least one covariant position in every
term where it appears at all, then it is to be summed over its entire
range of values.  Sometimes summations are indicated explicitly,
particularly when the range of summation is not clear from the context.}
\bge
   \begin{array}{ll}
   \qsupp{\mu} \bfxt = \qrowd \nixt &
   \mbox{for $\mu=1,\ldots,n_c$.}
   \end{array}
   \label{eq:consquan}
\ee
where the coefficients $\qrowd$ satisfy \label{pg:bm}
\bge
   \begin{array}{ll}
   0 = \qrowd \csup{i} (\nsup{\ast}) &
   \mbox{for $\mu=1,\ldots,n_c$.}
   \end{array}
   \label{eq:ccop}
\ee
It is important to note that the $\qrowd$ are constant coefficients,
independent of spatial position.  From Eq.~(\ref{eq:ccop}), it follows
that
\bgeas
   \sum_\bfx \qsupp{\mu} (\bfx,\tpdt)
   & = & \sum_\bfx \qrowd\nsup{i}(\bfxci,\tpdt) \\
   & = & \sum_\bfx \qrowd\left[\nixt + \csup{i}(\nsup{\ast})\right]
     =   \sum_\bfx \qsupp{\mu}\bfxt,
\eeas
so that the global sum of any conserved quantity is constant
in time.

Note that the conserved quantities naturally partition the set $S$ into
equivalence classes.  Two states belong to the same equivalence class if
they have the same values of all the conserved quantities (that is,
$s\sim s'$ iff $\qrowd\supscrpt{s}{\prime i} =
\qrowd\supscrpt{s}{i}$ for $\mu=1,\ldots,n_c$).  Collisions must map
states into other states of the same equivalence class.  This means that
the transition matrix, $a(s\rightarrow s')$ is block diagonal, in that
$a(s\rightarrow s')=0$ if $s\not\sim s'$.  More succinctly,
\bge
   a(s\rightarrow s')
   \left(
   \qrowd\supscrpt{s}{\prime i} - \qrowd\supscrpt{s}{i}
   \right) = 0.
   \label{eq:blocdiag}
\ee
Note that Eq.~(\ref{eq:ccop}) follows immediately from
Eqs.~(\ref{eq:collop}) and (\ref{eq:blocdiag}).

We observe in passing that some lattice gases also posess {\sl spurious
global conserved quantities}.  For example, it is easily seen that
lattice gases of single-speed particles on a Cartesian grid conserve all
quantities separately on both checkerboard sublattices.  Such spurious
global conserved quantities have no analog for continuum fluids, and
need to be considered carefully when using lattice gases to model
hydrodynamic phenomena.  (See, e.g., \cite{lebowitz,ernstb,ernstc}.)

\subsection{The Ensemble Average}

We now consider some statistical aspects of lattice gas theory.  Let us
suppose that we have prepared an ensemble of lattice gas simulations, on
grids of the same size, with initial conditions that are identical at
the hydrodynamic level but may differ (are sampled from some
distribution of initial conditions) at the kinetic level.  We may then
take averages across this ensemble.  We shall denote these ensemble
averages by angular brackets $\langle\rangle$.  We denote the ensemble
average of the quantity $\nsup{i}\bfxt$ by $\nnixt$; that is
\[
   \nnixt = \langle\nixt\rangle
\]
\label{pg:bj} Note that while the $\nsup{i}$'s are binary, the
$\nnsup{i}$'s take their values in the set of real numbers between zero
and one.

Next, the ensemble average of $a(s\rightarrow s')$ is defined to be
\label{pg:an}
\[
   A(s\rightarrow s') = \langle a(s\rightarrow s')\rangle,
\]
so that the ensemble average of Eq.~(\ref{eq:consprob}) is
\bge
   \sum_{s'} A(s\rightarrow s') = 1.
   \label{eq:aconsprob}
\ee
For deterministic lattice gases, $A(s\rightarrow s') = a(s\rightarrow
s')$, whereas for stochastic lattice gases, the elements of
$A(s\rightarrow s')$ are generally real numbers between $0$ and $1$.

Similarly, we can consider ensemble-averaged values of the conserved
quantities \label{pg:bo}
\bge
   \begin{array}{ll}
   \qqsupp{\mu}\bfxt
   \equiv\langle\qsupp{\mu}\bfxt\rangle
   =\qrowd\nnixt &
   \mbox{for $\mu=1,\ldots,n_c$.}
   \end{array}
   \label{eq:eecons}
\ee

Note that we now have three possible levels of description of the
lattice gas system.  At the finest level, the specification of the
$\nsup{i}\bfxt$ constitute a complete microscopic description of the
system.  Their time evolution can generally be obtained only by an
actual simulation of the lattice gas.  Often, however, precise knowledge
of each and every bit of the system is more information than one really
desires.  A coarser description, such as a closed set of {\sl kinetic}
equations for the ensemble averaged $\nnsup{i}\bfxt$ is often a more
appropriate description of the system.  Even this level of description,
however, is redundant for many purposes.  Therefore, at the coarsest
level, one might seek a closed set of {\sl hydrodynamic} equations for
the $\qqsupp{\mu}$.  The remainder of this paper will be concerned with
deriving these two reduced descriptions of the system.

\subsection{The Boltzmann Equation}

Toward the goal of obtaining a closed set of equations for the
$\nnsup{i}$, we take the ensemble average of Eq.~(\ref{eq:microdyn}).
We are immediately thwarted by the fact that the collision operator,
given by Eq.~(\ref{eq:collop}), is generally a nonlinear function of the
$\nsup{i}$.  As is well known, the average of a nonlinear function is
not in general expressible as a function of the averaged quantitites.
It also depends on the {\sl correlations} between the quantities -- in
this case between the incoming bits, $\nsup{i}$.

Thus, the simplest approximation that we can make to close the system of
equations for the $\nnsup{i}$ is to assume that the incoming bits,
$\nsup{i}$, are uncorrelated.  This is the discrete version of the
famous {\sl Boltzmann molecular chaos assumption}.  From this
assumption, it would follow that
\[
   \langle\csup{i}(\nsup{\ast})\rangle
   \approx \ccsup{i}(\langle\nsup{\ast}\rangle)
   = \ccsup{i}(\nnsup{\ast}),
\]
where, using Eq.~(\ref{eq:kronalg}) and Eq.~(\ref{eq:collop}),
$\ccsup{i}(\nnsup{\ast})$ is given by \label{pg:ar}
\bgea
   \ccsup{i}(\nnsup{\ast})
   & = &
   \sum_{s,s'} A(s\rightarrow s')
   (\supscrpt{s}{\prime i} - \supscrpt{s}{i})
   \prod_{j=1}^n
   \left(1-\supscrpt{s}{j}-\nnsup{j}+2\supscrpt{s}{j}\nnsup{j}\right)
   \nonumber \\
   & = &
   \sum_{s,s'} A(s\rightarrow s')
   (\supscrpt{s}{\prime i} - \supscrpt{s}{i})
   \prod_{j=1}^n
      \left(\nnsup{j}\right)^{\supscrpt{s}{j}}
      \left(1-\nnsup{j}\right)^{1-\supscrpt{s}{j}}.
   \label{eq:bcollop}
\eea
In this way, we get the {\sl lattice
Boltzmann equation},
\bge
   \nnsup{i}(\bfxci,\tpdt) =
   \nnixt + \ccsup{i}\left(\nnsup{\ast}\bfxt\right).
   \label{eq:boltzmann}
\ee

Physically speaking, the assumption of molecular chaos supposes that the
advection substep effectively decorrelates the different bits at each
site\footnote{For stochastic lattice gases, such decorrelation is
enhanced by the injection of stochasticity at each site at each time
step.}.  That is, it supposes that colliding particles have never had
any prior effect on each other.  This assumption is virtually never
strictly correct for a system of particles moving on a discrete lattice
in a finite number of dimensions.  By standard combinatorial arguments,
the reencounter probability for two particles executing a random walk on
a lattice is unity in one and two dimensions, is less than unity in
three or more dimensions, and falls to zero as the number of dimensions
goes to infinity.  One might thus expect that the molecular chaos
assumption becomes more valid as the number of spatial dimensions
increases.  Indeed, this is the case, and the molecular chaos assumption
can be thought of as a sort of mean-field theory.  In addition, in some
circumstances, it is possible for particles to set up coherent
structures that persist for long times.  Such structures, by their very
nature, invalidate the molecular chaos assumption in a rather dramatic
way.

The remainder of this paper is devoted to deriving the desired closed
set of hydrodynamic equations for the $\qqsupp{\mu}$.  We shall go about
this task in two stages.  First, in Section~\ref{sec:chen}, we shall
show how they can be derived under the molecular chaos assumption.  In
Section~\ref{sec:exact}, however, we shall abandon this assumption.  We
shall find that this has the effect of correcting, or {\sl
renormalizing}, the transport coefficients in the resulting hydrodynamic
equations.  For a large class of lattice gases of interest -- those
satisfying a condition known as {\sl semi-detailed balance} at lowest
order -- we shall show that it is possible to write an exact expression
for this correction as a diagrammatic series.

\newpage
\section{The Chapman-Enskog Analysis}
\setcounter{equation}{0}
\label{sec:chen}

\subsection{Asymptotic Ordering}

We shall now outline a perturbative analysis of the Boltzmann equation,
Eq.~(\ref{eq:boltzmann}).  With this analysis, we study the
hydrodynamics of systems which deviate slightly from local equilibrium
conditions.  To do this, we must first establish the asymptotic regime
that we are trying to study.  In this paper we use what is sometimes
called {\sl diffusion ordering} or {\sl Navier-Stokes ordering}.  This
ordering can be obtained formally by letting
$\bfc\rightarrow\epsilon\bfc$ and $\dt\rightarrow\epsilon^2\dt$ in the
dynamical equations, where $\epsilon$ is an expansion
parameter.\label{pg:ai}  Thus, we are taking $\dt \sim c^2$, as is
appropriate for diffusive or viscous processes.

Because the $\nnsup{i}$ are real numbers (as opposed to the $\nsup{i}$
which are bits), we are free to approximate them by smooth functions
that happen to coincide with them in value on the lattice points.  We
can then Taylor expand Eq.~(\ref{eq:boltzmann}), retaining terms to
order $\epsilon^2$.  We get
\[
     \epsilon^2\dt\frac{\pdv\nnsup{i}}{\pdv t}
   + \epsilon\bfci\cdot\nabla \nnsup{i}
   + \frac{\epsilon^2}{2}\bfci\bfci : \nabla\nabla \nnsup{i}
   = \ccsupo{i}{0} (\nnsup{\ast}),
   + \epsilon\ccsupo{i}{1} (\nnsup{\ast})
   + \epsilon^2\ccsupo{i}{2} (\nnsup{\ast}),
\]
where the double-dot notation ($:$) denotes two inner
products\footnote{This notation will be used only when no ambiguity can
arise from it, as is the case when at least one of the two dyads
involved is symmetric}.  Note that we have ordered the ensemble-averaged
collision operator in the expansion parameter $\epsilon$.  In what
follows, we shall assume that $C_0$ and $C_1$ respect the conservation
laws exactly, but that $C_2$ does not necessarily do so.  This will
allow us to consider lattice gases whose conservation laws are only
approximate. \label{pg:as}

The above equation can be written in the more suggestive form,
\bge
     \epsilon^2\frac{\pdv\nnsup{i}}{\pdv t}
   + \epsilon\nabla\cdot\left(\frac{\bfci}{\dt}\nnsup{i}\right)
   + \epsilon^2\nabla\nabla :
     \left(\frac{\bfci\bfci}{2\dt}\nnsup{i}\right)
   = \frac{1}{\dt}\ccsupo{i}{0}(\nnsup{\ast})
   + \frac{\epsilon}{\dt}\ccsupo{i}{1}(\nnsup{\ast})
   + \frac{\epsilon^2}{\dt}\ccsupo{i}{2}(\nnsup{\ast}).
   \label{eq:lboltza}
\ee
By contracting this with the constant $\qrowd$,
we get the $n_c$ conservation equations
\bge
     \epsilon\frac{\pdv\qqsupp{\mu}}{\pdv t}
   + \nabla\cdot\left[
     \left(
     \qrowd\frac{\bfci}{\dt}\nnsup{i}
     \right)
   + \epsilon\nabla\cdot
     \left(
     \qrowd\frac{\bfci\bfci}{2\dt}\nnsup{i}
     \right)
     \right] =
     \frac{\epsilon}{\dt}\qrowd\ccsupo{i}{2}(\nnsup{\ast}),
   \label{eq:lconslaws}
\ee
where $\mu = 1,\ldots,n_c$.  We can now clearly identify the quantity
in brackets as the flux corresponding to the conserved density
$\qqsupp{\mu}$, and the right-hand side as a source/sink term.

In what follows, we shall expand the $\nnsup{i}$ in a perturbation
series in powers of $\epsilon$ about an equilibrium state, \label{pg:bk}
\bge
   \nnsup{i} = \nnsupo{i}{0} +
   \epsilon \nnsupo{i}{1} +
   \epsilon^2 \nnsupo{i}{2} + \cdots
   \label{eq:ordering}
\ee
Here $\nnsupo{i}{0}$ is a local thermodynamic equilibrium.  In the next
section, we shall characterize these equilibria.  Then we shall derive
hydrodynamic equations for the system by considering its
near-equilibrium behavior.

\subsection{Semi-Detailed Balance and Equilibria}

An {\sl equilibrium distribution} for a given lattice gas is a
distribution on the state space of the system that is invariant under
the full dynamics.  A {\sl Boltzmann equilibrium} is a set of values
for the mean occupation numbers, $\nnsup{a}$, which is invariant under
the Boltzmann equation, Eq.~(\ref{eq:boltzmann}).  A Boltzmann
equilibrium can be associated with a distribution on the set of states
by independently sampling each bit $\nsup{a}$ with probability
$\nnsup{a}$.  If the Boltzmann equilibrium is spatially uniform, it
can be specified by a set of values for the $n$ mean occupation
numbers, $\nnsup{i}$; the associated distribution on the set of states
is given by independently sampling each bit $\nsup{a}$ with
probability $\nnsup{i(a)}$.

A Boltzmann equilibrium is defined to be {\sl stable} when it is also an
equilibrium in the more general sense; that is, when the full dynamics
of the system, Eq.~(\ref{eq:microdyn}), do not generate correlations
between the $\nsup{a}$.

We wish to study the dynamics of the system in the vicinity of a local
equilibrium.  Thus, we demand that the lowest-order terms in the mean
occupation numbers, $\nnsupo{a}{0}$, correspond to a Boltzmann
equilibrium with respect to the lowest-order part of the Boltzmann
equation, Eq.~(\ref{eq:lboltza}).  This condition is given by
\bge
   \ccsupo{i}{0}(\nnsupo{\ast}{0}(\bfx,t)) = 0.
   \label{eq:zorder}
\ee
We will also demand that the zero-order equilibrium $\nnsupo{i}{0}$ be
stable in the sense that the zero-order collision operator does not
generate correlations between the bits of the system.  However, we
shall allow the zero-order equilibrium parameters
$\nnsupo{i}{0}(\bfx,t)$ to have spatial dependence.  In this paper, we
shall henceforth restrict our attention to lattice gases for which
equilibria with these desired properties exist.  Fortunately, the
restriction that this places on the lattice gases that we may consider
can be stated as a simple sufficient condition on the lowest order
part of the transition matrix~\cite{fchc}.  To state this condition,
we first introduce a definition:

\begin{definition}
A lattice gas is said to obey {\sl detailed balance} if its transition
matrix satisfies
\bge
   A(s\rightarrow s') = A(s'\rightarrow s),
   \label{eq:db}
\ee
and it is said to obey {\sl semi-detailed balance} if its transition
matrix satisfies
\bge
   \sum_{s} A(s\rightarrow s') = 1.
   \label{eq:sdb}
\ee
\end{definition}

Note that semi-detailed balance is a weaker condition than detailed
balance, because Eqs.~(\ref{eq:db}) and (\ref{eq:aconsprob}) together
imply Eq.~(\ref{eq:sdb}).  Semi-detailed balance (coupled with
probability conservation) requires that the rows and columns of the
transition matrix all sum to unity.  Detailed balance additionally
requires that it be symmetric.

For a deterministic lattice gas, detailed balance requires that if the
transition matrix takes state $s$ to state $s'$, then it must also take
state $s'$ to state $s$.  In this case, semi-detailed balance is the
weaker condition that the final states are a permutation of the initial
states.

We now quote \cite{fchc} and prove a theorem on the existence of
stable Boltzmann equilibria.

\begin{thm}
\label{thm:sdb}
Stable and spatially uniform Boltzmann equilibria exist for any
lattice gas obeying semi-detailed balance.  These equilibria are
described by the Fermi-Dirac distribution,
\bge
   \nnsupo{i}{0} =
   \frac{1}{1 + \exp\left(-\sum_{\mu=1}^{n_c}
                          \alpsubp{\mu}
                          \qrowd
                    \right)},
   \label{eq:fd}
\ee
where the $\alpsubp{\mu}$ are $n_c$ arbitrary multipliers.
\end{thm}

\noindent {\bf Proof:} \,\,At first, Eq.~(\ref{eq:zorder}) appears to
impose $n$ conditions on the $n$ unknowns, $\nnsupo{i}{0}$, but the
$n_c$ restrictions imposed by Eq.~(\ref{eq:ccop}) mean that only $n-n_c$
of these conditions are independent, and therefore that we ought to
expect an $n_c$ parameter family of equilibria.  These will be the
constants, $\alpsubp{\mu}$ in Eq.~(\ref{eq:fd}). \label{pg:ac}

Taking the Fermi-Dirac distribution (\ref{eq:fd}) to define an
independent distribution on the bits of the system, the probability at
any site of a fixed state $s$ is given by
\bge
P_0 (s) =   \prod_{j=1}^n
      \left(N_0^j\right)^{\supscrpt{s}{j}}
      \left(1-N_0^j\right)^{1-\supscrpt{s}{j}}.
\label{eq:stateprobability}
\ee
In order to show that this distribution on states defines a stable
equilibrium, we must show that the distribution is unchanged by the
collision operator.  From the block diagonal property
(\ref{eq:blocdiag}) of the transition matrix, and the property of
semi-detailed balance, it is clear that it will suffice to prove that
the probability (\ref{eq:stateprobability}) is dependent only on the
equivalence class of $s$, and therefore on the quantities
$\qrowd\supscrpt{s}{i}$ for $\mu=1,\ldots,n_c$.  We have
\[
P_0 (s) =
 \frac{\exp\left(-\sum_{\mu=1}^{n_c}
 \alpsubp{\mu}\sum_{i}\qrowd (1-s^i)\right)}
 {\prod_{i}\left[1+\exp\left(-\sum_{\mu=1}^{n_c}
 \alpsubp{\mu}\qrowd\right)\right]}.
\]
This expression indeed only depends upon the equivalence class of the
state $s$, and therefore we have proven that the distribution on
states defined by Eq.~(\ref{eq:fd}) is a stable equilibrium.
$\Box$

The equilibrium distribution given by Eq.~(\ref{eq:fd}) is an
$n_c$-pa\-ram\-e\-ter family of solutions for the $\nnsupo{i}{0}$.  By
taking the parameters $\alpsubp{\mu}$ to be spatially dependent, we can
construct a family of Boltzmann equilibria which are not equilibria in
the more general sense, but which are still stable under the collision
operator.  These are precisely the type of spatially varying equilibria
which we desire for the lowest order means $\nnsupo{i}{0}$ of our
lattice gas.  Thus, in what follows we restrict our attention to lattice
gases that obey semi-detailed balance at lowest order.

Summarizing the constraints on the collision operator at each order,
we have the following: $C_0$ must respect the conservation laws and
obey semi-detailed balance; $C_1$ must respect the conservation laws,
but may violate semi-detailed balance; $C_2$ can violate either the
conservation laws or semi-detailed balance.

Finally, we note that since the parameters, $\alpsubp{\mu}$, are
arbitrary multipliers, any set of $n_c$ independent functions of them
would also suffice to parametrize the equilibrium.  In particular, a
natural and logical choice of parameters are the hydrodynamic
densities, $\qqsupp{\mu}$.  These can be related to the
$\alpsubp{\mu}$ by their definition (see Eq.~(\ref{eq:eecons})),
\bge
   \begin{array}{ll}
   \qqsupp{\mu}
   =
   \qrowd\nnsupo{i}{0}
   \left(\alpsubp{\ast}\right) &
   \mbox{for $\mu=1,\ldots,n_c$.}
   \end{array}
   \label{eq:eeconz}
\ee
Thus, the equilibrium distribution, $\nnsupo{i}{0}$ can be parametrized
solely by the equilibrium values of the $n_c$ conserved densities.

\subsection{The Fermi Metric}

In what follows, we shall need the first two derivatives of the
$\nnsupo{i}{0}$ with respect to the $\qqsupp{\mu}$, so we compute them
here by the chain rule.  First, by differentiating Eq.~(\ref{eq:fd})
with respect to $\alpsubp{\nu}$, we obtain
\bge
   \frac{\pdv\nnsupo{i}{0}}{\pdv\alpsubp{\nu}} =
   \nnsupo{i}{0}\left(1-\nnsupo{i}{0}\right)\qrow{\nu}{i}
   \label{eq:crregv}
\ee
(where there is no summation over $i$ because it appears only once on
the left).  Next, by differentiating Eq.~(\ref{eq:eeconz}) with respect
to $\qqsupp{\nu}$, we obtain
\[
   \mixten{\delta}{\mu}{\nu}
   = \sum_{\xi=1}^{n_c}
   \qrow{\mu}{j}
   \frac{\pdv\nnsupo{j}{0}}{\pdv\alpsubp{\xi}}
   \frac{\pdv\alpsubp{\xi}}{\pdv\qqsupp{\nu}}
   = \sum_{\xi=1}^{n_c}
   \supscrpt{g}{\mu\xi}
   \frac{\pdv\alpsubp{\xi}}{\pdv\qqsupp{\nu}},
\]
where we have defined the symmetric rank-two tensor,
\bge
   \supscrpt{g}{\mu\xi}\equiv
   \nnsupo{j}{0}\left(1-\nnsupo{j}{0}\right)\qrow{\mu}{j}\qrow{\xi}{j}.
   \label{eq:ffttwo}
\ee
We denote the inverse of this matrix by $\subscrpt{g}{\xi\nu}$ so that
\[
   \sum_{\xi=1}^{n_c}\supscrpt{g}{\mu\xi}\subscrpt{g}{\xi\nu} =
   \mixten{\delta}{\mu}{\nu}.
\]
Since $\subscrpt{g}{\xi\nu}$ is a symmetric second-rank tensor, we can
identify it as a metric on the space of hydrodynamic variables,
$\qqsupp{\nu}$.  We call it the {\sl Fermi metric}.  \label{pg:aw} In
terms of the Fermi metric, we have
\[
   \frac{\pdv\alpsubp{\xi}}{\pdv\qqsupp{\nu}} =
   \subscrpt{g}{\xi\nu}.
\]
Finally,  we can write
\bge
   \frac{\pdv\nnsupo{i}{0}}{\pdv\qqsupp{\mu}} =
   \sum_{\nu=1}^{n_c}
   \nnsupo{i}{0}\left(1-\nnsupo{i}{0}\right)\qrow{\nu}{i}
   \subscrpt{g}{\nu\mu}
   \label{eq:fdd}
\ee
(no sum on $i$).

In similar fashion, we compute the second derivative
\bge
   \frac{\pdv^2\nnsupo{i}{0}}{\pdv\qqsupp{\mu}\pdv\qqsupp{\nu}} =
   \nnsupo{i}{0}(1-\nnsupo{i}{0})(1-2\nnsupo{i}{0})
   \qrow{\xi}{i}\qrow{\eta}{i}
   \subscrpt{g}{\xi\mu}\subscrpt{g}{\eta\nu} +
   2\nnsupo{i}{0}(1-\nnsupo{i}{0})
   \qrow{\xi}{i}
   \subscrpt{g}{\xi\eta}\mixten{\Gamma}{\eta}{\mu\nu},
   \label{eq:fddd}
\ee
where we have defined the {\sl Fermi connection}, \label{pg:af}
\bgea
\mixten{\Gamma}{\eta}{\mu\nu}
   & = &
   \frac{1}{2}\supscrpt{g}{\eta\xi}
   \left(\frac{\pdv\subscrpt{g}{\xi\mu}}{\pdv\qqsupp{\nu}} +
         \frac{\pdv\subscrpt{g}{\xi\nu}}{\pdv\qqsupp{\mu}} -
         \frac{\pdv\subscrpt{g}{\mu\nu}}{\pdv\qqsupp{\xi}}\right)
   \nonumber \\
   & = &
  -\frac{1}{2}\subscrpt{g}{\mu\xi}\subscrpt{g}{\nu\zeta}
   \nnsupo{j}{0}\left(1-\nnsupo{j}{0}\right)
   \left(1-2\nnsupo{j}{0}\right)
   \qrow{\eta}{j}\qrow{\xi}{j}\qrow{\zeta}{j}.
   \label{eq:fftthree}
\eea

Next, we introduce a characteristic lattice spacing $c$ to define the
dimensionless lattice vectors, \label{pg:zz}
\[
   \bfesup{i}\equiv\frac{\bfci}{c},
\]
and consider the completely symmetric outer product of $k$ of these
vectors, $\bigotimes^k\bfesup{i}$.  It will be useful to include these
outer products in the above sums.  Thus, we define the {\sl generalized
Fermi metric},
\bge
   \supscrpt{{\bf g}(k)}{\mu\xi}\equiv
   \nnsupo{j}{0}\left(1-\nnsupo{j}{0}\right)
   \qrow{\mu}{j}\qrow{\xi}{j}
   \left(\bigotimes^k\bfesup{j}\right),
   \label{eq:gffttwo}
\ee
and the {\sl generalized Fermi connection} \label{pg:ag},
\bge
   \mixten{\bfgamma (k)}{\eta}{\mu\nu} \equiv
   -\frac{1}{2}\subscrpt{g}{\mu\xi}\subscrpt{g}{\nu\zeta}
   \nnsupo{j}{0}\left(1-\nnsupo{j}{0}\right)
   \left(1-2\nnsupo{j}{0}\right)
   \qrow{\eta}{j}\qrow{\xi}{j}\qrow{\zeta}{j}
   \left(\bigotimes^k\bfesup{j}\right).
   \label{eq:gfftthree}
\ee
Note that $\supscrpt{{\bf g}(0)}{\mu\xi}=\supscrpt{g}{\mu\xi}$ and
$\mixten{\bfgamma (0)}{\eta}{\mu\nu}=\mixten{\Gamma}{\eta}{\mu\nu}$.
Once again, to raise and lower the indices of these objects, we use the
Fermi metric, $\subscrpt{g}{\nu\mu}$, as a metric tensor; thus, e.g.,
$\mixten{{\bf g}(2)}{\xi}{\nu} = \supscrpt{{\bf
g}(2)}{\xi\mu}\subscrpt{g}{\mu\nu}$.

In passing, we note that since the $\bfesup{j}$ are obviously
independent of the $\qqsupp{\ast}$, we have
\[
\mixten{\bfgamma (k)}{\eta}{\mu\nu} =
   \frac{1}{2}\supscrpt{g}{\eta\xi}
   \left(\frac{\pdv\subscrpt{{\bf g}(k)}{\xi\mu}}{\pdv\qqsupp{\nu}} +
         \frac{\pdv\subscrpt{{\bf g}(k)}{\xi\nu}}{\pdv\qqsupp{\mu}} -
         \frac{\pdv\subscrpt{{\bf g}(k)}{\mu\nu}}{\pdv\qqsupp{\xi}}\right)
\]
for all $k$.  That is, we have defined a set of metrics and their
associated connections.  For each $k$, the members of this set comprise
a completely symmetric tensor of rank $k$.  As we shall show, this
structure is very useful for the problem at hand.

{}From Eqs.~(\ref{eq:fdd}) and (\ref{eq:fddd}), we get
\[
   \frac{\pdv}{\pdv\qqsupp{\mu}}
   \left[
     \qrow{\eta}{i}\nnsupo{i}{0}
     \left(\bigotimes^k\bfesup{i}\right)
   \right] =
   \mixten{{\bf g}(k)}{\eta}{\mu},
\]
and
\[
   \frac{\pdv^2}{\pdv\qqsupp{\mu}\pdv\qqsupp{\nu}}
   \left[
     \qrow{\eta}{i}\nnsupo{i}{0}
     \left(\bigotimes^k\bfesup{i}\right)
   \right] = 2\left[
   \mixten{{\bf g}(k)}{\eta}{\xi}
   \mixten{\Gamma}{\xi}{\mu\nu} -
   \mixten{\bfgamma (k)}{\eta}{\mu\nu}
   \right].
\]
Note that when $k=0$ these reduce to the identities $\pdv\qqsupp{\eta}
/\pdv\qqsupp{\mu} = \mixten{\delta}{\eta}{\mu}$ and
$\pdv^2\qqsupp{\eta} / \pdv\qqsupp{\mu}\pdv\qqsupp{\nu} = 0$,
respectively.

\subsection{Zero-Order Conservation Equations}

We can now examine the conservation equation, Eq.~(\ref{eq:lconslaws}),
at ${\cal O}(1)$.  We have
\bge
   \begin{array}{ll}
   \nabla\cdot
   \left(
   \qrowd\frac{\bfci}{\dt}\nnsupo{i}{0}
   \right) = 0 &
   \mbox{for $\mu=1,\ldots,n_c$.}
   \end{array}
   \label{eq:conszero}
\ee
Using Eq.~(\ref{eq:fdd}), this can be written solely in terms of the
conserved densities and their gradients.  We find
\[
   \begin{array}{ll}
   \frac{c}{\dt}
   \sum_{\nu=1}^{n_c} \mixten{{\bf g}(1)}{\mu}{\nu}
   \cdot\nabla\qqsupp{\nu} = 0 &
   \mbox{for $\mu=1,\ldots,n_c$.}
   \end{array}
\]

\subsection{The Linearized Boltzmann Equation}
\label{ssec:linearboltzmann}

We return to Eqs.~(\ref{eq:lboltza}) and (\ref{eq:ordering}).  At ${\cal
O}(\epsilon)$ we find
\bge
   \nabla\cdot\left(\frac{\bfci}{\dt} \nnsupo{i}{0}\right) =
   \frac{1}{\dt}\left[
   \jij \nnsupo{j}{1} + \ccsupo{i}{1}(\nnsupo{\ast}{0})
   \right],
   \label{eq:eqfornone}
\ee
where we have defined the Jacobian matrix of the lowest-order collision
operator at equilibrium, \label{pg:az}
\[
   \jij \equiv
   \left.\frac{\pdv \ccsupo{i}{0}}{\pdv \nnsup{j}}
   \right|_{N = \subscrpt{N}{0}}.
\]
By differentiating Eq.~(\ref{eq:bcollop}), we can write this directly in
terms of the lowest-order transition matrix\footnote{That is, the
transition matrix corresponding to the lowest-order collision operator,
$C_0$.}, $A_0(s\rightarrow s')$,
\bge
 \jij = \sum_{s,s'} A_0(s\rightarrow s')
   (\supscrpt{s}{\prime i} - \supscrpt{s}{i})
   (2\supscrpt{s}{k} - 1)
   \prod_{k\neq j}^n
   \left(\nnsupo{k}{0}\right)^{\supscrpt{s}{k}}
   \left(1-\nnsupo{k}{0}\right)^{1-\supscrpt{s}{k}}.
   \label{eq:jacexp}
\ee
Note that the $\qrowd$'s comprise the components of $n_c$ null left
eigenvectors of $\jij$, since
\[
   \qrowd\jij
   =  \left.\frac{\pdv}{\pdv \nnsup{j}}
         \left(\qrowd \ccsupo{i}{0}\right)
         \right|_{N = \subscrpt{N}{0}}= 0.
\]

In what follows, we denote the eigenvalues of $J$ by $\lamsupp{\mu}$
\label{pg:aj}.
The corresponding right (left) eigenvectors are indexed as contravariant
(covariant) vectors, and enumerated with a subscript (superscript).
\label{pg:bn} Thus,
\bge
   \jij \qcold = \lamsupp{\mu} \qcol{\mu}{i}
   \label{eq:eigcol}
\ee
and
\bge
   \qrowd \jij = \lamsupp{\mu} \qrow{\mu}{j}.
   \label{eq:eigrow}
\ee
Note that, for $\mu = 1,\ldots,n_c$, this coincides with the definition
of the $\qrowd$ introduced in Eq.~(\ref{eq:consquan}).

The modes enumerated $1,\ldots,n_c$, correspond to null eigenvalues of
$J$, and will be called {\sl hydrodynamic modes}; those modes enumerated
$n_c+1,\ldots,n$ will be called {\sl kinetic modes}.  We shall often
write $H$ for the set $\{1,\ldots,n_c\}$, and $K$ for the set
$\{n_c+1,\ldots,n\}$.  \label{pg:ax} For lattice gases of interest, the
kinetic eigenvalues satisfy $\lamsupp{\nu}<0$; in this case we define
the system to be {\sl linearly stable}.  Since the eigenvalues of $J$
set the time scale for the equation, we see that for a linearly stable
lattice gas, the kinetic modes decay away rapidly, while the
hydrodynamic modes persist for long times.
For the remainder of this paper we restrict attention to lattice
equilibria which are linearly stable.

Postmultiplying Eq.~(\ref{eq:eigrow}) by $\qcold$, premultiplying
Eq.~(\ref{eq:eigcol}) by $\qrowd$, and subtracting, we get
\[
   0 = (\lamsupp{\beta} - \lamsupp{\eta})
       \qrow{\eta}{i} \qcol{\beta}{i},
\]
so that right and left eigenvectors corresponding to different
eigenvalues are orthogonal.  Thus, they may be chosen so that
\bge
   \mixten{\delta}{\mu}{\nu} = \qrow{\mu}{i}\qcol{\nu}{i}.
   \label{eq:ortho}
\ee
By including $\bigotimes^k\bfesup{i}$ in Eq.~(\ref{eq:ortho}), we can
define a {\sl generalized Kronecker delta} in the same spirit that we
generalized the Fermi metric and connection.  \label{pg:ay} Thus,
\bge
   \mixten{\bfdelta (k)}{\mu}{\nu}
   \equiv
   \qrow{\mu}{j}
   \left(
   \bigotimes^k \bfesup{j}
   \right)
   \qcol{\nu}{j},
   \label{eq:hten}
\ee
so that $\mixten{\bfdelta (0)}{\mu}{\nu}=\mixten{\delta}{\mu}{\nu}$.
Also, note that the indices of the generalized Fermi metric and
connection, as well as those of the generalized Kronecker delta, can now
be extended to run over the kinetic modes as well as the hydrodynamic
modes, by simply using the kinetic left eigenvectors in
Eqs.~(\ref{eq:gffttwo}), (\ref{eq:gfftthree}) and (\ref{eq:hten}),
respectively.

Finally, we note that it is possible to write an explicit, closed
expression for the right hydrodynamic eigenvectors.  From
Eqs.~(\ref{eq:zorder}), (\ref{eq:fd}), and (\ref{eq:crregv}), it follows
immediately that
\[
   0 =
   \frac{\partial}{ \partial \alpha_\mu}  C_0^i (\nnsupo{\ast}{0}) =
   \jij \frac{\partial N^j_0}{\partial \alpha_\mu} =
   \jij \left[q^{\mu}_{j} N^j_0 (1 - N^j_0)\right],
\]
whence we identify the right hydrodynamic eigenvectors,
\[
   q^i_{\mu} = q^\mu_i N^i_0 (1 -N^i_0).
\]

\subsection{First-Order Solution}

Consider Eq.~(\ref{eq:eqfornone}) for the $\nnsupo{i}{1}$.  Since $J$ is
a singular matrix, we must verify that the equation is consistent.  The
consistency requirement is found by premultiplying the equation by the
null left eigenvectors.  We obtain the requirement
\[
   \nabla\cdot\left(\qrowd\frac{\bfci}{\dt}\nnsupo{i}{0}\right)
   = \frac{1}{\dt}\qrowd\left(\jij\nnsupo{j}{1}
                              + \ccsupo{i}{1}(\nnsupo{\ast}{0})
                          \right)=0
\]
for $\mu\in H$.  Note that this consistency requirement is precisely the
zeroth-order conservation equation, Eq.~(\ref{eq:conszero}).  It follows
that $\nabla\cdot\left(\bfci\nnsupo{i}{0}\right) -
\ccsupo{i}{1}(\nnsupo{\ast}{0})$ has no components in the null space of
$J$, since we have assumed linear stability.  Thanks to the
completeness and orthonormality of the eigenvectors, this expression
can be written as
\[
   \nabla\cdot\left(\bfci\nnsupo{i}{0}\right) -
   \ccsupo{i}{1}(\nnsupo{\ast}{0})
   = \sum_{\nu\in K} \etasupp{\nu} \qcol{\nu}{i},
\]
where
\[
   \begin{array}{ll}
   \etasupp{\nu} \equiv \qrow{\nu}{j}
   \left[
   \nabla\cdot\left(\bfcsup{j} \nnsupo{j}{0}\right) -
                    \ccsupo{j}{1}(\nnsupo{\ast}{0})
   \right] &
   \mbox{for $\nu\in K$.}
   \end{array}
\]
The solution for $\nnsupo{i}{1}$ can then be written down immediately
\[
   \nnsupo{i}{1} =
   \sum_{\nu\in H}\thetasupp{\nu} \qcol{\nu}{i} +
   \sum_{\nu\in K}\frac{\etasupp{\nu}}{\lamsupp{\nu}}\qcol{\nu}{i},
\]
where the $\thetasupp{\nu}$ are arbitrary.  We fix the solution by
assuming, without loss of generality, that $\thetasupp{\nu}=0$.  That
is, we assume that the first order contribution does not affect the
definitions of the conserved densities.  The final result for the
$\nnsupo{i}{1}$ is then
\bge
   \nnsupo{i}{1}
   = \sum_{\nu\in K}
         \frac{\qcol{\nu}{i}\qrow{\nu}{j}}{\lamsupp{\nu}}
         \left[
         \nabla\cdot\left(\bfcsup{j} \nnsupo{j}{0}\right) -
         \ccsupo{j}{1}(\nnsupo{\ast}{0})
         \right],
   \label{eq:none}
\ee
so that we have
\bge
   \supscrpt{N}{i} =
   \nnsupo{i}{0} + \epsilon
   \sum_{\nu\in K}
         \frac{\qcol{\nu}{i}\qrow{\nu}{j}}{\lamsupp{\nu}}
         \left[
         \nabla\cdot\left(\bfcsup{j} \nnsupo{j}{0}\right) -
         \ccsupo{j}{1}(\nnsupo{\ast}{0})
         \right] +
   {\cal O}(\epsilon^2).
   \label{eq:cebde}
\ee

\subsection{First-Order Conservation Equations}
\label{ssec:foce}

We now write the conservation equation, Eq.~(\ref{eq:lconslaws}),
retaining terms to ${\cal O}(\epsilon)$.  We get
\[
   \begin{array}{ll}
     \frac{\pdv\qqsupp{\mu}}{\pdv t}
   + \nabla\cdot\left[
     \left(\qrowd
     \frac{\bfci}{\dt}
     \nnsupo{i}{1}\right)
   + \nabla\cdot\left(\qrowd
     \frac{\bfci\bfci}{2\dt}
     \nnsupo{i}{0}\right)\right] =
     \frac{1}{\dt}\qrowd\ccsupo{i}{2} (\nnsupo{\ast}{0}) &
   \mbox{for $\mu\in H$.}
   \end{array}
\]
Substituting Eq.~(\ref{eq:none}) for the $\nnsupo{i}{1}$, we get, after
some manipulation, the following closed set of equations for the
conserved densities:
\bge
   \begin{array}{ll}
   \frac{\pdv\qqsupp{\mu}}{\pdv t} +
   \nabla\cdot\advco{\mu} =
   \sum_{\xi\in H}
   \nabla\cdot
   \left(
   \cdop{\mu}{\xi}\cdot\nabla\qqsupp{\xi}
   \right) +
   \supscrpt{\cal S}{\mu} &
   \mbox{for $\mu\in H$,}
   \end{array}
   \label{eq:hyd}
\ee
where the advection, diffusion and source coefficients are given by
\label{pg:ao}
\bge
   \advco{\mu}(\qqsupp{\ast})\equiv
   \frac{c}{\dt}\sum_{\nu\in K}
   \frac{\mixten{\bfdelta (1)}{\mu}{\nu}
         \supsub{\cal C}{\nu}{1}
         \left(\qqsupp{\ast}\right)}
        {(-\lamsupp{\nu})},
   \label{eq:hydadv}
\ee
\bge
   \cdop{\mu}{\xi} (\qqsupp{\ast}) \equiv
   \frac{c^2}{\dt}
     \left[
     \sum_{\nu\in K}
     \frac{\mixten{\bfdelta (1)}{\mu}{\nu}
           \otimes
           \mixten{{\bf g}(1)}{\nu}{\xi}}
          {(-\lamsupp{\nu})}
   - \frac{1}{2}\mixten{{\bf g}(2)}{\mu}{\xi}
     \right],
   \label{eq:hyddif}
\ee
\bge
   \supscrpt{\cal S}{\mu} (\qqsupp{\ast}) \equiv
   \frac{1}{\dt}\supsub{\cal C}{\mu}{2}
   (\qqsupp{\ast}),
   \label{eq:hydsour}
\ee
respectively, and where
\bge
   \begin{array}{ll}
   \supsub{\cal C}{\mu}{n}(\qqsupp{\ast}) \equiv
   \qrow{\mu}{i}\ccsupo{i}{n}
   \left(\nnsupo{\ast}{0}(\qqsupp{\ast})\right) &
   \mbox{for $n=1,2$.}
   \end{array}
   \label{eq:calcdef}
\ee
The compact form of this result makes it straightforward to compute
the transport coefficients of any lattice gas.  Note that knowledge of
the conserved quantities and the concomitant hydrodynamic modes is
sufficient to predict the form of this equation.

\subsection{Ordering the Conserved Quantities}

There is one additional technicality that we need to discuss before
presenting examples of this formalism.  In many situations of interest,
the conserved quantities themselves are ordered in the expansion
parameter, $\epsilon$.  For example, in an incompressible fluid, the
hydrodynamic density is assumed to vary by ${\cal O}(\epsilon^2)$ from a
constant background value, and the hydrodynamic velocity is assumed to
be ${\cal O}(\epsilon)$ (low Mach number).

Consider the general ordering,
\bge
\qqsupp{\mu} = \qqsuppo{\mu}{0} +
               \epsilon\qqsuppo{\mu}{1} +
               \epsilon^2\qqsuppo{\mu}{2} + \cdots,
\label{eq:consord}
\ee
where it is assumed that the zero-order value, $\qqsuppo{\mu}{0}$, is
always independent of position and time.  We can then expand the
Fermi-Dirac equilibrium as follows:
\bgeas
\lefteqn{\nnsupo{i}{0}(\qqsupp{\ast})} \\
  & = &
  \nnsupo{i}{0}(\qqsuppo{\ast}{0}) +
  \sum_{\mu\in H}
  \left.\frac{\pdv\nnsupo{i}{0}}{\pdv\qqsupp{\mu}}\right|_0
  \left(\epsilon\qqsuppo{\mu}{1} + \epsilon^2\qqsuppo{\mu}{2}\right) \\
  & & + \frac{1}{2}\sum_{\mu , \nu\in H}
  \left.\frac{\pdv\nnsupo{i}{0}}{\pdv\qqsupp{\mu}\pdv\qqsupp{\nu}}\right|_0
  \left(\epsilon^2\qqsuppo{\mu}{1}\qqsuppo{\nu}{1}\right) + \cdots \\
  & = &
  \nnsupo{i}{00} +
  \epsilon\sum_{\xi\in H}
  \nnsupo{i}{00} (1-\nnsupo{i}{00}) \qrow{\xi}{i}
  \subscrpt{g}{\xi\mu}\qqsuppo{\mu}{1}\\
  & & +\epsilon^2\Biggl[\sum_{\xi\in H}
  \nnsupo{i}{00} (1-\nnsupo{i}{00}) \qrow{\xi}{i}
  \subscrpt{g}{\xi\mu}\qqsuppo{\mu}{2} +
  \sum_{\xi, \eta\in H}
  \nnsupo{i}{00} (1-\nnsupo{i}{00}) \qrow{\xi}{i}\subscrpt{g}{\xi\eta}
  \mixten{\Gamma}{\eta}{\mu\nu}\qqsuppo{\mu}{1}\qqsuppo{\nu}{1}\\
  & & \;\;\;\;\;\;\;\;\; + \frac{1}{2}
  \sum_{\xi , \eta\in H}
  \nnsupo{i}{00} (1-\nnsupo{i}{00}) (1-2\nnsupo{i}{00})
  \qrow{\xi}{i}\qrow{\eta}{i}
  \subscrpt{g}{\xi\mu}\subscrpt{g}{\eta\nu}\qqsuppo{\mu}{1}\qqsuppo{\nu}{1}
  \Biggr] + \cdots,
\eeas
where we have defined the lowest-order Chapman-Enskog equilibrium,
\[
  \nnsupo{i}{00}\equiv\nnsupo{i}{0}(\qqsuppo{\ast}{0}),
\]
and we note that the  Fermi metrics and connections in the
above expression are now defined in terms of $\nnsupo{i}{00}$, rather
than $\nnsupo{i}{0}$.

To make it possible to incorporate this sort of ordering into the above
formalism, we return to the Chapman-Enskog solution for the Boltzmann
distribution, Eq.~(\ref{eq:cebde}), insert the above expansion of
$\nnsupo{i}{0}(\qqsupp{\ast})$, multiply by the $k$-fold outer product
of the $\bfei$ vectors and contract with $\qrowd$.  In the interests of
simplifying the algebra a bit, we shall assume that $\ccsupo{i}{1} = 0$
in this subsection; we do, however, continue to include $C_2$ in our
analysis.  The assumption that $C_1=0$ is not in any way essential, but
it serves to keep the algebra in check.  We get
\bgeas
\qrowd\left(\bigotimes^k\bfei\right)
       \nnsup{i}(\qqsupp{\ast})
 & = & \qrowd\left(\bigotimes^k\bfei\right)
       \nnsupo{i}{00} +
       \sum_{\xi\in H}
       \mixten{{\bf g}(k)}{\mu}{\xi}
       \left(
       \epsilon\qqsuppo{\xi}{1} + \epsilon^2\qqsuppo{\xi}{2}
       \right) \\
 & + & \epsilon^2\sum_{\xi, \eta\in H}
       \left[
       \sum_{\zeta\in H}
       \mixten{{\bf g}(k)}{\mu}{\zeta}\mixten{\Gamma}{\zeta}{\xi\eta} -
       \mixten{\bfgamma (k)}{\mu}{\xi\eta}
       \right]
       \qqsuppo{\xi}{1}\qqsuppo{\eta}{1} \\
 & + & \epsilon^2 c\sum_{\nu\in K}\sum_{\xi\in H}
       \frac{\mixten{\bfdelta (k)}{\mu}{\nu}
             \otimes
             \mixten{{\bf g}(1)}{\nu}{\xi}}
            {\lamsupp{\nu}}
       \cdot\nabla\qqsuppo{\xi}{1}.
\eeas
This result can be directly inserted in the conservation equation,
Eq.~(\ref{eq:lconslaws}), retaining terms to ${\cal O}(\epsilon^2)$.
After cancelling a factor of $\epsilon^2$ throughout the equation, we
get
\bge
   \begin{array}{ll}
   \frac{\pdv\qqsuppo{\mu}{1}}{\pdv t} +
   \nabla\cdot\advco{\mu} =
   \sum_{\xi\in H}
   \nabla\cdot
   \left(
   \cdop{\mu}{\xi}\cdot\nabla\qqsuppo{\xi}{1}
   \right)  +
   \supscrpt{\cal S}{\mu} &
   \mbox{for $\mu\in H$,}
   \end{array}
   \label{eq:hydocc}
\ee
where we have defined the advection coefficient,
\bgea
   \lefteqn{\advco{\mu} (\qqsuppo{\ast}{0}) \equiv
   \frac{c}{\dt}\sum_{\xi\in H}
   \Biggl\{
   \mixten{{\bf g}(1)}{\mu}{\xi}
   \left(
   \frac{1}{\epsilon}\qqsuppo{\xi}{1} + \qqsuppo{\xi}{2}
   \right) +} \nonumber \\
   & &
   \sum_{\eta\in H}
   \left[
   \sum_{\zeta\in H}
   \mixten{{\bf g}(1)}{\mu}{\zeta}\mixten{\Gamma}{\zeta}{\xi\eta} -
   \mixten{\bfgamma (1)}{\mu}{\xi\eta}
   \right]
   \qqsuppo{\xi}{1}\qqsuppo{\eta}{1}
   \Biggr\},
   \label{eq:hydoccadv}
\eea
where the diffusion coefficient and source term are still given by
Eqs.~(\ref{eq:hyddif}) and (\ref{eq:hydsour}), respectively, and where
all quantities are now understood to be evaluated at $\nnsupo{i}{00}$.
Note that the diffusion coefficient is now a function only of
$\qqsuppo{\ast}{0}$, and hence strictly independent of space and time.
The advection coefficient, on the other hand, is generally quadratic in
$\subscrpt{Q}{1}$ and linear in $\subscrpt{Q}{2}$.

Also note that the advection operator has an ${\cal O}(1/\epsilon)$ term
and an ${\cal O}(1)$ term.  If the ${\cal O}(1/\epsilon)$ term does not
vanish, it is the dominant term in the equation.  In this situation, the
hydrodynamic equation reduces to the zero-order conservation equation,
\bge
   \nabla\cdot
   \left[
   \frac{c}{\dt}
   \sum_{\xi\in H}
   \mixten{{\bf g}(1)}{\mu}{\xi}\qqsuppo{\xi}{1}
   \right] = 0.
   \label{eq:hydocceps}
\ee

Finally, note that knowledge of the conservation laws and concomitant
hydrodynamic modes is sufficient to predict the form of this equation
{\sl and} compute the advection coefficient.  Only the computation of
the diffusion coefficient requires knowledge of the kinetic modes in
this ordering scheme.

\newpage
\section{Examples of the Boltzmann / \newline Chapman-Enskog Analysis}
\setcounter{equation}{0}
\label{sec:exboce}

In this section, we present four examples of the Boltzmann /
Chapman-Enskog formalism described in the previous two sections.  In
each case, we work out the form of the hydrodynamic equation, and the
transport coefficient(s) predicted by the theory.

The first example is a one-dimensional, stochastic, diffusive lattice
gas (1D3P).  The second is a lattice gas for Burgers'
equation~\cite{bmbcdl} in one dimension.  The third is a two-dimensional
lattice gas (2D4P) giving rise to coupled diffusion equations.  Finally,
we consider lattice fluid models; we consider a general class of such
models, and  describe in detail a particular model, the FHP-I lattice
gas fluid.

Later in this paper, after we present the complete lattice gas kinetic
theory, we shall return to these examples, discuss experimentally
observed discrepencies from the Boltzmann theory, and show how they
are explained by the new theory.  As we shall see, each example has
its own unique and interesting features in this regard.

\subsection{The 1D3P Lattice Gas}
\label{ssec:odtp}

As a first example, we consider a diffusive lattice gas model in one
dimension ($D=1$).  \label{pg:zy} The model has three bits per site
($n=3$), corresponding to the presence or absence of left-moving,
stationary, and right-moving particles, respectively.  These bits are
denoted by the respective elements of the set $B=\{-,0,+\}$.  Collisions
occur only if exactly two particles enter a site.  If we denote the
two-particle states by $\widehat{+}\equiv\{-,0\}$,
$\widehat{0}\equiv\{-,+\}$, and $\widehat{+}\equiv\{0,-\}$, then the
nontrivial elements of the state transition table can be written
\bgc
\begin{tabular}{|c|c||c|c|c|}
\hline
\multicolumn{2}{|c||}{$a(s\rightarrow s')$} & \multicolumn{3}{c|}{$s'$} \\
\cline{3-5}
\multicolumn{2}{|c||}{} & $\widehat{+}$ &
\raisebox{-1.8pt}{$\widehat{0}$} & \raisebox{-1.8pt}{$\widehat{-}$}\\
\hline
\hline
    & $\widehat{+}$ & $1-\nsup{p}$ & $\nsup{p}(1-\nsup{r})$ &
$\nsup{p}\nsup{r}$ \\
\cline{2-5}
$s$ & \raisebox{-1.8pt}{$\widehat{0}$} & $\nsup{p}\nsup{r}$ &
$1-\nsup{p}$ & $\nsup{p}(1-\nsup{r})$ \\
\cline{2-5}
    & \raisebox{-1.8pt}{$\widehat{-}$}& $\nsup{p}(1-\nsup{r})$ &
$\nsup{p}\nsup{r}$ & $1-\nsup{p}$ \\
\hline
\end{tabular}
\ec
The bits $n^p$ and $n^r$ are random bits\footnote{Note that $r$ and $p$
are not indices here, but simply labels for the random bits.}
\label{pg:bg} which are sampled separately at each lattice site and at
each timestep with average values $\langle\nsup{p}\rangle = 2p$ and
$\langle\nsup{r}\rangle = 1/2$.  Here, the parameter $p\in
[0,\frac{1}{2}]$ may be thought of as the probability of collision from,
e.g., $\widehat{+}$ to $\widehat{0}$.  The value of the bit $\nsup{p}$
effectively determines whether or not a collision will occur, and that
of $\nsup{r}$ determines which of the two possible outcomes will result.

Note that these collisions obey semi-detailed balance, since the columns
of the above table sum to unity.  Also note that they conserve particles
($n_c=1$); it is the particle density that will obey the macroscopic
diffusion equation.  The coefficients $\qrow{1}{\ast}$ for the conserved
quantity are
\bge
   \qrow{1}{-} = \qrow{1}{0} = \qrow{1}{+} = 1.
   \label{eq:odtpqrow}
\ee
The collision operator is given by Eq.~(\ref{eq:collop}),
\bge
\csup{i}(\nsup{\ast}) = \nsup{p} [\overline{\nsup{i}} \nsup{i+1} \nsup{i+2}
           - \overline{\nsup{r}} \nsup{i} \overline{\nsup{i+1}} \nsup{i+2}
           - \nsup{r} \nsup{i} \nsup{i+1} \overline{\nsup{i+2}}],
\label{eq:collision1}
\ee
for each $i\in B$, where $\overline{\nsup{i}}\equiv 1-\nsup{i}$
denotes the complement of a bit, and where the addition of integers to
$i$ is understood to increment $i$ through the set $B$ in cyclic
fashion.  According to Eq.~(\ref{eq:ccop}), we
observe that
\[
   0 = \csup{-} + \csup{0} + \csup{+}.
\]

We now consider the ensemble average of this system.
The ensemble-averaged state transition table is
\bgc
\begin{tabular}{|c|c||c|c|c|}
\hline
\multicolumn{2}{|c||}{$A(s\rightarrow s')$} & \multicolumn{3}{c|}{$s'$} \\
\cline{3-5}
\multicolumn{2}{|c||}{} & $\widehat{+}$ &
\raisebox{-1.8pt}{$\widehat{0}$} & \raisebox{-1.8pt}{$\widehat{-}$} \\
\hline
\hline
    & $\widehat{+}$ & $1-2p$ & $p$ & $p$ \\
\cline{2-5}
$s$ & \raisebox{-1.8pt}{$\widehat{0}$} & $p$ & $1-2p$ & $p$ \\
\cline{2-5}
    & \raisebox{-1.8pt}{$\widehat{-}$} & $p$ & $p$ & $1-2p$ \\
\hline
\end{tabular}
\ec
and the ensemble-averaged collision operator in the Boltzmann
approximation is given by Eq.~(\ref{eq:bcollop}) which reads
\[
\ccsup{i}(\nnsup{\ast}) =
               2p \nnsup{i+1} \nnsup{i+2}
              - p \nnsup{  i} \nnsup{i+1}
              - p \nnsup{  i} \nnsup{i+2}.
\]
We do not order the collision operator in this lattice gas, so that
$C^i = C^i_0$.  Using Eq.~(\ref{eq:odtpqrow}), we calculate the local
Fermi-Dirac equilibrium from Eq.~(\ref{eq:fd}),
\bge
   \nnsupo{-}{0} =
   \nnsupo{0}{0} =
   \nnsupo{+}{0} =
   \frac{1}{1+e^{-\alpha}} \equiv f.
   \label{eq:equilibrium1}
\ee
where $\alpha$ parametrizes the distribution function, and $f$ is the
mean occupation number.  Note that the total density is given in terms
of $\alpha$ (and $f$) by
\[
   Q = \frac{3}{1+e^{-\alpha}} = 3f,
\]
which is the analog of Eq.~(\ref{eq:eeconz}).

We now perform the Chapman-Enskog perturbative analysis.  The Jacobian
of the collision operator at equilibrium is
\[
J = \left(\begin{array}{ccc}
          -2pf & pf & pf \\
          pf & -2pf & pf \\
          pf & pf & -2pf
          \end{array}
    \right).
\]
This matrix has eigenvalues
\[
   \lamsupp{1} = 0
\]
and
\[
   \lamsupp{2} = \lamsupp{3} = -3pf,
\]
with corresponding left eigenvectors
\bge
\begin{array}{l}
   \qsupp{1}
   =     \left(\begin{array}{ccc} +1 & +1 & +1 \end{array}\right) \\
   \qsupp{2}
   =     \left(\begin{array}{ccc} -1 &  \phantom{+}0 & +1 \end{array}\right) \\
   \qsupp{3}
   =     \left(\begin{array}{ccc} -1 & +2 & -1 \end{array}\right),
\end{array}
\label{eq:odtprev}
\ee
and right eigenvectors
\[
\begin{array}{ccc}
   \qsubp{1}
   = \frac{1}{3}
 \left(\begin{array}{r} +1 \\ +1 \\ +1 \end{array}\right) &
   \qsubp{2}
   = \frac{1}{2}
 \left(\begin{array}{r} -1 \\  0 \\ +1 \end{array}\right) &
   \qsubp{3}
   = \frac{1}{6}
 \left(\begin{array}{r} -1 \\ +2 \\ -1 \end{array}\right).
\end{array}
\]
It is easy to check that these eigenvectors satisfy
Eq.~(\ref{eq:ortho}).

We may now construct the Fermi metric.  There is only one hydrodynamic
mode, so $\supscrpt{g}{\mu\nu}$ is a 1-by-1 matrix.  From
Eq.~(\ref{eq:ffttwo}) we see that its element is
\[
\supscrpt{g}{11} = 3f(1-f),
\]
and hence
\[
\subscrpt{g}{11} = \frac{1}{3f(1-f)}.
\]
To get the generalized Fermi metric, we write $\supscrpt{\bfc}{i} =
c\supscrpt{\bfe}{i}$ where
\[
\begin{array}{ccc}
  \supscrpt{\bfe}{-} = -\hat{\bfx}, &
  \supscrpt{\bfe}{0} = 0,           &
  \supscrpt{\bfe}{+} = +\hat{\bfx},
\end{array}
\]
with $\hat{\bfx}$ being a unit vector, and  $c$  the lattice
spacing.  Using Eq.~(\ref{eq:odtprev}) in Eq.~(\ref{eq:gffttwo}), it
follows that
\[
\begin{array}{ccc}
  \supscrpt{{\bf g}(1)}{11} = 0, &
  \supscrpt{{\bf g}(1)}{21} = 2f(1-f)\hat{\bfx}, &
  \supscrpt{{\bf g}(1)}{31} = 0,
\end{array}
\]
and
\[
\begin{array}{ccc}
  \supscrpt{{\bf g}(2)}{11} = +2f(1-f)\hat{\bfx}\hat{\bfx}, &
  \supscrpt{{\bf g}(2)}{21} = 0, &
  \supscrpt{{\bf g}(2)}{31} = -2f(1-f)\hat{\bfx}\hat{\bfx}.
\end{array}
\]
Using $\subscrpt{g}{11}$ to lower indices, we find
\[
\begin{array}{ccc}
  \mixten{{\bf g}(1)}{1}{1} = 0, &
  \mixten{{\bf g}(1)}{2}{1} = \frac{2}{3}\hat{\bfx}, &
  \mixten{{\bf g}(1)}{3}{1} = 0
\end{array}
\]
and
\[
\begin{array}{ccc}
  \mixten{{\bf g}(2)}{1}{1} = +\frac{2}{3}\hat{\bfx}\hat{\bfx}, &
  \mixten{{\bf g}(2)}{2}{1} = 0, &
  \mixten{{\bf g}(2)}{3}{1} = -\frac{2}{3}\hat{\bfx}\hat{\bfx}.
\end{array}
\]
Finally, the generalized Kronecker delta, $\mixten{\bfdelta
(1)}{\mu}{\nu}$, given by Eq.~(\ref{eq:hten}), has components
\[
\begin{array}{cc}
  \mixten{\bfdelta (1)}{1}{2} = \hat{\bfx}, &
  \mixten{\bfdelta (1)}{1}{3} = 0.
\end{array}
\]

The conserved quantity, $Q$, is not ordered, so we can use the results
of Subsection~\ref{ssec:foce}.  The collision operator was not
ordered, so $\supsub{\cal C}{\nu}{1}=\supsub{\cal C}{\nu}{2} = 0$;
Eq.~(\ref{eq:hyd}) thus tells us that there is no advection or source
term in the final hydrodynamic equation.  The diffusivity is given by
Eq.~(\ref{eq:hyddif}),
\bgeas
\cdop{1}{1} & = & \frac{c^2}{\dt}
                  \left[
                  \sum_{\nu\in K}
                  \frac{\mixten{\bfdelta (1)}{1}{\nu}
                        \otimes
                        \mixten{{\bf g}(1)}{\nu}{1}}
                       {(-\lamsupp{\nu})}
                - \frac{1}{2}\mixten{{\bf g}(2)}{1}{1}
                  \right] \\
            & = & \frac{c^2}{3\dt}\left( \frac{2}{-\lambda^{2}} -1\right)
                  \hat{\bfx}\hat{\bfx} \\
            & = & {\cal D} \hat{\bfx}\hat{\bfx}.
\eeas
Writing $\nabla = \hat{\bfx}\frac{\pdv}{\pdv x}$, Eq.~(\ref{eq:hyd})
gives us the hydrodynamic equation,
\[
  \frac{\pdv f}{\pdv t} =
  \frac{\pdv}{\pdv x}
    \left(
       {\cal D} \frac{\pdv f}{\pdv x}
    \right),
\]
where the scalar diffusivity is
\[
  {\cal D}\equiv\frac{c^2}{3\dt}\left( \frac{2}{-\lambda^{2}} -1\right)
  =\frac{c^2}{3\dt}\left( \frac{2}{3pf} -1\right).
\]
Since $p$ lies in $[0,\frac{1}{2}]$ and $f$ lies in $[0,1]$, it follows
that ${\cal D}$ is always positive.

\subsection{The Burgers' Equation Lattice Gas}
\label{ssec:belg}

The next example is a lattice gas for the one-dimensional ($D=1$)
Burgers' equation~\cite{bmbcdl}.  This model has two bits per site
($n=2$) corresponding to particles moving left and right.  The bits are
denoted by the elements of the set $B=\{-,+\}$.  Collisions occur only
when exactly one particle enters a site, from either direction.  The
result of a collision is then one particle leaving to the left (state
$\{ - \}$) with probability $(1-a)/2$, or to the right (state $\{ +
\}$) with probability $(1+a)/2$, regardless of the direction of the
incoming particle.

The state transition table is thus given by
\bgc
\begin{tabular}{|c|c||c|c|}
\hline
\multicolumn{2}{|c||}{$a(s\rightarrow s')$} & \multicolumn{2}{c|}{$s'$} \\
\cline{3-4}
\multicolumn{2}{|c||}{} & $-$ & $+$ \\
\hline
\hline
 $ s $ & $-$ & $1-\nsup{r}$ & $\nsup{r}$ \\
\cline{2-4}
    & $+$ & $1-\nsup{r}$ & $\nsup{r}$ \\
\hline
\end{tabular}
\ec
where $\nsup{r}$ is a random bit with mean $\langle n^r \rangle
=(1+a)/2$.  Note that particles are conserved, so
\[
   \qrow{1}{-} = \qrow{1}{+} = 1.
\]
The collision operator is given by Eq.~(\ref{eq:collop}),
\[
\csup{\pm}(\nsup{\ast}) =
   \pm\left[\nsup{r} \overline{\nsup{+}} \nsup{-}
         -  \overline{\nsup{r}} \nsup{+} \overline{\nsup{-}} \right].
\]
We observe that, in accordance with Eq.~(\ref{eq:ccop}),
\[
   0 = \csup{-} + \csup{+}.
\]
The ensemble-averaged state transition table is
\bgc
\begin{tabular}{|c|c||c|c|}
\hline
\multicolumn{2}{|c||}{$A(s\rightarrow s')$} & \multicolumn{2}{c|}{$s'$} \\
\cline{3-4}
\multicolumn{2}{|c||}{} & $-$ & $+$ \\
\hline
\hline
  s & $-$ & $\frac{1-a}{2}$ & $\frac{1+a}{2}$ \\
\cline{2-4}
    & $+$ & $\frac{1-a}{2}$ & $\frac{1+a}{2}$ \\
\hline
\end{tabular}
\ec
and the ensemble-averaged collision operator in the Boltzmann
approximation is given by Eq.~(\ref{eq:bcollop}), which reads
\[
\ccsup{\pm}(\nnsup{\ast}) =
   \pm\left[\frac{1+a}{2}(1-\nnsup{+})\nnsup{-}
          - \frac{1-a}{2}\nnsup{+}(1-\nnsup{-})\right].
\]
Note that the collisions do not satisfy semi-detailed balance unless $a
= 0$.  We therefore restrict attention to small values of the bias
($a\sim {\cal O}(\epsilon)$), and order the Boltzmann collision operator
as follows:
\[
\ccsupo{\pm}{0}(\nnsup{\ast}) =
   \pm\frac{1}{2}(\nnsup{-}-\nnsup{+}),
\]
\[
\ccsupo{\pm}{1}(\nnsup{\ast}) =
   \pm\frac{a}{2}(\nnsup{-}+\nnsup{+})\mp a\nnsup{-}\nnsup{+}.
\]
Note that both $\ccsupo{\pm}{0}$ and $\ccsupo{\pm}{1}$ conserve
particles exactly, and that $\ccsupo{\pm}{0}$ satisfies semi-detailed
balance.

The local Fermi-Dirac equilibrium is given by
\[
   \nnsupo{\pm}{0} =
   \frac{1}{1+e^{-\alpha}} \equiv f.
\]
where $\alpha$ parametrizes the distribution function, and $f$ is the
mean occupation number.   The total density is given in terms
of $\alpha$ (and $f$) by
\[
   Q = \frac{2}{1+e^{-\alpha}} = 2f.
\]
The Jacobian of the collision operator at equilibrium is
\[
J = \left(\begin{array}{cc}
          -\frac{1}{2} & +\frac{1}{2} \\
          +\frac{1}{2} & -\frac{1}{2}
          \end{array}
    \right).
\]
This matrix has eigenvalues
\[
   \lamsupp{1} = 0,  \; \;   \lamsupp{2} = -1,
\]
with corresponding left eigenvectors
\[
\begin{array}{l}
   \qsupp{1}
   =      \left(\begin{array}{cc} +1 & +1 \end{array}\right) \\
   \qsupp{2}
   =      \left(\begin{array}{cc} +1 & -1 \end{array}\right),
\end{array}
\]
and right eigenvectors
\[
\begin{array}{cc}
   \qsubp{1}
   = \frac{1}{2}
\left(\begin{array}{r} +1 \\ +1 \end{array}\right) &
   \qsubp{2}
   = \frac{1}{2}
\left(\begin{array}{r} +1 \\ -1 \end{array}\right).
\end{array}
\]
Once again, it is easy to check that these eigenvectors satisfy
Eq.~(\ref{eq:ortho}).

Again, there is only one hydrodynamic mode.  The Fermi metric components
are given by
\[
\supscrpt{g}{11} = 2f(1-f),
\]
\[
\subscrpt{g}{11} = \frac{1}{2f(1-f)},
\]
\[
\begin{array}{cc}
  \supscrpt{{\bf g}(1)}{11} = 0, &
  \supscrpt{{\bf g}(1)}{21} = 2f(1-f)\hat{\bfx},
\end{array}
\]
\[
\begin{array}{cc}
  \supscrpt{{\bf g}(2)}{11} = 2f(1-f)\hat{\bfx}\hat{\bfx}, &
  \supscrpt{{\bf g}(2)}{21} = 0,
\end{array}
\]
\[
\begin{array}{cc}
  \mixten{{\bf g}(1)}{1}{1} = 0, &
  \mixten{{\bf g}(1)}{2}{1} = \hat{\bfx},
\end{array}
\]
\[
\begin{array}{cc}
  \mixten{{\bf g}(2)}{1}{1} = \hat{\bfx}\hat{\bfx}, &
  \mixten{{\bf g}(2)}{2}{1} = 0,
\end{array}
\]
and the generalized Kronecker delta, $\mixten{\bfdelta (1)}{\mu}{\nu}$,
can again be calculated,
\[
  \mixten{\bfdelta (1)}{1}{2} = \hat{\bfx}.
\]

Once again, the conserved quantity, $Q$, is not ordered, so we can use
the results of Subsection~\ref{ssec:foce}.  This time, however, the
collision operator is ordered.  The first-order collision operator
respects the conserved quantity, as required, so $\supsub{\cal
C}{1}{1}=0$.  The component $\supsub{\cal C}{2}{1}$, on the other hand
is nonzero, so there will be an advective term in the hydrodynamic
equation.  From Eq.~(\ref{eq:calcdef}), we find
\bgeas
  \supsub{\cal C}{2}{1} & = & a (\nnsupo{+}{0} + \nnsupo{-}{0} -
                                      2\nnsupo{+}{0}\nnsup{-}{0}) \\
                        & = & 2a f(1-f) \\
                        & = & aQ\left(1-\frac{Q}{2}\right).
\eeas
The advection coefficient is then given by Eq.~(\ref{eq:hydadv}),
\bgeas
   \advco{\mu}(\qqsupp{\ast})
   & = &
   \frac{c}{\dt}\sum_{\nu\in K}
   \frac{\mixten{\bfdelta (1)}{\mu}{\nu}
         \supsub{\cal C}{\nu}{1}
         \left(\qqsupp{\ast}\right)}
        {(-\lamsupp{\nu})} \\
   & = &
   \frac{a c}{\dt} Q\left(1-\frac{Q}{2}\right)\hat{\bfx},
\eeas
and the diffusivity is given by Eq.~(\ref{eq:hyddif}),
\bgeas
\cdop{1}{1} & = & \frac{c^2}{\dt}
                  \left[
                  \sum_{\nu\in K}
                  \frac{\mixten{\bfdelta (1)}{1}{\nu}
                        \otimes
                        \mixten{{\bf g}(1)}{\nu}{1}}
                       {(-\lamsupp{\nu})}
                - \frac{1}{2}\mixten{{\bf g}(2)}{1}{1}
                  \right] \\
            & = & \frac{c^2}{2\dt}\hat{\bfx}\hat{\bfx}.
\eeas
Writing $\nabla = \hat{\bfx}\frac{\pdv}{\pdv x}$, Eq.~(\ref{eq:hyd})
gives us the hydrodynamic equation,
\[
  \frac{\pdv Q}{\pdv t} +
  \frac{\pdv {\cal A}}{\pdv x} =
  \frac{\pdv}{\pdv x}
    \left(
       {\cal D} \frac{\pdv Q}{\pdv x}
    \right),
\]
where the scalar advection coefficient is
\[
  {\cal A}\equiv\frac{a c}{\dt} Q\left(1-\frac{Q}{2}\right),
\]
and the scalar diffusivity is
\[
  {\cal D}\equiv\frac{c^2}{2\dt}.
\]
Finally, if we make the change of variables, $u\equiv\frac{a c}{\dt}
(1-Q)$, this becomes Burgers' equation in the more familiar form
\[
  \frac{\pdv u}{\pdv t} +
  u\frac{\pdv u}{\pdv x} =
  {\cal D} \frac{\pdv^2 u}{\pdv x^2}.
\]

\subsection{The 2D4P Lattice Gas}

Next, we consider a diffusive lattice gas model~\cite{bmbcdltdfp} in two
spatial dimensions ($D=2$).  The lattice for this model is the standard
two-dimensional cartesian lattice.  The lattice gas has four bits per
site ($n=4$), corresponding to the presence or absence of particles
moving along each of the four unit vectors in the lattice, which we will
refer to as east, north, west, and south.  These bits are denoted by the
elements of the set $B=\{1,2,3,4\}$, respectively.  Collisions occur if
and only if exactly two particles enter a site at right angles, and are
effected by taking the complement of all four bits at such a site.  For
convenience, we will use a diagrammatic notation for the set of states
$S =\{s \subseteq B\}$.  In this diagrammatic notation, each nonzero bit
in a state $s$ is denoted by a line segment emanating from a common
vertex.  Thus, for example, the state $s =\{2,3\}$ would be denoted by
the symbol $\vertex0110$.  Using this notation, the nontrivial elements
of the state transition table can be written
\bgc
\begin{tabular}{|c|c||c|c|c|c|}
\hline
\multicolumn{2}{|c||}{$a(s\rightarrow s')$} & \multicolumn{4}{c|}{$s'$} \\
\cline{3-6}
\multicolumn{2}{|c||}{} & $\vertex1100$ & $\vertex0110$ &
$\vertex0011$ & $\vertex1001$ \\
\hline
\hline
    & $\vertex1100$ & $0$ & $0$ & $1$ & $0$ \\
\cline{2-6}
$s$ & $\vertex0110$ & $0$ & $0$ & $0$ & $1$ \\
\cline{2-6}
    & $\vertex0011$ & $1$ & $0$ & $0$ & $0$ \\
\cline{2-6}
    & $\vertex1001$ & $0$ & $1$ & $0$ & $0$ \\
\hline
\end{tabular}
\ec

This lattice gas obeys semi-detailed balance, since the
columns of the above collision table sum to unity.  In fact, the
lattice gas obeys detailed balance, since the transition matrix is
symmetric.  This collision rule separately conserves the number of
east/west-moving particles and the number of north/south-moving
particles ($n_c=2$).  Thus, the coefficients $\qrow{1}{\ast}$ and
$\qrow{2}{\ast}$ for the conserved quantities are given by
\bgea
   \qrow{1}{1} & = & \qrow{1}{3} = 1 \nonumber \\
   \qrow{1}{2} & = & \qrow{1}{4} = 0 \nonumber \\
   \qrow{2}{2} & = & \qrow{2}{4} = 1 \nonumber \\
   \qrow{2}{1} & = & \qrow{2}{3} = 0.
   \label{eq:tdfpqrow}
\eea
The collision operator is given by Eq.~(\ref{eq:collop}),
\bgeas
\csup{i}(\nsup{\ast})
  & = &
  \overline{\nsup{i}}\,\overline{\nsup{i+1}}\nsup{i+2}\nsup{i+3} +
  \overline{\nsup{i}}\nsup{i+1}\nsup{i+2}\overline{\nsup{i+3}} \\
  & & -
  \nsup{i}\nsup{i+1}\overline{\nsup{i+2}}\,\overline{\nsup{i+3}} -
  \nsup{i}\overline{\nsup{i+1}}\,\overline{\nsup{i+2}}\nsup{i+3} \\
  & = &
  \left(\nsup{i+2} - \nsup{i}\right)
  \left(\nsup{i+1} + \nsup{i+3} - \nsup{i+1}\nsup{i+3}\right),
\eeas
where $i\in B$ and the addition of integers to $i$ is taken modulo 4.
Since the lattice gas is deterministic, we have $A(s\rightarrow s')$ =
$a(s\rightarrow s')$, and the ensemble-averaged collision operator in
the Boltzmann approximation is given by $\ccsup{i} = \csup{i}$; that is
\[
\ccsup{i}(\nnsup{\ast})=\ccsup{i}_0(\nnsup{\ast})
  =
  \left(\nnsup{i+2} - \nnsup{i}\right)
  \left(\nnsup{i+1} + \nnsup{i+3} - \nnsup{i+1}\nnsup{i+3}\right).
\]
Using Eq.~(\ref{eq:tdfpqrow}), we calculate the local Fermi-Dirac
equilibrium from Eq.~(\ref{eq:fd}),
\bgeas
 \nnsupo{1}{0}=\nnsupo{3}{0}
 & = & \frac{1}{1+e^{-\alpsubp{1}}}\equiv\mu \\
 \nnsupo{2}{0}= \nnsupo{4}{0}
 & = & \frac{1}{1+e^{-\alpsubp{2}}}\equiv\nu,
\eeas
where $\alpsubp{1}$ and $\alpsubp{2}$, or equivalently $\mu$ and
$\nu$, parametrize the equilibrium distribution function.  The
east/west and north/south densities are given by
\bgeas
   \qqsupp{1} & = & \frac{2}{1+e^{-\alpsubp{1}}} = 2\mu \\
   \qqsupp{2} & = & \frac{2}{1+e^{-\alpsubp{2}}} = 2\nu.
\eeas

We now perform the Chapman-Enskog perturbative analysis.  The Jacobian
of the collision operator at equilibrium is
\[
J = \left(\begin{array}{cccc}
          -\Lambda(\nu) & 0 & +\Lambda(\nu) & 0 \\
          0 & -\Lambda(\mu) & 0 & +\Lambda(\mu) \\
          +\Lambda(\nu) & 0 & -\Lambda(\nu) & 0 \\
          0 & +\Lambda(\mu) & 0 & -\Lambda(\mu)
          \end{array}
    \right),
\]
where
\[
\Lambda(z)\equiv 2z(1-z).
\]
This matrix has eigenvalues
\bgeas
   \lamsupp{1} & = & \lamsupp{2} = 0 \\
   \lamsupp{3} & = & -2\Lambda(\mu) \\
   \lamsupp{4} & = & -2\Lambda(\nu),
\eeas
with corresponding left eigenvectors
\bge
\begin{array}{l}
   \qsupp{1}
   =      \left(\begin{array}{rrrr} +1 & 0 & +1 & 0 \end{array}\right) \\
   \qsupp{2}
   =      \left(\begin{array}{rrrr} 0 & +1 & 0 & +1 \end{array}\right) \\
   \qsupp{3}
   =      \left(\begin{array}{rrrr} +1 & 0 & -1 & 0 \end{array}\right) \\
   \qsupp{4}
   =      \left(\begin{array}{rrrr} 0 & +1 & 0 & -1 \end{array}\right),
\end{array}
\label{eq:tdfprev}
\ee
and right eigenvectors
\[
\begin{array}{cccc}
   \qsubp{1}
   = \frac{1}{2}
\left(\begin{array}{r} +1 \\ 0 \\ +1 \\ 0 \end{array}\right) &
   \qsubp{2}
   = \frac{1}{2}
\left(\begin{array}{r} 0 \\ +1 \\ 0 \\ +1 \end{array}\right) &
   \qsubp{3}
   = \frac{1}{2}
\left(\begin{array}{r} +1 \\ 0 \\ -1 \\ 0 \end{array}\right) &
   \qsubp{4}
   = \frac{1}{2}
\left(\begin{array}{r} 0 \\ +1 \\ 0 \\ -1 \end{array}\right).
\end{array}
\]

We now construct the Fermi metric.  There are two hydrodynamic
modes, so $\supscrpt{g}{\mu\nu}$ is represented by a 2-by-2 matrix.
{}From Eq.~(\ref{eq:ffttwo}) we see that
\[
\supscrpt{g}{\mu\nu} =
  \supscrpt{
  \left(
  \begin{array}{cc}
  \Lambda(\mu) & 0 \\
  0 & \Lambda(\nu)
  \end{array}
  \right)}{\mu\nu}
\]
and hence
\[
\subscrpt{g}{\mu\nu} =
  \subscrpt{
  \left(
  \begin{array}{cc}
  \frac{1}{\Lambda(\mu)} & 0 \\
  0 & \frac{1}{\Lambda(\nu)}
  \end{array}
  \right)}{\mu\nu}.
\]
To get the generalized Fermi metric we write $\supscrpt{\bfc}{i} =
c\supscrpt{\bfe}{i}$   with
\bgeas
  \supscrpt{\bfe}{1} = +\hat{\bfx}  & \; \; \; \; &
  \supscrpt{\bfe}{2} = +\hat{\bfy} \\
  \supscrpt{\bfe}{3} = -\hat{\bfx}  & \; \; \; \; &
  \supscrpt{\bfe}{4} = -\hat{\bfy},
\eeas
where $\hat{\bfx}$ and $\hat{\bfy}$ are the unit vectors in the $x$ and
$y$ directions, respectively.  Using Eq.~(\ref{eq:tdfprev}) in
Eq.~(\ref{eq:gffttwo}), it follows that the useful nonvanishing
components of ${\bf g}(1)$ are
\bgeas
  \supscrpt{{\bf g}(1)}{31} & = & \Lambda(\mu)\hat{\bfx} \\
  \supscrpt{{\bf g}(1)}{42} & = & \Lambda(\nu)\hat{\bfy},
\eeas
while those of ${\bf g}(2)$ are
\bgeas
  \supscrpt{{\bf g}(2)}{11} & = & \Lambda(\mu)\hat{\bfx}\hat{\bfx} \\
  \supscrpt{{\bf g}(2)}{22} & = & \Lambda(\nu)\hat{\bfy}\hat{\bfy}.
\eeas
Finally, the generalized Kronecker delta, $\mixten{\bfdelta
(1)}{\mu}{\nu}$, given by Eq.~(\ref{eq:hten}), has components
\[
\begin{array}{cc}
  \mixten{\bfdelta (1)}{1}{3} = \hat{\bfx}, &
  \mixten{\bfdelta (1)}{2}{4} = \hat{\bfy}.
\end{array}
\]

The conserved quantities are not ordered, so we can use the results of
Subsection~\ref{ssec:foce}.
The collision operator is not
ordered for this lattice gas, so $\supsub{\cal C}{\nu}{1} =
\supsub{\cal C}{\nu}{2} = 0$
and there are no advection or source terms in the hydrodynamic
equations.  The diffusivity is given by Eq.~(\ref{eq:hyddif}),
\bgeas
\cdop{\mu}{\xi}
            & = & \frac{c^2}{\dt}
                  \left[
                  \sum_{\nu\in K}
                  \frac{\mixten{\bfdelta (1)}{\mu}{\nu}
                        \otimes
                        \mixten{{\bf g}(1)}{\nu}{\xi}}
                       {(-\lamsupp{\nu})}
                - \frac{1}{2}\mixten{{\bf g}(2)}{\mu}{\xi}
                  \right] \\
            & = & \mixten{\left(\begin{array}{cc}
                  {\cal D}(\nu, \mu)\hat{\bfx}\hat{\bfx} & 0 \\
                  0 & {\cal D}(\mu, \nu)\hat{\bfy}\hat{\bfy}
                  \end{array}\right)}{\mu}{\xi},
\eeas
where we have defined the scalar diffusivities,
\[
  {\cal D}(\mu,\nu)\equiv
		\frac{c^2}{2\dt}\left(\frac{2}{(-\lamsupp{3})}-1\right)
               = \frac{c^2}{2\dt}\left(\frac{1}{2\mu (1-\mu )}-1\right)
\]
and
\[
  {\cal D}(\nu, \mu)\equiv
		\frac{c^2}{2\dt}\left(\frac{2}{(-\lamsupp{4})}-1\right)
               = \frac{c^2}{2\dt}\left(\frac{1}{2\nu (1-\nu )}-1\right).
\]
Writing $\nabla = \hat{\bfx}\frac{\pdv}{\pdv x} +
\hat{\bfy}\frac{\pdv}{\pdv y}$, Eq.~(\ref{eq:hyd}) gives us the
pair of hydrodynamic equations,
\bgeas
  \frac{\pdv \mu}{\pdv t} & = &
  \frac{\pdv}{\pdv x}
    \left(
       {\cal D}(\nu, \mu) \frac{\pdv \mu}{\pdv x}
    \right) \\
  \frac{\pdv \nu}{\pdv t} & = &
  \frac{\pdv}{\pdv y}
    \left(
       {\cal D}(\mu, \nu) \frac{\pdv \nu}{\pdv y}
    \right).
\eeas
Since $\mu$ and $\nu$ lie in $[0,1]$, it follows that the diffusivities
are always positive.

Note that the diffusivity ${\cal D} (\mu, \nu)$ of north/south particles
depends only on the density $\mu$ of east/west particles, and vice
versa.  In the context of the Boltzmann theory, this is because
north/south particles scatter only from east/west particles, and vice
versa.  In Section~\ref{sec:examples} we compute the effects of
correlations on these diffusivities, and find that both renormalized
diffusivities depend upon both particle densities; this is the reason
for including the functional dependence on both densities in ${\cal D}$.

\subsection{Lattice Gas Fluids}

Finally, we consider a class of lattice gases that have been widely
used in recent years for the simulation of incompressible
Navier-Stokes fluids~\cite{fhp,fchc}.  Such models exist in dimensions
two, three, and higher.  They conserve mass and momentum, as is
appropriate for the Navier-Stokes equations.

\begin{figure}
\centering
\begin{picture}(300,150)(0,0)
\put(75.,150.){\makebox(0,0){In}}
\put(225.,150.){\makebox(0,0){Out}}
\put(225.,90.){\makebox(0,0){\footnotesize or}}
\put(180.,90.){
\begin{picture}(50,50)(0,0)
\put(0.,0.){\vector(2,3){12.5}}
\put(0.,0.){\vector(-2,-3){12.5}}
\put(0.,0.){\circle*{5.}}
\end{picture}}
\put(270.,90.){
\begin{picture}(50,50)(0,0)
\put(0.,0.){\vector(-2,3){12.5}}
\put(0.,0.){\vector(2,-3){12.5}}
\put(0.,0.){\circle*{5.}}
\end{picture}}
\put(225.,30.){
\begin{picture}(50,50)(0,0)
\put(0.,0.){\vector(1,0){25.}}
\put(0.,0.){\vector(-2,3){12.5}}
\put(0.,0.){\vector(-2,-3){12.5}}
\put(0.,0.){\circle*{5.}}
\end{picture}}
\put(75.,90.){
\begin{picture}(50,50)(0,0)
\put(25.,0.){\vector(-1,0){22.5}}
\put(-25.,0.){\vector(1,0){22.5}}
\put(0.,0.){\circle*{5.}}
\end{picture}}
\put(75.,30.){
\begin{picture}(50,50)(0,0)
\put(25.,0.){\vector(-1,0){22.5}}
\put(-12.5,21.65){\vector(2,-3){11.1132}}
\put(-12.5,-21.65){\vector(2,3){11.1132}}
\put(0.,0.){\circle*{5.}}
\end{picture}}
\end{picture}
\caption{FHP-I Collision Rules}
\label{fig:fhp}
\end{figure}
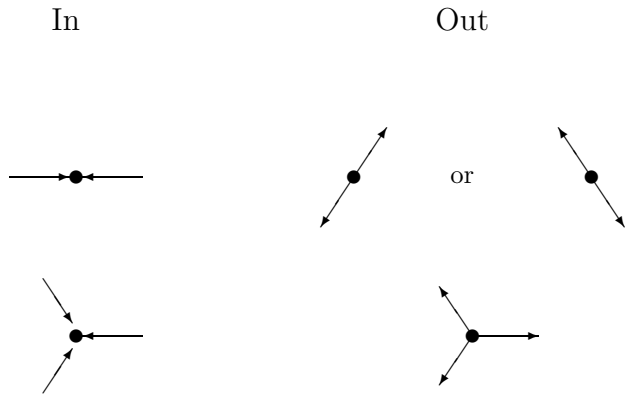

The nontrivial collision rules for one of the simplest models of this
type, known as the FHP-I model, are shown in Fig.~\ref{fig:fhp}.  The
FHP-I model is defined in two dimensions on a hexagonal lattice.
Note that the two-body collisions have two possible outcomes.  We may
choose between these by a random bit, $\supscrpt{n}{r}$, with mean
$\frac{1}{2}$.  Variants of this model exist with other three-body and
four-body collisions, with rest particles, etc.  We use the above model
because it is simple but nevertheless fully illustrative for our
purposes.  Inclusion of the three-body symmetric collision is
essential because there would otherwise be three conserved components of
momentum.

We will now describe the general class of lattice gas fluids and carry
out the Chapman-Enskog analysis for systems of this type.  Following the
general analysis we will return to the specific example of the FHP-I
lattice gas.  We consider a general lattice gas with a collision rule
preserving particle number and momentum in each direction on a lattice
in $D$ dimensions.  \label{pg:at} We assume that bit $i$ at a site
represents the presence or absence of a particle of unit
mass\footnote{The restriction to unit mass particles is made only to
simplify this presentation, and is not in any way essential.  Many
lattice gas fluid models allow for particles of different masses.} and
momentum $\bfci /
\dt$.  We can then write the (ensemble-averaged) mass and momentum
densities as
\[
\rho = \sum_{i = 1}^n \nnsup{i},
\]
and
\[
\bfu = \sum_{i = 1}^n \frac{\bfci}{\dt} \nnsup{i},
\]
respectively.  The full set of ensemble-averaged conserved quantities
for this problem is thus
\[
Q = \left(\begin{array}{c}\rho \\ \bfu\end{array}\right).
\]
Note that this is a $(D+1)$-component column vector, since there are $D$
components of conserved momentum density and one conserved mass density.
We shall abuse notation by sometimes using $\rho$ and $\bfu$ as indices;
thus, we separate the above equation into the components, $Q^\rho =
\rho$, and $Q^\bfu = \bfu$.  Note that $Q^\bfu$ refers to $D$ distinct
components of $Q$.

We can at once identify the left hydrodynamic eigenvectors of the
system.  They are $\qrow{\rho}{i} = 1$ and $\qrow{\bfu}{i} =
\frac{\bfci}{\dt}$.  To simplify the presentation, we use
natural lattice units ($c=\dt=1$) throughout this subsection; we
can always reintroduce $c$ and $\dt$ later by dimensional analysis
considerations.  Thus, we write
\[
\begin{array}{ccc}
\qrow{\rho}{i} = 1 &
\mbox{and} &
\qrow{\bfu}{i} = \bfei.
\end{array}
\]

For an incompressible fluid, the conserved densities are ordered in
the expansion parameter $\epsilon$ as follows~\cite{ll},
\[
\rho = \subscrpt{\rho}{0} + \epsilon^2\subscrpt{\rho}{2}
\]
\[
\bfu = \epsilon\subscrpt{\bfu}{1}.
\]
The second of these equations states that the Mach number is of the same
order as the smallness parameter used in the Chapman-Enskog analysis;
that is, the Mach number scales as the Knudsen number.  The first equation
says that the density fluctuations are smaller still -- they go as the
Mach number squared.

The ordering of the conserved quantities, Eq.~(\ref{eq:consord}), for
this system is then,
\bgeas
Q & = &
    \left(\begin{array}{c} \rho \\ \bfu\end{array} \right)
  = \left(\begin{array}{c} \subscrpt{\rho}{0} \\ 0 \end{array}\right) +
    \epsilon
    \left(\begin{array}{c} 0 \\ \subscrpt{\bfu}{1} \end{array}\right) +
    \epsilon^2
    \left(\begin{array}{c} \subscrpt{\rho}{2} \\ 0 \end{array}\right) \\
  & = & \subscrpt{Q}{0} +
        \epsilon\subscrpt{Q}{1} +
        \epsilon^2\subscrpt{Q}{2}.
\eeas
Thus, the zero-order Fermi-Dirac equilibrium is found by considering
only $\qsup{\rho}$ in Eq.~(\ref{eq:fd}).  We get
\[
\nnsupo{i}{00} = \frac{1}{1+e^{-\alpha}} \equiv f,
\]
where $f$ is the mean occupation number
of the zero-order equilibrium.  Note that $f$ is strictly constant --
independent of spatial position and time.

To proceed, it is necessary to impose some requirements on the lattice.
Let us assume that the lattice is a regular lattice, and that all
tensors formed from outer products of the lattice vectors are isotropic
through the fourth rank~\cite{henonvis}.  That is, we demand that the
lattice vectors be such that
\bgea
\sum_{i=1}^n \bfei &=& 0 \nonumber \\
\sum_{i=1}^n \bfei\bfei &=& \frac{n}{D} \bfid \nonumber \\
\sum_{i=1}^n \bfei\bfei\bfei &=& 0 \nonumber \\
\sum_{i=1}^n \bfei\bfei\bfei\bfei &=& \frac{n}{D(D+2)}\bfomega,
\label{eq:iso}
\eea
where we have defined
\[
  {\bfomega}_{ijkl} =
  \subscrpt{\delta}{ij}\subscrpt{\delta}{kl} +
  \subscrpt{\delta}{ik}\subscrpt{\delta}{jl} +
  \subscrpt{\delta}{il}\subscrpt{\delta}{jk}.
\]

We can now construct the Fermi metric.  There are $D+1$ hydrodynamic
modes, so $\supscrpt{g}{\mu\nu}$ is a $(D+1)$-by-$(D+1)$ matrix.  From
Eqs.~(\ref{eq:ffttwo}) and (\ref{eq:iso}), we see that it is
\[
\supscrpt{g}{\mu\nu} =
   nf(1-f)
   \supscrpt{
   \left(\begin{array}{cc}
   1 & 0 \\
   0 & \frac{1}{D}\bfid
   \end{array}\right)}{\mu\nu}.
\]
The inverse metric is then
\[
\subscrpt{g}{\mu\nu} =
   \frac{1}{nf(1-f)}
   \subscrpt{
   \left(\begin{array}{cc}
   1 & 0 \\
   0 & D\bfid
   \end{array}\right)}{\mu\nu}.
\]

Since the conserved quantities are ordered, we are going to need the
Fermi connection as well.  From Eqs.~(\ref{eq:fftthree}) and
(\ref{eq:iso}), we see that it is
\[
\mixten{\Gamma}{\eta}{\mu\nu} =
   -\frac{(1-2f)}{2nf(1-f)}
   \supscrpt{
   \left(\begin{array}{cc}
   \subscrpt{
   \left(\begin{array}{cc}
   1 & 0 \\
   0 & D\bfid
   \end{array}\right)}{\mu\nu} &
   \subscrpt{
   \left(\begin{array}{cc}
   0 & \bfid \\
   \bfid & 0
   \end{array}\right)}{\mu\nu}
   \end{array}\right)
   }{\eta},
\]
where we have used a row of matrices to represent the variation of the
three indices; we shall occasionally use this representation for
third-rank quantities when it does not cause ambiguity.

Next, we can compute the components of the generalized Fermi metric with
hydrodynamic indices.  From Eqs.~(\ref{eq:gffttwo}) and
(\ref{eq:gfftthree}), we have
\[
\supscrpt{{\bf g}(k)}{\mu\nu} =
   f(1-f)
   \supscrpt{
   \left(\begin{array}{cc}
   \sum_i\bigotimes^{k  }\bfesup{i} & \sum_i\bigotimes^{k+1}\bfesup{i} \\
   \sum_i\bigotimes^{k+1}\bfesup{i} & \sum_i\bigotimes^{k+2}\bfesup{i}
   \end{array}\right)}{\mu\nu}.
\]
Likewise, the components of the generalized Fermi connection are
\bgeas
\lefteqn{\mixten{\bfgamma (k)}{\eta}{\mu\nu} =
   -\frac{(1-2f)}{2n^2f(1-f)}} \\
   & &
   \;\;\;\;\;\;\;\;
   \left(
   \subscrpt{
   \left(\begin{array}{cc}
   \sum_i\bigotimes^{k  }\bfesup{i} &
   D\sum_i\bigotimes^{k+1}\bfesup{i} \\
   D\sum_i\bigotimes^{k+1}\bfesup{i} &
   D^2\sum_i\bigotimes^{k+2}\bfesup{i}
   \end{array}\right)}{\mu\nu}
   \right. \\
   & &
   \;\;\;\;\;\;\;\;\;\;\;\;\;\;\;\;\;\;\;\;\;
   \supscrpt{
   \left.
   \subscrpt{
   \left(\begin{array}{cc}
   \sum_i\bigotimes^{k+1}\bfesup{i} &
   D\sum_i\bigotimes^{k+2}\bfesup{i} \\
   D\sum_i\bigotimes^{k+2}\bfesup{i} &
   D^2\sum_i\bigotimes^{k+3}\bfesup{i}
   \end{array}\right)}{\mu\nu}
   \right)
   }{\eta}.
\eeas
All of the components of these objects which we will need can be
evaluated from Eqs.~(\ref{eq:iso}).

Eq.~(\ref{eq:hydocc}) now tells us the form
that the hydrodynamic equations will take.  We first examine the
equation for conservation of mass.  Setting the index $\mu$ to $\rho$,
we see that the ${\cal O}(1/\epsilon)$ advection term survives, and all
the ${\cal O}(1)$ terms vanish; the next surviving terms are ${\cal
O}(\epsilon)$.  Thus, looking to Eq.~(\ref{eq:hydocceps}), we get the
equation
\[
   \nabla\cdot\left[\frac{c}{\dt}
                    \mixten{{\bf g}(1)}{\rho}{\xi}
                    \qqsuppo{\xi}{1}
              \right] = 0,
\]
or, upon simplification,
\bge
   \nabla\cdot\subscrpt{\bfu}{1} = 0,
   \label{eq:incompc}
\ee
to ${\cal O}(\epsilon^2)$.  We recognize this as expressing the
condition that the velocity field must have zero divergence in the
incompressible limit.

Turning attention next to the equation for conservation of momentum, we
set the index $\mu$ to $\bfu$.  This time we find that the ${\cal
O}(1/\epsilon)$ part of the advection term vanishes, so the hydrodynamic
equation is given by the ${\cal O}(1)$ terms in Eq.~(\ref{eq:hydocc}).
For this situation, Eq.~(\ref{eq:hydoccadv}) for the advection
coefficient reads
\bgeas
   \lefteqn{\advco{\bfu} (\subscrpt{\rho}{0})} \\
   & = &
   \mixten{{\bf g}(1)}{\bfu}{\rho}\qqsuppo{\rho}{2} +
   \left[
   \mixten{{\bf g} (1)}{\bfu}{\rho}\mixten{\Gamma}{\rho}{\bfu\bfu} +
   \mixten{{\bf g} (1)}{\bfu}{\bfu}\mixten{\Gamma}{\bfu}{\bfu\bfu} -
   \mixten{\bfgamma (1)}{\bfu}{\bfu\bfu}
   \right] :
   \subscrpt{\bfu}{1}\subscrpt{\bfu}{1}. \\
   & = &
   P \bfid + g(f)\subscrpt{\bfu}{1}\subscrpt{\bfu}{1},
\eeas
where we have defined the factor
\[
   g(f) \equiv \frac{(1-2f)}{nf(1-f)}
\]
and the pressure
\bge
   P \equiv \frac{\subscrpt{\rho}{2}}{D} - \frac{g(f)\supsub{u}{2}{1}}{D+2},
   \label{eq:eos}
\ee
written in lattice units.

Next we turn our attention to the diffusivity tensor,
$\cdop{\bfu}{\bfu}$, given by Eq.~(\ref{eq:hyddif}).  We have
\bgeas
  \mixten{\bfdelta (1)}{\bfu}{\nu} & = &
    \qrow{\bfu}{j}\bfesup{j}\qcol{\nu}{j} = \sum_{j=1}^n
    \qcol{\nu}{j}\bfesup{j}\bfesup{j} \\
  \mixten{{\bf g} (1)}{\nu}{\bfu} & = &
    \subscrpt{g}{\bfu\bfu}\supscrpt{{\bf g}(1)}{\bfu\nu} =
    \frac{D}{n}\sum_{j=1}^n\qrow{\nu}{j}\bfesup{j}\bfesup{j} \\
  \mixten{{\bf g} (2)}{\bfu}{\bfu} & = &
    \subscrpt{g}{\bfu\bfu}\supscrpt{{\bf g}(2)}{\bfu\bfu} =
    \frac{D}{n}\sum_{j=1}^n\bfesup{j}\bfesup{j}\bfesup{j}\bfesup{j} =
    \frac{1}{D+2}\bfomega,
\eeas
so Eq.~(\ref{eq:hyddif}) becomes
\bge
   \supsub{\left(\cdop{u_i}{u_l}\right)}{j}{k} =
   \frac{c^2}{(D+2)\dt}
     \left\{
     \frac{D(D+2)}{n}
     \supsub{\bfe}{m}{i}\supsub{\bfe}{m}{j}
     \supsub{\bfe}{p}{k}\supsub{\bfe}{p}{l}
     \left[
     \sum_{\nu\in K}
     \frac{\qcol{\nu}{m}\qrow{\nu}{p}}
          {(-\lamsupp{\nu})}
     \right]
   - \frac{1}{2}\bfomega
     \right\},
   \label{eq:fhphyddif}
\ee
where we have made it clear that $\cdop{\bfu}{\bfu}$ is a fourth-rank
object by explicity writing the four spatial indices, $i$, $j$, $k$, and
$l$, and where there is an implicit sum over $m$ and $p$.  Now, since
the principal fourth-rank tensor constructed from our lattice vectors
(see Eq.~(\ref{eq:iso})) is isotropic, and since we have done nothing
else to break isotropy, this fourth-rank tensor must be isotropic as
well.  That means that it must be of the form
\bge
   \supsub{\left(\cdop{u_i}{u_l}\right)}{j}{k} =
   \nu    \subscrpt{\delta}{il}\subscrpt{\delta}{jk} +
   \alpha \subscrpt{\delta}{ij}\subscrpt{\delta}{lk} +
   \beta  \subscrpt{\delta}{ik}\subscrpt{\delta}{lj},
   \label{eq:isoiso}
\ee
so that the diffusive term on the right-hand side of the hydrodynamic
equation becomes
\[
   \subscrpt{\nabla}{j}
   \supsub{\left(\cdop{u_i}{u_l}\right)}{j}{k}
   \subscrpt{\nabla}{k} u_l =
   \nu \nabla^2 u_i +
   \subscrpt{\nabla}{i}
   \left[\left(\alpha + \beta\right)\nabla\cdot\bfu\right].
\]
Note that the second term on the right vanishes due to the
incompressibility condition, Eq.~(\ref{eq:incompc}), and $\nu$ emerges
as the shear viscosity in the hydrodynamic equation,
Eq.~(\ref{eq:hydocc}), which can now be written
\bge
   \frac{\pdv\subscrpt{\bfu}{1}}{\pdv t} +
   g(f)\subscrpt{\bfu}{1}\cdot\nabla\subscrpt{\bfu}{1} =
   -\nabla P + \nu \nabla^2 \bfu.
   \label{eq:ns}
\ee
With the exception of the pathological factor $g(f)$ in the inertial
term, which we will address below, we recognize this as the
incompressible Navier-Stokes equation.

To get a closed expression for the viscosity, $\nu$, we take the
following traces of Eq.~(\ref{eq:isoiso}):
\bgeas
   \supsub{\left(\cdop{u_i}{u_i}\right)}{j}{j} & = &
      D\alpha + D\beta + D^2\nu \\
   \supsub{\left(\cdop{u_i}{u_j}\right)}{j}{i} & = &
      D\alpha + D^2\beta + D\nu \\
   \supsub{\left(\cdop{u_i}{u_j}\right)}{i}{j} & = &
      D^2\alpha + D\beta + D\nu.
\eeas
These three equations may be solved for $\nu$ to yield
\[
   \nu = \frac{(D+1)\supsub{\left(\cdop{u_i}{u_i}\right)}{j}{j} -
                    \supsub{\left(\cdop{u_i}{u_j}\right)}{j}{i} -
                    \supsub{\left(\cdop{u_i}{u_j}\right)}{i}{j}}
              {D(D-1)(D-2)}.
\]
Inserting Eq.~(\ref{eq:fhphyddif}), after a bit of algebra we get the
following result for the viscosity of a lattice fluid:
\bge
   \nu = \frac{c^2}{(D+2)\dt}
         \left(
         \frac{D}{n(D-1)}\sum_{\nu\in K}
         \frac{\qcol{\nu}{m}
               \left(\bfesup{m}\cdot\bfesup{n}\right)^2
               \qrow{\nu}{n}}
              {(-\lamsupp{\nu})} -
         \frac{1}{2}
         \right).
   \label{eq:visvis}
\ee
We can compare this result with that of H\'{e}non who
writes~\cite{henonvis}
\[
   \nu = \frac{c^2}{(D+2)\dt}
         \left(
         \lambda_H -
         \frac{1}{2}
         \right),
\]
where the quantity $\lambda_H$ is given by
\[
   \frac{1}{\lambda_H} \equiv
   \frac{1}{2n}\left(\frac{D}{D-1}\right)
   \sum_{s,s'}A(s\rightarrow s')
   f^{p-1} (1-f)^{n-p-1}
   \left[{\bf Y}(s) - {\bf Y}(s')\right] :
   \left[{\bf Y}(s) - {\bf Y}(s')\right],
\]
where ${\bf Y}(s)$ is the traceless part of ${\bf X}(s)$
\[
   {\bf Y}(s) \equiv {\bf X}(s) - \frac{p}{D}\bfid,
\]
and ${\bf X}(s)$ is in turn given by
\[
   {\bf X}(s) \equiv \sum_j \supscrpt{s}{j}\bfesup{j}\bfesup{j}
\]
with trace
\[
   p \equiv {\rm Tr}[{\bf X}(s)] = \sum_j \supscrpt{s}{j}.
\]
Note that our analysis has yielded an alternative expression for
H\'{e}non's quantity,
\[
   \lambda_H =
         \frac{D}{n(D-1)}\sum_{\nu\in K}
         \frac{\qcol{\nu}{m}
               \left(\bfesup{j}\cdot\bfesup{k}\right)^2
               \qrow{\nu}{n}}
              {(-\lamsupp{\nu})}.
\]

So far, we have assumed only that the lattice vectors satisfy
Eqs.~(\ref{eq:iso}), and that the particles each have unit mass and
speed.  To evaluate Eq.~(\ref{eq:visvis}) for the shear viscosity,
however, it is necessary to specify a particular collision rule.

Thus, we now specialize to the FHP-I lattice gas, in two dimensions
($D=2$), with six bits per site ($n=6$), and with lattice vectors given
by
\[
   \bfesup{j} = \cos\left(\frac{2\pi j}{6}\right)\hat{\bf x} +
                \sin\left(\frac{2\pi j}{6}\right)\hat{\bf y}.
\]
We then have
\[
   \left(\bfesup{j}\cdot\bfesup{k}\right)^2 =
   \cos^2\left(\frac{2\pi (j-k)}{6}\right) =
   \left(\begin{array}{cccccc}
         1 & \mbox{\scriptsize 1/4} & \mbox{\scriptsize 1/4} &
         1 & \mbox{\scriptsize 1/4} & \mbox{\scriptsize 1/4}\\
         \mbox{\scriptsize 1/4} & 1 & \mbox{\scriptsize 1/4} &
         \mbox{\scriptsize 1/4} & 1 & \mbox{\scriptsize 1/4}\\
         \mbox{\scriptsize 1/4} & \mbox{\scriptsize 1/4} & 1 &
         \mbox{\scriptsize 1/4} & \mbox{\scriptsize 1/4} & 1\\
         1 & \mbox{\scriptsize 1/4} & \mbox{\scriptsize 1/4} &
         1 & \mbox{\scriptsize 1/4} & \mbox{\scriptsize 1/4}\\
         \mbox{\scriptsize 1/4} & 1 & \mbox{\scriptsize 1/4} &
         \mbox{\scriptsize 1/4} & 1 & \mbox{\scriptsize 1/4}\\
         \mbox{\scriptsize 1/4} & \mbox{\scriptsize 1/4} & 1 &
         \mbox{\scriptsize 1/4} & \mbox{\scriptsize 1/4} & 1
         \end{array}
   \right)^{jk}.
\]

The collision rules are illustrated in Fig.~\ref{fig:fhp}.  We form the
Jacobian of the collision operator from Eq.~(\ref{eq:jacexp}).  Because
the collision rules are invariant under cyclic interchange of the
lattice vectors, the Jacobian is a circulant matrix, so it is
particularly easy to write down its eigenvalues and eigenvectors.  (For
more detail on this point, see \cite{swolf}.)  We find
\bgeas
   \lamsupp{1} & = & 0 \\
   \lamsupp{2} & = & 0 \\
   \lamsupp{3} & = & 0 \\
   \lamsupp{4} & = & -3f(1-f)^3 \\
   \lamsupp{5} & = & -6f^2(1-f)^2 \\
   \lamsupp{6} & = & -3f(1-f)^3,
\eeas
with corresponding left eigenvectors
\bgeas
   \qsubp{1}
   &=&
     \left(\begin{array}{rrrrrr} \mbox{\scriptsize +1} &
\mbox{\scriptsize +1} & \mbox{\scriptsize +1} & \mbox{\scriptsize +1} &
\mbox{\scriptsize +1} & \mbox{\scriptsize +1} \end{array}\right) \\
   \qsubp{2}
   &=& \frac{\sqrt{3}}{2}
     \left(\begin{array}{rrrrrr}  \mbox{\scriptsize 0} &
\mbox{\scriptsize +1} & \mbox{\scriptsize +1} &  \mbox{\scriptsize 0} &
\mbox{\scriptsize -1} & \mbox{\scriptsize -1} \end{array}\right) \\
   \qsubp{3}
   &=& \frac{1}{2}
     \left(\begin{array}{rrrrrr} \mbox{\scriptsize +2} &
\mbox{\scriptsize +1} & \mbox{\scriptsize -1} & \mbox{\scriptsize -2} &
\mbox{\scriptsize -1} & \mbox{\scriptsize +1} \end{array}\right) \\
   \qsubp{4}
   &=& \frac{1}{2}
     \left(\begin{array}{rrrrrr} \mbox{\scriptsize +2} &
\mbox{\scriptsize -1} & \mbox{\scriptsize -1} & \mbox{\scriptsize +2} &
\mbox{\scriptsize -1} & \mbox{\scriptsize -1} \end{array}\right) \\
   \qsubp{5}
   &=&
     \left(\begin{array}{rrrrrr} \mbox{\scriptsize +1} &
\mbox{\scriptsize -1} & \mbox{\scriptsize +1} & \mbox{\scriptsize -1} &
\mbox{\scriptsize +1} & \mbox{\scriptsize -1} \end{array}\right) \\
   \qsubp{6}
   &=& \frac{\sqrt{3}}{2}
     \left(\begin{array}{rrrrrr}  \mbox{\scriptsize 0} &
\mbox{\scriptsize +1} & \mbox{\scriptsize -1} &  \mbox{\scriptsize 0} &
\mbox{\scriptsize +1} & \mbox{\scriptsize -1} \end{array}\right),
\eeas
and right eigenvectors
\[
\begin{array}{ccc}
   \qsupp{1}
   = \frac{1}{6}
   \left(\begin{array}{r} \mbox{\scriptsize +1} \\ \mbox{\scriptsize +1}
\\ \mbox{\scriptsize +1} \\ \mbox{\scriptsize +1} \\ \mbox{\scriptsize
+1} \\ \mbox{\scriptsize +1} \end{array}\right) &
   \qsupp{2}
   = \frac{\sqrt{3}}{6}
   \left(\begin{array}{r}  \mbox{\scriptsize 0} \\ \mbox{\scriptsize +1}
\\ \mbox{\scriptsize +1} \\  \mbox{\scriptsize 0} \\ \mbox{\scriptsize
-1} \\ \mbox{\scriptsize -1} \end{array}\right) &
   \qsupp{3}
   = \frac{1}{6}
   \left(\begin{array}{r} \mbox{\scriptsize +2} \\ \mbox{\scriptsize +1}
\\ \mbox{\scriptsize -1} \\ \mbox{\scriptsize -2} \\ \mbox{\scriptsize
-1} \\ \mbox{\scriptsize +1} \end{array}\right)
\end{array}
\]
\[
\begin{array}{ccc}
   \qsupp{4}
   = \frac{1}{6}
   \left(\begin{array}{r} \mbox{\scriptsize +2} \\ \mbox{\scriptsize -1}
\\ \mbox{\scriptsize -1} \\ \mbox{\scriptsize +2} \\ \mbox{\scriptsize
-1} \\ \mbox{\scriptsize -1} \end{array}\right) &
   \qsupp{5}
   = \frac{1}{6}
   \left(\begin{array}{r} \mbox{\scriptsize +1} \\ \mbox{\scriptsize -1}
\\ \mbox{\scriptsize +1} \\ \mbox{\scriptsize -1} \\ \mbox{\scriptsize
+1} \\ \mbox{\scriptsize -1} \end{array}\right) &
   \qsupp{6}
   = \frac{\sqrt{3}}{6}
   \left(\begin{array}{r}  \mbox{\scriptsize 0} \\ \mbox{\scriptsize +1}
\\ \mbox{\scriptsize -1} \\  \mbox{\scriptsize 0} \\ \mbox{\scriptsize
+1} \\ \mbox{\scriptsize -1} \end{array}\right).
\end{array}
\]

It is now a simple matter to plug these quantities into
Eq.~(\ref{eq:visvis}) to obtain the shear viscosity\footnote{We now
restore the dimensioned quantities, $c$ and $\Delta t$.}
\[
   \nu = \frac{c^2}{8\dt}
         \left(
         \frac{1}{(-\lamsupp{4})} +
         \frac{1}{(-\lamsupp{6})} -
         1
         \right)
       = \frac{c^2}{\dt}
         \left(
         \frac{1}{12f(1-f)^3} -
         \frac{1}{8}
         \right),
\]
which is the expected result for the viscosity of the FHP-I lattice gas
under the Boltzmann approximation~\cite{swolf,fchc}.

To conclude this subsection, we return to consider the pathological
factor, $g(f)$, that appears in front of the inertial term of the
Navier-Stokes equation, Eq.~(\ref{eq:ns}).  Note that the convective
derivative operator, $\partial_t + \bfu\cdot\nabla$, is Galilean
invariant, since it retains its form under a Galilean transformation,
$\bfx\rightarrow\bfx' + {\bf V}t'$ and $t\rightarrow t'$.  Thus, the
presence of the $g(f)$ factor in the inertial term is reflective of a
breakdown of Galilean invariance.  As has been pointed out by numerous
authors (see, e.g., \cite{fhp,swolf,fchc}), this is not surprising since
the lattice itself constitutes a preferred Galilean frame of reference.

In practical simulations of incompressible fluids, this factor is not a
problem, since it is constant and can be removed by a simple rescaling
of either the dependent or independent variables.  Similarly, the rather
unphysical equation of state, Eq.~(\ref{eq:eos}), is not a problem,
since the equation of state is irrelevant in the incompressible limit.
Efforts to extend lattice gas methods to treat compressible fluids,
however, must deal with these problems.  Techniques are known for doing
this, but they are outside the scope of this paper.

\newpage
\section{Exact Analysis}
\setcounter{equation}{0}
\label{sec:exact}

We will now proceed to develop an exact description of hydrodynamic
behaviour in the scaling limit, dropping the molecular chaos assumption
and including effects due to correlations.  As mentioned in
Section~\ref{sec:intro}, the diagrammatic formalism we develop here is
similar in many ways to the analogous formalism for continuum kinetic
theory.  Unlike continuous systems, however, the discrete nature of
lattice gases allows us to explicitly write the complete set of terms
which contribute correlations over a finite time interval as a sum over
a finite number of diagrams.  The discretization of lattice gases also
changes the nature of the vertices in correlation diagrams.  In a
lattice gas system, the vertices represent correlation interactions at a
single lattice site, and can be simply calculated from the
time-development equation.  There are a finite number of distinct vertex
types, corresponding to correlated particles arriving at and departing
{}from a single lattice point at a single timestep.

In Subsection~\ref{ssec:exactnot} we generalize our notation slightly to
deal with arbitrary sets of particles on the lattice.  We begin
developing the diagrammatic formalism for lattice gases in Subsection
\ref{ssec:exactcor} by discussing several alternate descriptions of
correlations in ensembles.  In Subsection~\ref{ssec:exactdyn} we express
a renormalized version of the Jacobian matrix $\jij$ in terms of an
infinite series.  The terms in this series are factorized into
independent contributions from each lattice site in
Subsection~\ref{ssec:factcollop}, and written in diagrammatic notation
in Subsection~\ref{ssec:diagramm}.  In Subsection~\ref{ssec:approxx} we
describe how several useful approximations, such as the ring
appproximation, can be calculated in our formalism as a sum over a
restricted class of diagrams.  In Subsection~\ref{ssec:eigrnrm}, we
prove that a fairly wide class of lattice gases have the property that
the only effect of correlations is to modify the eigenvalues of the
$J$-matrix in terms of which the transport coefficients are described.
Finally, in Subsection~\ref{ssec:correlationsandcollision}, we describe
the effects of the higher-order collision terms $C_1,C_2$ on the exact
hydrodynamic equations.

In Sections~\ref{sec:examples} and \ref{sec:approximations}, we apply
the techniques of this section to the lattice gases described in
Section~\ref{sec:exboce}.  The reader may find it helpful to refer to
Sections~\ref{sec:examples} and \ref{sec:approximations} for concrete
examples of the formalism while reading this section.

\subsection{General Notation}
\label{ssec:exactnot}

In this subsection, we develop a slightly more general system of
notation suitable for describing the exact dynamics of a lattice gas.
This notation is similar to the matrix form of notation used in
Section 2; however, we now wish to consider the space of all
configurations of the full system, rather than simply the set of
states at a single lattice point.

Recall that we can refer to an arbitrary bit of the system by an index
$a \in {\cal B}$.  We now introduce an {\sl advection operator},
\label{pg:ap} $\advop{b}{c}$, which acts on the space of bits in the
entire system.  This operator is an $N$ by $N$ permutation matrix.  It
connects bit $b$ with bit $c$ if and only if the particle represented by
bit $c$ moves into bit $b$ during the advection substep.  That is,
\[
   \advop{b}{c} =
   \left\{\begin{array}{ll}
            1 & \mbox{if $b=a\left(i(c),\bfx(c)+\bfcsup{i(c)}\right)$}\\
            0 & \mbox{otherwise.}
          \end{array}
   \right.
\]
In terms of this operator, the exact dynamical equation for the lattice
gas, Eq.~(\ref{eq:microdyn}), may be written
\bge
   \nsup{b}(\tpdt) =
   \advop{b}{c}
   \left(
      \nsup{c}\left(t\right)+
      \csup{i(c)}
         \left(
         \nsup{a(\ast,\bfx (c))}\left(t\right)
         \right)
   \right),
   \label{eq:microdynm}
\ee
and the lattice Boltzmann equation, Eq.~(\ref{eq:boltzmann}), may be
written,
\bge
   \nnsup{b}(\tpdt) =
   \advop{b}{c}
   \left(
      \nnsup{c}\left(t\right)+
      \ccsup{i(c)}
         \left(
         \nnsup{a(\ast,\bfx (c))}\left(t\right)
         \right)
   \right).
   \label{eq:boltzmanna}
\ee

We denote the ensemble mean of an arbitrary product
of the $\nsup{a}$'s by
\[
\nnsup{\alpha} = \left\langle \prod_{a \in \alpha} \nsup{a}
\right\rangle, \;\;
\alpha \subseteq {\cal B}.
\]
Henceforth, we use the Greek letters $\alpha, \beta, \ldots$ to denote
subsets of the set ${\cal B}$ and the letters $\mu, \nu, \ldots$ to
denote subsets of $B$.  We will sometimes use a roman index to denote an
index set with a single element, as in $\nnsup{a} = \nnsup{\{a\}}$.
Additionally, for quantities subscripted or superscripted by a single
set, we will sometimes replace the set by its elements, as in
$\nnsup{abc}=\nnsup{\{ a,b,c \}}$.  As a final point of notation, an
index set with a circumflex \label{pg:ab} is assumed to have at least
two elements; i.e., $|\widehat{\alpha}| \geq 2$.

Finally, we generalize the advection operator ${\cal A}$ to be a
permutation matrix $\advop{\alpha}{\beta}$ acting on the $2^N$
dimensional space of subsets of ${\cal B} $.  For a fixed set of bits
$\beta = \{b_1, \ldots, b_q\} \subseteq {\cal B} $, if we take $\alpha
= \{a_1,
\ldots, a_q\}$ to be the set of particles which $\beta$ goes to under
advection; i.e., $a_j=a(i(b_j), \bfx(b_j) +
\bfcsup{i(b_j)})$, then
\[
   \advop{\alpha'}{\beta} =
   \left\{ \begin{array}{ll}
             1, & \alpha' = \alpha \\
             0, & \alpha' \neq \alpha
           \end{array}
   \right.
\]
Thus, for example, $\advop{a}{b}$ is 1 when $a=a(i(b), \bfx(b) +
\bfcsup{i(b)})$, and 0 otherwise, in agreement with the previous notation.

\subsection{Representations of Correlations}
\label{ssec:exactcor}

An ensemble is generally defined to be a distribution on the space of
possible configurations of the entire lattice gas system.  In this
subsection, we discuss several alternative descriptions of the
probability distribution describing an ensemble.

Given a set $S$ of boolean variables $S = \{\nsup{1}, \nsup{2}, \ldots,
\nsup{N}\}$, a probability distribution on $S$ can be described in
several equivalent ways.  The most familiar description is given by
assigning a probability to each possible set of values for the
$\nsup{i}$'s; i.e., given any set $\alpha\subseteq \{1,2,\ldots,N\}$,
we define the probability that the corresponding set of $n$'s are
equal to 1 and the rest are 0 to be \label{pg:bl}
\[
   \ppsup{\alpha} =
   \mbox{Probability that $(\nsup{1}, \nsup{2}, \ldots, \nsup{N}) =
   (\chisup{\alpha}(1), \chisup{\alpha}(2), \ldots, \chisup{\alpha}(N))$},
\]
where
\[
   \chisup{\alpha}(i) =
     \left\{
       \begin{array}{ll}
         1, & i \in \alpha \\
         0, & i \not\in \alpha
       \end{array}
     \right.
\]
Since there are $2^N$ such subsets $\alpha$, and since
\[
\sum_\alpha \ppsup{\alpha} = 1
\]
is the only constraint, the space of probability distributions on these
$N$ variables is $(2^N - 1)$-dimensional.  (In fact, it is a $(2^N -
1)$-dimensional simplex.)

An equivalent description of a probability distribution on $S$ can be
given by defining the means $\nnsup{\alpha}$ for each possible product of
elements of $S$.  (Note that $(\nsup{i})^2 = \nsup{i}$, so that the mean
of any product of elements of $S$ is equal to $\nnsup{\alpha}$ for some
$\alpha$.) \label{pg:bi}

In terms of the $\ppsup{\alpha}$'s, the means can be expressed as
\bge
\nnsup{\alpha} = \sum_{\beta \supseteq \alpha} \ppsup{\beta.}
\label{eq:probabilitytomean}
\ee
This relationship can be inverted to get
\bge
   \ppsup{\alpha} =
   \sum_{\beta \supseteq \alpha} (-1)^{|\beta| - |\alpha|}
   \nnsup{\beta}.
\label{eq:meantoprobability}
\ee
The space of allowed values for the means is also $2^N-1$ dimensional,
since $\nnsup{\emptyset} = 1$.  These two descriptions of a distribution
are equivalent, in the sense that the information contained in either
description is exactly sufficient to completely specify the
distribution.  In fact, Eqs.~(\ref{eq:probabilitytomean}) and
(\ref{eq:meantoprobability}) show that the probabilities and the
multipoint means are related by a linear transformation.

Probability distributions in which the $\nsup{i}$ are
distributed independently have means given by
\[
\nnsup{\alpha} = \prod_{a \in \alpha} \nnsup{a}.
\]
The space of {\sl independent} distributions on $N$ variables is clearly
$N$ dimensional, and is parameterized by $\nnsup{a}$, $a \in \{1, 2,
\ldots, N\}$.  The Fermi-Dirac equilibrium (\ref{eq:fd}) is an example
of an independent distribution.

A third description of a distribution on $S$ can be given in terms of
{\sl connected correlation functions}, or CCF's~\cite{ZJ}.  For each
$\alpha \subseteq\{1,2,\ldots, N\}$, there is a CCF, which we denote
$\ggamsup{\alpha}$. \label{pg:ae} It is easiest to define the
CCF's implicitly by expressing the means in terms of the CCF's
through the equation
\bge
   \nnsup{\alpha} =
   \fsup{\alpha}(\ggamsup{\ast}) =
   \sum_{\bfzeta \in \pi(\alpha)}
   \ggamsup{\zeta_1} \ggamsup{\zeta_2} \ldots \ggamsup{\zeta_q},
   \label{eq:meanccf}
\ee
\label{pg:au} where $\pi(\alpha)$ is the set of all partitions of
$\alpha$ into disjoint subsets, $\zeta_1, \ldots, \zeta_q$.  Explicitly,
\bgeas
   \nnsup{a} & = & \ggamsup{a} \\
   \nnsup{ab} & = & \ggamsup{ab} + \ggamsup{a} \ggamsup{b} \\
   \nnsup{abc} & = & \ggamsup{abc} + \ggamsup{a}\ggamsup{bc} +
                       \ggamsup{b}\ggamsup{ac} +
                       \ggamsup{c}\ggamsup{ab} + \ggamsup{a}
                       \ggamsup{b} \ggamsup{c.} \\
   {\rm etc...} & &
\eeas
We will refer to $\nnsup{\alpha}$ ($\ggamsup{\alpha}$) as an $n$-mean
($n$-CCF), when $|\alpha| = n$.

Observing that the defining equation for an $n$-mean in terms of CCF's
contains on the right-hand side exactly one $n$-CCF, and otherwise
only $m$-CCF's with $m < n$, we see that the above set of equations can be
inverted by induction on $n$, to get a functional relationship of the form
\bge
  \ggamsup{\alpha} = \gsup{\alpha}(\nnsup{\ast}),
  \label{eq:ccfmean}
\ee
\label{pg:av} where $g$ and $f$ are inverses.  Explicitly, we have
\bgeas
   \ggamsup{a} & = & \nnsup{a} \\
   \ggamsup{ab} & = & \nnsup{ab} - \nnsup{a} \nnsup{b} \\
   \ggamsup{abc} & = &\nnsup{abc} - \nnsup{a}\nnsup{bc} -
                        \nnsup{b}\nnsup{ac} -
                        \nnsup{c}\nnsup{ab} + 2 \nnsup{a} \nnsup{b}
                        \nnsup{c}. \\
   {\rm etc...} & &
\eeas

Again, the description of a probability distribution in terms of CCF's
is completely equivalent to the discriptions in terms of
$\nnsup{\alpha}$'s and $\ppsup{\alpha}$'s.  There are $2^N - 1$
independent $\ggamsup{\alpha}$'s, as $\ggamsup{\emptyset}$ is not
defined.  Note that the relationship between the $\ggamsup{\alpha}$'s
and the $\nnsup{\alpha}$'s (or the $\ppsup{\alpha}$'s) is nonlinear.

The main reason that CCF's will be a useful description for us is that
in an independent distribution, all $n$-CCF's are 0, for $n > 1$.  Thus,
the distance of a distribution from one which is independent is measured
by the quantities $\ggamsup{\widehat{\alpha}}$ (recall that an index set
with a circumflex ($\;\widehat{}\;$) is constrained to have more than one
element).  In the subsequent analysis, we will find both the
$\nnsup{\alpha}$ and $\ggamsup{\alpha}$ notations to be useful, and we
will use the functions $f$ and $g$ to move between the two descriptions.

\subsection{Exact Dynamics}
\label{ssec:exactdyn}

We will now proceed to rewrite the exact dynamical equation for the
lattice gas, Eq.~(\ref{eq:microdynm}), in a form similar to that of the
lattice Boltzmann equation.  For most of this section we will assume
that the collision operator respects the conservation laws exactly and
obeys semi-detailed balance, so that $C^i = C^i_0$ and $C^i_1 = C^i_2 =
0$; in Subsection~\ref{ssec:correlationsandcollision} we will discuss
the effect of correlations when we include nonzero $C_1^i$ and $C_2^i$.

Recall that the collision operator in the lattice Boltzmann equation
(\ref{eq:boltzmanna}) can be linearized, as in Eq.~(\ref{eq:eqfornone}),
to give
\bge
\nnsup{b}(\tpdt)=
   \advop{b}{c} \left(N^c (t)+
	\epsilon                      \jop{i(c)}{j}
   \nnsup{a(j,\bfx (c))}_1 (t)\right).
   \label{eq:eqfornonea}
\ee
In order to describe the macroscopic behavior of the system we need only
include the effect of the collision operator $C_0$ up to the order
$\epsilon$ term associated with the Jacobian, since as we have seen in
Section~\ref{sec:chen} the higher order terms associated with the
collision operator have no effect upon the hydrodynamic equations of the
system.  In fact, the hydrodynamic equations derived in
Section~\ref{sec:chen}, and the associated advective and diffusive
transport coefficients, depend upon only the Jacobian matrix $\jij$,
through its eigenvalues and eigenvectors.  What we now wish to show is
that if we drop the Boltzmann molecular chaos assumption and analyze the
exact ensemble-averaged equation of motion for a lattice gas, in the
scaling limit we will get an equation identical to
Eq.~(\ref{eq:eqfornonea}) in form, but with a {\sl renormalized}
$J$-matrix.  The exact transport coefficients can then be expressed in
terms of the eigenvalues and eigenvectors of the renormalized $J$-matrix
using precisely the same expressions as in the Boltzmann analysis.
Furthermore, we find that for a large class of lattice gases, the
eigenvectors of the renormalized $J$-matrix are unchanged; only the
eigenvalues of the matrix undergo renormalization due to correlations.

We begin with the exact time-development equation,
Eq.~(\ref{eq:microdynm}).  By taking the ensemble average of the product
of this equation over all $a$ in an arbitrary set $\alpha \subseteq
{\cal B}$, we can write the exact equation for an arbitrary multipoint
mean at time $\tpdt$ in terms of multipoint means at time $t$.  We have
\bge
\nnsup{\alpha}(\tpdt) =
\left\langle \prod_{a \in \alpha} \nsup{a} (\tpdt) \right\rangle =
\sum_\beta \advop{\alpha}{\beta}
\left\langle \prod_{b \in \beta}
\left[\nsup{b} (t) + \csup{i(b)}(\nsup{\ast}(\bfx(b), t))\right]
\right\rangle.
\label{eq:exactm}
\ee
To express the right-hand side in terms of multipoint means, it will be
convenient to rewrite this equation in a more compact notation.  For a
set $\beta \subseteq {\cal B}$, let us define $L_\beta$ to be the subset
of points in $L$ which contain at least one particle in the set $\beta$;
that is,
\[
   L_\beta = \{ \bfy \in L : \bfx(b) = \bfy\;\; {\rm for\; some}\;\; b \in
   \beta\}.
\]
Similarly, we define $\beta_\bfx$ to be the set of $i$'s corresponding
to the particles in $\beta$ at the point $\bfx$; that is,
\[
   \beta_\bfx = \{ i \in B: a(i, \bfx) \in\beta\}.
\]
We can now factorize \label{pg:ad} the product appearing on the
right-hand side of Eq.~(\ref{eq:exactm}) into contributions from each of
the points in $L_\beta$, by writing
\[
   \prod_{b \in \beta}
   \left[\nsup{b} (t) + \csup{i(b)}(\nsup{\ast}(\bfx(b), t))\right] =
   \prod_{\bfx \in L_\beta}
   \prod_{i \in \beta_\bfx}
   \left[\nixt + \csup{i}(\nsup{\ast}\bfxt)\right].
\]
The innermost product on the right now depends only on quantities at a
single site, $\bfx$.

The functions $\csup{i}(\nsup{\ast})$ can be expressed as polynomials in
the $\nsup{i}$'s of the form
\[
   \csup{i}(\nsup{\ast}) = \sum_{\nu \subseteq B} \mixten{k}{i}{\nu}
                            \prod_{j \in \nu} \nsup{j},
\]
where the $\mixten{k}{i}{\nu}$ are coefficients which may depend only on
random bits at each lattice site, and which are constant for
deterministic lattice gases.  \label{pg:bc} Thus, we can write
\[
   \prod_{i \in \mu} [\nsup{i} + \csup{i}(\nsup{\ast})] =
   \mixten{v}{\mu}{\nu} \prod_{j \in \nu}\nsup{j},
\]
where the quantities $\mixten{v}{\mu}{\nu}$ may contain random bits at
each site.  Taking the ensemble average over any such random bits, we
get the {\it mean vertex coefficients} $\mixten{V}{\mu}{\nu}$
\bge
\mixten{V}{\mu}{\nu} = \langle\mixten{v}{\mu}{\nu} \rangle.
\label{eq:vertexmean1}
\ee

The state transition probabilities $A (s \rightarrow s')$ may be
interpreted as elements of a collision matrix on the space of
probabilities, $\ppsup{s}$, in the sense that the post-collision
probability of a state $s'$ is given by
\bge
\sum_s A (s \rightarrow s') \ppsup{s}.
\label{eq:cmosops}
\ee
Similarly, the matrix $\vop{\mu}{\nu}$ can be interpreted as a collision
matrix on the space of means.  \label{pg:bq} Using
Eqs.~(\ref{eq:probabilitytomean}), (\ref{eq:meantoprobability}), and
(\ref{eq:cmosops}), the matrix $V$ can be related to $A$ through the
equation
\bge
\mixten{V}{\mu}{\nu} = \sum_{s' \supseteq \mu}
\sum_{s \subseteq \nu}(-1)^{| \nu | - | s |} A (s \rightarrow s'),
\label{eq:vertexmean2}
\ee
where we have identified the state $s$ with the set of bits which are 1
in that state ($s \subseteq B$).  Clearly, the $2^{2n}$ matrix elements
$\vop{\mu}{\nu}$ depend only on the sets $\mu$ and $\nu$, and on the
form of the collision operator.  In particular, they do not depend on
$\bfx$, or on the values of the $\nsup{a}$'s.  Note that
$\vop{\emptyset}{\nu} =
\mixten{\delta}{\emptyset}{\nu}$, regardless of the specific
lattice gas or collision rule.

Eq.~(\ref{eq:exactm}) can now be rewritten in the form
\bge
\nnsup{\alpha}(\tpdt) =
   \advop{\alpha}{\beta} \kop{\beta}{\gamma} \nnsup{\gamma} (t),
   \label{eq:mdabs}
\ee
where $K$ is an operator expressing the complete collision part of the
time development, given by \label{pg:be}
\bge
   \kop{\beta}{\gamma}
   = \prod_{\bfx \in L_\beta} \vop{\beta_\bfx}{\gamma_\bfx}.
   \label{eq:dmprod}
\ee
We shall now transform the exact equation of motion,
Eq.~(\ref{eq:mdabs}), into an equation of motion for the CCF's.  Using
the functions $f$ and $g$ from Eqs.~(\ref{eq:meanccf}) and
(\ref{eq:ccfmean}) to convert from means to CCF's and back,
Eq.~(\ref{eq:mdabs}) can be rewritten as
\[
   \ggamsup{\alpha}(\tpdt) =
   \gsup{\alpha}(\advop{\ast}{\beta} \kop{\beta}{\gamma}
   \fsup{\gamma}(\ggamsup{\ast})).
\]
However, note that from the definitions of $f$ and $g$, a permutation on
the bit labels can be performed before or after calculating means from
CCF's or vice versa, without changing the result.  Thus, $g$ and ${\cal
A}$ commute, and this equation can be rewritten as
\bge
   \ggamsup{\alpha}(\tpdt) =
   \advop{\alpha}{\beta} \phisup{\beta}(\ggamsup{\ast}),
   \label{eq:gamtd}
\ee
where
\[
   \phisup{\beta}(\ggamsup{\ast}) \equiv
   \gsup{\beta}(\kop{\ast}{\gamma} \fsup{\gamma}(\ggamsup{\ast})).
\]

Since we have chosen to expand around an equilibrium which is an
independent distribution, all of the quantities
$\ggamsup{\widehat{\alpha}}$ are of order $\epsilon$ (recall $|
\widehat{\alpha}| \geq 2$).  To perform a complete analysis of
the system in the scaling limit, we need only keep terms of first order
in $\epsilon$ in these quantities, and hence in the expression
$\phisup{\widehat{\beta}}(\ggamsup{\ast})$.  This is essentially because
the conservation equation (\ref{eq:lconslaws}) is linear, and is
unchanged by the inclusion of correlations; the effect of correlations
appears only at order $\epsilon$ in Eq.~(\ref{eq:lboltza})\footnote{The
ordering of $\ggamsup{\widehat{\alpha}}$ at ${\cal O}(\epsilon)$ will be
assumed throughout this paper.  Of course, there will be ${\cal
O}(\epsilon^2)$ contributions as well.  These terms are irrelevant in
the scaling limit unless there is a divergence in their coefficients.
We know of no lattice gases which have such a divergence at ${\cal
O}(\epsilon^2)$ that do not already diverge at ${\cal O}(\epsilon)$.}.
We can therefore linearize Eq.~(\ref{eq:gamtd}) in an analogous fashion
to the linearization of the Boltzmann equation in
Eq.~(\ref{eq:eqfornonea}), to get
\bge
   \ggamsup{\widehat{\alpha}}(\tpdt) =
   \advop{\widehat{\alpha}}{\widehat{\beta}} \left(
   \epsilon \ckop{\widehat{\beta}}{a}
   \nnsup{a}_1 (t) +
   \ckop{\widehat{\beta}}{\widehat{\gamma}}
   \ggamsup{\widehat{\gamma}}(t)\right),
   \label{eq:cdlin}
\ee
where \label{pg:bf}
\bge
   \ckop{\beta}{\gamma} = \subscrpt{\left.
   \frac{\pdv \phisup{\beta}}{\pdv \ggamsup{\gamma}}
   \right|}{0} =  \subscrpt{\left.
   \frac{\pdv \gsup{\beta}}{\pdv \nnsup{\sigma}}
   \right|}{0} \kop{\sigma}{\tau} \subscrpt{\left.
   \frac{\pdv \supscrpt{f}{\tau}}{\pdv \ggamsup{\gamma}}
   \right|}{0}.
   \label{eq:dexp}
\ee
Similarly, when we include the effects of correlations to order
$\epsilon$ in Eq.~(\ref{eq:gamtd}) for $\alpha= \{a\}$, the dynamical
equation for $N^a = \Gamma^a$ becomes
\bge
\nnsup{a}(\tpdt)  =
   \advop{a}{ b}
     \left\{
        N^b (t)+
	\epsilon\left[ \ckop{b}{c} -\mixten{\delta}{b}{c}\right]
        \nnsup{c}_1 (t) +
        \ckop{b}{\widehat{\gamma}} \ggamsup{\widehat{\gamma}}(t)
     \right\}.
  \label{eq:exact1}
\ee
Note that if we set
$\ggamsup{\widehat{\alpha}} = 0$ in this equation, we get back the
linearized Boltmann equation (\ref{eq:eqfornonea}), since
\bge
 \ckop{b}{c} -\mixten{\delta}{b}{c} = \mixten{\delta}{\bfx(b)}{\bfx(c)}
 \mixten{J}{i (b)}{i (c)}.
 \label{eq:fubar}
\ee
Inserting Eq.~(\ref{eq:cdlin}), we can now write the exact equation of
motion for the quantities $\nnsup{a}$ in the form of an infinite series,
\bge
\nnsup{a}(\tpdt)  =
   \advop{a}{ b} \left(N^b (t)+
	\epsilon \cjop{b}{c}
	\nnsup{c}_1 (t)\right),
  \label{eq:rbolt}
\ee
where
\bge
   \cjop{b}{c} =
\mixten{\delta}{\bfx(b)}{\bfx(c)}
\mixten{J}{i (b)}{i (c)} +
  \ckop{b}{\widehat{\alpha}}
   (\advop{\widehat{\alpha}}{\widehat{\beta}} \ckop{\widehat{\beta}}{c} +
   \advop{\widehat{\alpha}}{\widehat{\beta}}
   \ckop{\widehat{\beta}}{\widehat{\gamma}}
   (\advop{\widehat{\gamma}}{\widehat{\delta}}\ckop{\widehat{\delta}}{c} +
   \ldots)).
   \label{eq:rj}
\ee
\label{pg:bb} We now have an expression for the mean occupation number
of a certain bit of the system at position $\bfx$ and time $t + \Delta
t$, written as an infinite sum of terms, each of which is a function of
the quantities $N_0$ and $N_1$ at nearby lattice sites $\bfx'$ and at
previous time steps $t'$.  As we consider terms in this series with more
and more factors of ${\cal A}{\cal K}$, the positions and times at which
these quantities are evaluated will differ from $\bfx$ and $t$ by
greater amounts.  However, for any given term in the series, the means
$N_0^{a (i,\bfx')}(t')$ and $N_1^{a (i,\bfx')}(t')$ can be replaced by
$N_0^{a (i,\bfx)}(t)$ and $N_1^{a (i,\bfx)}(t)$, and the expression
(\ref{eq:rj}) will only change by a quantity of order $\epsilon$, since
spatial derivatives are ordered as $\epsilon$, and temporal derivatives
are ordered as $\epsilon^2$.  Such a modification for a finite number of
terms does not change the behavior of the system in the hydrodynamic
limit.  In fact, it follows that whenever the sum of terms in
Eq.~(\ref{eq:rj}) converges on a scale which goes to zero in the
hydrodynamic limit, we can drop all the spatial and temporal variations
in the single-particle means.  Thus, Eq.~(\ref{eq:rbolt}) can be
rewritten in precisely the form of Eq.~(\ref{eq:eqfornonea}), where $J$
is replaced by the renormalized matrix \label{pg:ba}
\bge
   \tjop{i}{j} (\bfx) =
   \sum_{\bfy\in L}
   \cjop{a(i, \bfx)}{a(j, \bfy)},
   \label{eq:rjt}
\ee
with all ${\cal K}$'s in ${\cal J}$ evaluated at the point $\bfx$ and
the time $t$.

It is important to note that the above argument breaks down when the sum
(\ref{eq:rj}) is divergent.  In this case, the effects of large scale
variations in the $\nnsup{a(i,\bfx)}$'s must be considered.  In general,
for lattice gases where the sum (\ref{eq:rj}) is divergent, one must be
quite careful about the analysis.  For certain lattice gases, however,
particularly systems in which the conserved quantities are ordered, we
are interested in expanding around an equilibrium which is spatially
invariant (for example, the FHP-I lattice gas).  In this case, the
zero-order means can be replaced by their universal values; however, one
must still treat the spatial variation of the first-order means
carefully.

Now that we have rewritten the exact dynamical equation in a form
commensurate with the original form of the lattice Boltzmann equation,
the renormalized transport coefficients for the theory can be
related to the eigenvalues of the matrix $\tjij$ in the same way that
the original (Boltzmann) transport coefficients were related to the
eigenvalues of the matrix $\jij$.  Thus, if we can compute the matrix
$\tjij$ exactly, we can also compute the exact renormalized transport
coefficients.  Most of the rest of this paper is devoted to methods of
calculating and approximating the matrix $\tjij$, and applications to
specific lattice gases.

\subsection{Factorization of Collision Operator}
\label{ssec:factcollop}

In the next two subsections we will show that expression~(\ref{eq:rj})
for ${\cal J}$ can be written in a diagrammatic notation, allowing us
to perform a perturbative calculation of ${\cal J}$ by summing over
``Feynman diagram''-like objects, where the contribution from each
diagram is just the product of factors associated with its vertices.
The principal observation which allows this reduction is the fact that
the collision operator $\ckop{\alpha}{\beta}$ is factorizable.  We
devote this subsection to demonstrating the exact form of this
factorization.
\begin{thm}
\label{thm:factor}
For fixed $\alpha$ and $\beta$, $\ckop{\alpha}{\beta}$ can be broken
down into a product of contributions from distinct vertices.
Explicitly,
\bge
   \ckop{\alpha}{\beta} =
   \prod_{\bfx \in L_\alpha}\cvop{\alpha_\bfx}{\beta_\bfx},
   \label{eq:dv}
\ee
where the correlation vertex coefficients (CVC's), ${\cal V}$, are
defined by \label{pg:br}
\bge
   \cvop{\alpha_\bfx}{\beta_\bfx} =
   \sum_{\mu, \nu}
   (-1)^{|\alpha_\bfx \setminus \mu|}
   \left(
      \prod_{i \in (\alpha_\bfx \setminus \mu)} \nnsupo{i}{0}
   \right)
   \left(
      \prod_{j \in (\nu \setminus \beta_\bfx)} \nnsupo{j}{0}
   \right) \vop{\mu}{\nu},
   \label{eq:vdef}
\ee
with the sum taken over $\mu \subseteq \alpha_\bfx$ and $\nu \supseteq
\beta_\bfx,
\nu \subseteq B$.
\end{thm}

\noindent {\bf Proof:} \,\, To prove the theorem, we need only compute the
derivatives of the means and CCF's with repect to one another, evaluated
at the equilibrium point.  From the definitions (\ref{eq:meanccf}) and
(\ref{eq:ccfmean}), it is clear that when $\beta \not \subseteq \alpha$,
$\pdv \gsup{\alpha} / \pdv \nnsup{\beta} = \pdv \fsup{\alpha} / \pdv
\ggamsup{\beta} = 0$.  When $\beta \subseteq \alpha$, it is also fairly
straightforward to compute from Eq.~(\ref{eq:meanccf}),
\bge
   \subscrpt{\left. \frac{\pdv \fsup{\alpha}}{\pdv \ggamsup{\beta}}
   \right|}{0} =
   \prod_{a \in (\alpha \setminus \beta)}
   \nnsupo{i(a)}{0} =
   \prod_{\bfx \in L_\alpha}\prod_{i \in (\alpha_\bfx \setminus \beta_\bfx)}
   \nnsupo{i}{0}.
   \label{eq:dfc}
\ee
We claim that similarly, when $\beta \subseteq \alpha$, the derivatives $\pdv
g^{\alpha} / \pdv  N^{\beta}$ are given by
\begin{lem}
\bge
   \subscrpt{\left. \frac{\pdv \gsup{\alpha}}{\pdv \nnsup{\beta}}
   \right|}{0} =
   \prod_{a \in \alpha \setminus \beta}
   (-\nnsupo{i(a)}{0}) =
   \prod_{\bfx \in L_\alpha}\prod_{i \in \alpha_\bfx \setminus \beta_\bfx}
   (-\nnsupo{i}{0}).
   \label{eq:dgm}
\ee
\end{lem}
\noindent {\bf Proof of Lemma:} \,\,From Eqs.~(\ref{eq:ccfmean}) and
(\ref{eq:meanccf}), one has
\[
   \subscrpt{\left. \frac{\pdv \gsup{\alpha}}{\pdv \nnsup{\beta}}
   \right|}{0} =
   \mixten{\delta}{\alpha}{\beta} -
   \sum_{\gamma \subseteq \alpha \setminus \beta}
   (\prod_{a \in \gamma}  \nnsupo{i(a)}{0})
   \subscrpt{\left.
   \frac{\pdv \gsup{\alpha \setminus \gamma}}{\pdv \nnsup{\beta}}
   \right|}{0}.
\]
Applying this equation repeatedly, we arrive at an expression of the form
\[
   \subscrpt{\left.
   \frac{\pdv \gsup{\alpha}}{\pdv \nnsup{\beta}}
   \right|}{0} =
   h(|\alpha \setminus \beta|)
   \left(\prod_{i \in \alpha \setminus \beta} \nnsupo{i}{0}\right),
\]
where the function $h(n)$ satisfies $h(0) = 1$ and
\[
   \begin{array}{cc}
   h(n) = -\sum_{m = 1}^n
          \left(
            \begin{array}{c}
            n \\ m
            \end{array}
          \right) h(n-m),  &
   n\neq 0.
   \end{array}
\]
Assuming that $h(n') = (-1)^{n'}$ for $n' < n$, we have
\[
   h(n) = -\sum_{k = 0}^{n-1}
          \left(
            \begin{array}{c}
            n \\ k
            \end{array}
          \right) (-1)^k
        = (-1)^n,
\]
so by induction, we have proven the lemma.$\Box$

Substituting Eqs.~(\ref{eq:dmprod}), (\ref{eq:dfc}), and (\ref{eq:dgm})
into Eq.~(\ref{eq:dexp}), we get
\[
   \ckop{\alpha}{\beta} =
   \sum_{\gamma \subseteq \alpha, \zeta \supseteq \beta}
   \left(\prod_{\bfx \in L_\alpha}
         \prod_{i \in (\alpha_\bfx \setminus \gamma_\bfx)}
         (-\nnsupo{i}{0})
   \right)
   \left(\prod_{\bfx \in L_\gamma}
         \vop{\gamma_\bfx}{\zeta_\bfx}
   \right)
   \left(\prod_{\bfx \in L_\zeta}
         \prod_{j \in (\zeta_\bfx \setminus \beta_\bfx)}
         \nnsupo{j}{0}
   \right).
\]
Since $\gamma \subseteq \alpha$, clearly $L_\gamma\subseteq L_\alpha$.  The
fact that $\vop{\emptyset}{\nu} = 0$ when $\nu \neq 0$ implies that all
terms with $L_\zeta \not\subseteq L_\gamma$ vanish, so we can restrict the
sum over $\zeta$ to only those $\zeta$ with $L_\zeta\subseteq L_\alpha$.
In this case for $\bfx\not\in L_\gamma$, clearly
$\vop{\gamma_\bfx}{\zeta_\bfx} = \vop{\emptyset}{\emptyset} = 1$
or $\vop{\gamma_\bfx}{\zeta_\bfx} = 0$, and
for $\bfx\not\in L_\zeta$, clearly $\zeta_\bfx \setminus \beta_\bfx =
\emptyset$, so we can replace all the products with products over $\bfx
\in L_\alpha$, giving
\bge
   \ckop{\alpha}{\beta} =
   \sum_{\gamma ,\zeta}
   \prod_{\bfx \in L_\alpha}
   \prod_{i \in (\alpha_\bfx \setminus \gamma_\bfx)} (-\nnsupo{i}{0})
   \vop{\gamma_\bfx}{\zeta_\bfx}
   \prod_{i \in (\zeta_\bfx \setminus \beta_\bfx)}
   \nnsupo{i}{0},
   \label{eq:dsp}
\ee
where the sum is over all $\gamma \subseteq \alpha$ and all $\zeta$
satisfying $L_\zeta \subseteq L_\alpha$ and $\zeta \supseteq \beta$.
For each $\bfx$, however, this means that $\gamma_\bfx$ and $\zeta_\bfx$
are summed over all $\gamma_\bfx\subseteq\alpha_\bfx$ and $\zeta_\bfx
\supseteq \beta_\bfx$.  Since the $\gamma_\bfx$ and $\zeta_\bfx$ are
independent for different $\bfx$, a distributive rule can be applied to
Eq.~(\ref{eq:dsp}), giving $\ckop{\alpha}{\beta}$ in exactly the form
stated in Theorem
\ref{thm:factor}, so the proof is complete.$\Box$

\subsection{Diagrammatics}
\label{ssec:diagramm}

Using the result from the previous subsection, it is possible to express
every term in ${\cal J}$ in
diagrammatic form.  A generic term in $\cjop{a}{b}$ is of the form
\bge
\ckop{\alpsupp{k}}{\betsupp{k}}
\advop{\betsupp{k}}{\alpsupp{k-1}}
\ckop{\alpsupp{k-1}}{\betsupp{k-1}}
\cdots
\advop{\betsupp{2}}{\alpsupp{1}}
\ckop{\alpsupp{1}}{\betsupp{1}}
\label{eq:term}
\ee
with $\alpsupp{i}$ and $\betsupp{i}$ fixed (i.e., not summed over), and
$|\alpsupp{i}|, |\betsupp{i}| \geq 2$, except for the endpoints where
$\alpsupp{k} = \{a\}$ and $\betsupp{1} = \{b\}$.

We define a {\it diagram} $T$ by an integer $k(T)$, which we refer to as
the {\it length} of the diagram $T$, \label{pg:bd} and a function
$\alpha_{T}(\tau)$, where for each $\tau\in\{0,\ldots,k\}$,
$\alpha_{T}(\tau) \subseteq {\cal B}$.  Geometrically, we associate each
$a\in\alpha_{T}(\tau)$ with a {\sl virtual particle} (VP) moving from
$(\bfx (a),\tau)$ to $(\bfx (a) +
\bfcsup{i(a)}, \tau+1)$ on the lattice $\Lambda_{k+1}=
L\times\{0,\ldots,k+1\}$.  We refer to $\alpha_{T}(\tau)$ as the set of
{\sl outgoing} VP's for the diagram $T$. \label{pg:ak}

It is natural to define a corresponding set of {\sl incoming} VP's for
$\tau > 0$ by $\beta_{T}(\tau) = \{b: a(i(b),\bfx (b)-\bfcsup{i(b)}) \in
\alpha_{T}(\tau-1)\}.$ We also define $\sigma_{T}(\tau) = |\alpha_{T}(\tau)|$
to be the total number of outgoing VP's for each value of $\tau$.
Finally, given a diagram $T$, we can define a weight function
\[
   W(T) = \prod_{\bfx \in L} \prod_{1 \leq \tau \leq k(T)}
          \cvop{\alpha_{T}(\tau)_\bfx}{\beta_{T}( \tau)_\bfx},
\]
by taking the product of ${\cal V}$ over all vertices.

The term (\ref{eq:term}) can now be represented by the diagram $T$ with
$\alpha_{T}(\tau) =\alpsupp{\tau}$, where for consistency $\alpsupp{0}$
is defined to be the unique set with
$\advop{\betsupp{1}}{\alpsupp{0}}=1$.  When $\tau \neq 0$,
$\advop{\betsupp{\tau+1}}{\alpsupp{\tau}} = 1$, so $\beta_T(\tau) =
\betsupp{\tau}$ for all $\tau$.  It follows that the contribution from
the term (\ref{eq:term}) is exactly given by $W(T)$.  Thus, we can
rewrite expression (\ref{eq:rj}) for $\cjop{a}{b}$ as a sum over
diagrams
\bge
   \cjop{a}{b} = \sum_{k=1}^\infty \sum_{T\in \ctop{a}{b}(k)} W(T),
   \label{eq:caljsum}
\ee
where in general we define the set of diagrams $\ctop{\alpha}{\beta}(k)$ by
\bgeas
\ctop{\alpha}{\beta}(k)  & = & \{T:k = k (T), \;
\sigma_T (l) > 1, \;\mbox{for}\;1\leq
l < k,  \;
\alpha_{T} (k) = \alpha,\beta_{T} (1) =  \beta\}.
\eeas

Note that $\cvop{\emptyset}{\nu}=\mixten{\delta}{\emptyset}{\nu}$, so
any diagram with incoming VP's at $\bfxt$ but no outgoing VP's has
weight zero.  For many lattice gases certain other vertex factors
$\cvop{\mu}{\nu}$ vanish also; diagrams with such vertices can be
dropped from the sum~(\ref{eq:caljsum}).  From Eq.~(\ref{eq:rjt}),
$\tilde{J}$ can now be written as a sum over diagrams in the same
fashion as ${\cal J}$,
\bge
   \tjij (\bfx) =  \sum_{k = 1}^{ \infty}
   \sum_{T \in \ctop{i}{j}(\bfx, k)} W(T),
   \label{eq:jsum}
\ee
where the set of diagrams to be summed over is given by
\[
\ctop{i}{j}(\bfx, k) =
\bigcup_{b:i (b) = j} \ctop{a (i,\bfx)}{b}(k).
\]
We will find it useful later to generalize this set of diagrams to
the sets of diagrams
\bgeas
   \ctop{\mu}{\nu}(\bfx, k)  & = & \{T:k = k (T), \;
   \sigma_T (l) > 1, \;\mbox{for}\;1\leq
   l < k,  \;
   | L_{\alpha_{T} (0)} |=1,\;L_{\alpha_{T}(k)} = \{\bfx\}, \\
   & &\hspace{.3in} \mu (\alpha_{T} (k)) = \mu, \;
   \mu (\alpha_{T} (0)) =  \nu
   \},
\eeas
where we have used the notation \label{pg:zx}
\[
\mu (\{a_1, \ldots, a_j\}) = \{i (a_1),  \ldots, i (a_j)\}.
\]

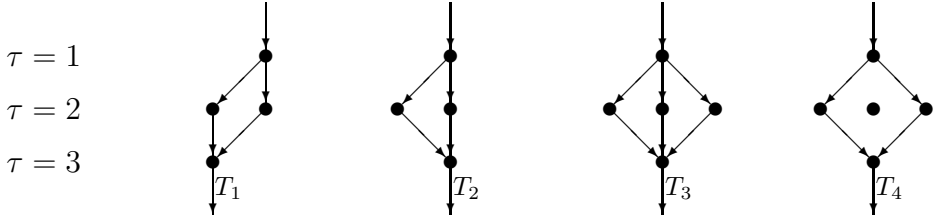
\begin{figure}
\centering
\begin{picture}(380,125)(0,0)
\put(69.92,45.){\makebox(0,0){\footnotesize $T_1$}}
\put(159.98,45.){\makebox(0,0){\footnotesize $T_2$}}
\put(240.16,45.){\makebox(0,0){\footnotesize $T_3$}}
\put(319.96,45.){\makebox(0,0){\footnotesize $T_4$}}
\put(0.,55.){\makebox(0,0){$\tau = 3$}}
\put(0.,75.){\makebox(0,0){$\tau = 2$}}
\put(0.,95.){\makebox(0,0){$\tau = 1$}}
\put(60.04,35.){
\begin{picture}(20,80)(0,0)
\put(20.,80.){\vector(0,-1){17.5}}
\put(20.,60.){\vector(0,-1){17.5}}
\put(20.,60.){\vector(-1,-1){18.2322}}
\put(0.,40.){\vector(0,-1){17.5}}
\put(20.,40.){\vector(-1,-1){18.2322}}
\put(0.,20.){\vector(0,-1){20.}}
\put(20.,60.){\circle*{5.}}
\put(20.,40.){\circle*{5.}}
\put(0.,40.){\circle*{5.}}
\put(0.,20.){\circle*{5.}}
\end{picture}}
\put(129.96,35.){
\begin{picture}(20,80)(0,0)
\put(20.,80.){\vector(0,-1){17.5}}
\put(20.,60.){\vector(0,-1){17.5}}
\put(20.,60.){\vector(-1,-1){18.2322}}
\put(0.,40.){\vector(1,-1){18.2322}}
\put(20.,40.){\vector(0,-1){17.5}}
\put(20.,20.){\vector(0,-1){20.}}
\put(20.,60.){\circle*{5.}}
\put(20.,40.){\circle*{5.}}
\put(0.,40.){\circle*{5.}}
\put(20.,20.){\circle*{5.}}
\end{picture}}
\put(210.14,35.){
\begin{picture}(40,80)(0,0)
\put(20.,80.){\vector(0,-1){17.5}}
\put(20.,60.){\vector(-1,-1){18.2322}}
\put(20.,60.){\vector(0,-1){17.5}}
\put(20.,60.){\vector(1,-1){18.2322}}
\put(0.,40.){\vector(1,-1){18.2322}}
\put(20.,40.){\vector(0,-1){17.5}}
\put(40.,40.){\vector(-1,-1){18.2322}}
\put(20.,20.){\vector(0,-1){20.}}
\put(20.,60.){\circle*{5.}}
\put(0.,40.){\circle*{5.}}
\put(20.,40.){\circle*{5.}}
\put(40.,40.){\circle*{5.}}
\put(20.,20.){\circle*{5.}}
\end{picture}}
\put(289.94,35.){
\begin{picture}(40,80)(0,0)
\put(20.,80.){\vector(0,-1){17.5}}
\put(20.,60.){\vector(-1,-1){18.2322}}
\put(20.,60.){\vector(1,-1){18.2322}}
\put(0.,40.){\vector(1,-1){18.2322}}
\put(40.,40.){\vector(-1,-1){18.2322}}
\put(20.,20.){\vector(0,-1){20.}}
\put(20.,60.){\circle*{5.}}
\put(0.,40.){\circle*{5.}}
\put(20.,40.){\circle*{5.}}
\put(40.,40.){\circle*{5.}}
\put(20.,20.){\circle*{5.}}
\end{picture}}
\end{picture}
\caption{$k=3$ Diagrams for a 1D Lattice Gas with Three Bits/Site}
\label{fig:diags}
\end{figure}

\noindent {\bf Example:} \, As a simple example  of the diagrammatic
notation, consider the allowed diagrams for the 1D3P lattice gas
considered in Section 2.5.  The complete set of diagrams needed to
compute the $k=3$ correction to $\tjop{0}{0}$ is the set of diagrams
$T_1$ -- $T_4$ represented in Fig.~\ref{fig:diags}, along with the
diagrams achieved by reflecting $T_1$ and $T_2$ across the
$x$-axis\footnote{Throughout, we shall depict diagrams for
one-dimensional lattice gases with a vertical time axis.}.  The weight
of diagram $T_2$, for example, is $W(T_2) = \cvop{0}{ \widehat{-}}\cvop{
+}{ -} \cvop{0}{0}
\cvop{\widehat{+}}{0}$. We will compute the vertex factors
$\cvop{\mu}{\nu}$ for this lattice gas in Section~\ref{ssec:odtpex},
and we will see that the contribution from diagram $T_3$ in fact
vanishes.

\subsection{Approximations}
\label{ssec:approxx}

We have so far managed to write the exact formula for the hydrodynamic
equations in the scaling limit only in terms of an infinite formal
series.  The natural next question to confront is whether this series
can be summed.  We would like to know whether the series is finite, and
if we cannot sum the full series, at least we would like to find a set
of reasonable approximations which we can make to truncate the series to
one which is summable.  The questions of convergence are rather
difficult, and we will not address them here in full generality; in
general, the convergence properties of the series depend on the form of
the conserved quantities in the system.  A variety of methods for
performing partial sums of infinite series of diagrams while retaining
physically important terms have been applied to related problems in
quantum field theory and quantum many-body theory~\cite{IZ,FW}.  We will
describe here several particular approximation methods which are useful
for the kind of series which arise for known lattice gases.

\subsubsection{Short-$\tau$ and Small-$\ell$ Truncations}
The simplest useful approximations involve truncating the sum
(\ref{eq:caljsum}) to a finite number of terms by putting an upper bound
on either the number of timesteps or the number of distinct nontrivial
vertices allowed in each diagram.  In the first case, the expression for
the renormalized $J$-matrix is
\[
   \tjop{(\tau)i}{j} (\bfx)
   = \sum_{k=1}^\tau\sum_{T \in \ctop{i}{j}(\bfx,k)} W(T),
\]
where the diagrams summed over are the same as those summed in
Eq.~(\ref{eq:jsum}).  Since for each fixed value of $k$ there are a
finite number of allowed diagrams, this sum is finite.  We refer to this
approximation as the {\sl short}-$\tau$ {\sl approximation}.  In the
second case, we allow $k$ to be arbitrary, but allow only diagrams where
the total number of vertices $(\bfx, k')$ with nonempty outgoing sets
$\alpha_T( k')_\bfx$ is less than or equal to some fixed number $\ell$.
We denote the sum restricted to these diagrams by $\tjop{[\ell]i}{j}$.
Again, there are only a finite number of such diagrams in this sum,
which means that this sum must also be finite.  This approximation is
analogous to the weak-coupling expansions in quantum field theory,
although in this case the coupling constants $\cvop{\mu}{\nu}$ are
usually not particularly small.  The short-$\tau$ and small-$\ell$
truncations give good consecutive approximations for many lattice gases.
In either of these two approximations, the Boltzmann approximation can
be recovered, by taking $\tau = 1$ or $\ell = 1$.

\subsubsection{BBGKY Truncations}
Another good class of approximations, in which a reduced but still
infinite set of diagrams is summed, corresponds to truncations of the
BBGKY hierarchy of equations.  Such an approximation involves neglecting
$q$-CCF's with $q > n$ for some fixed value of $n$.  In our diagrammatic
formalism, this amounts to restricting the sum to diagrams with
$\sigma_T(k') \leq n$ for $1 \leq k' \leq k$.  For example, with $n=2$,
diagram $T_3$ of Fig.~\ref{fig:diags} would be neglected.  Whereas the
computational complexity of including all diagrams in the complete sum
grows exponentially in $k$, that of the truncated BBGKY approximations
grow polynomially, and are therefore computationally more tractable.

\subsubsection{The Ring Approximation}
The $n=2$ version of the BBGKY approximation is closely related to the
{\sl ring approximation}.  The ring approximation is made by neglecting
interactions between two propagating correlated quantities except at the
initial and final vertices of a diagram.  It is generally possible to
calculate a closed-form expression for the infinite sum of diagrams
corresponding to this approximation.  Furthermore, it is usually fairly
easy to calculate the asymptotic form of this approximation as $k
\rightarrow\infty$.  This calculation often captures the most
significant part of the long-term renormalization effects.  In
particular, for certain lattice gases which model two dimensional fluid
systems, the ring approximation diverges logarithmically in $|L|$, which
is in agreement with predictions from other theoretical
frameworks~\cite{Ern}, and also with observed behaviour~\cite{Kad}.

In Section 7, we will apply the different approximation methods
described here to the 1D3P lattice gas and compare the results from
these approximations with experimental results.

\subsection{Eigenvalue Renormalization}
\label{ssec:eigrnrm}

Within the framework of the formalism developed in the previous
subsections, we can now demonstrate that for a large class of lattice
gas models, the effects of correlations are to renormalize only the
eigenvalues of the $J$-matrix, and not to change the eigenvectors.
This result follows from a pair of fairly simple theorems.
\begin{thm}
The finite matrix $\ctij = \tjtij - \jij$ of corrections to the
$J$-matrix can be restricted to be a matrix in the space of kinetic
eigenvectors of $J$; i.e., if $\nu \in H$ then
\[
   \qsub{i}^\nu (\tjop{(\tau)i}{j} - \jij) =
   (\tjop{(\tau)i}{j} - \jij) \qsup{j}_\nu = 0.
\]
\label{t:erone}
\end{thm}
\noindent {\bf Proof:} \,\,
{}From Eqs.~(\ref{eq:rj}) and (\ref{eq:dv}), it will clearly suffice to
show that $\qsub{i}^\nu \cvop{i}{\widehat{\mu}} = \cvop{\widehat{\mu}}{j}
\qsup{j}_\nu = 0$ for every $\widehat{\mu}\subseteq B$ with
$|\widehat{\mu}| \geq 2$.

We first demonstrate this sufficient condition for the right
hydrodynamic eigenvectors $q^i_\nu$.  We showed in Section
\ref{ssec:linearboltzmann} that $q^i_\nu$ was a right eigenvector of the
$J$ matrix with eigenvalue 0 by differentiating the identity $C^i
(\nnsupo{\ast}{0})= 0$, which holds for any equilibrium, with respect to
the parameters $\alpha_\nu$ of the equilibrium.  We can similarly show
that $q^i_\nu$ is a null right eigenvector of $\cvop{\widehat{\mu}}{i}$, by
using the stability of the Boltzmann equilibria.  The stability of the
local Boltzmann equilibrium tells us that at each lattice site,
\[
g^{\widehat{\mu}} (\sum_{\rho \subseteq B}
\vop{*}{\rho} \prod_{i \in \rho}^{}  N^i_0) = 0.
\]
Differentiating this equation with respect to the  parameter
$\alpha_\nu$ of the local Boltzmann equilibrium gives
\[
0 = \sum_{ \xi \subseteq \widehat{\mu}, \rho \subseteq B}
\left(\prod_{j \in \widehat{\mu}\setminus \xi} -N^j_0 \right)
\vop{\xi}{\rho}
\left[ \sum_{i \in \rho} (\prod_{k \in \rho \setminus{i}}
 N^k_0) q^i_\nu \right]
= \cvop{\widehat{\mu}}{i}
\qsup{i}_\nu,
\]
as desired.

Next, consider the left hydrodynamic eigenvector $q^\nu_i$.  From
Eq.~(\ref{eq:vdef}), we have
\[
   \cvop{i}{\widehat{\mu}} =
   \sum_{\nu \supseteq \widehat{\mu}}
   (\prod_{j \in \nu \setminus \widehat{\mu}} \nnsupo{j}{0} ) \vop{i}{\nu}.
\]
Since $\qsub{i}^\nu\nnsup{i}$ is a conserved quantity for any values of
$\nnsup{i}$, we have
\[
   \qsub{i}^\nu
   \sum_{\mu \subseteq B}
   \vop{i}{\mu}
   \left(\prod_{j \in  \mu} \nnsupo{j}{0}
   \right)
   = \qsub{i}^\nu\nnsupo{i}{0}
\]
for arbitrary $\nnsupo{i}{0}$, so $\qsub{i}^\nu\vop{i}{\widehat{\mu}}
=\qsub{i}^\nu\cvop{i}{\widehat{\mu}}=0$ for all
$\widehat{\mu}\subseteq  B$ with $|\widehat{\mu}| \geq 2$.$\Box$

\begin{thm}
If there exists a symmetry $\Sigma$ of a lattice gas which can be
expressed as a combination of a permutation $\dsop{\bfx}{\bfy}$ on the
lattice $L$ fixing a point $\bfx$ and an independent permutation
$\bsop{i}{j}$ on the bit set $B$, where the zero-order Boltzmann
equilibrium at $\bfx$ satisfies $\bsop{i}{j}N^j_0 (\bfx)= N^i_0 (\bfx)$,
then $\bar{\Sigma}$ commutes with $J$ and $\tjt$; i.e.,
\[
   \bsop{i}{j}\jop{j}{k}(\bfx) -  \jop{i}{j}(\bfx)\bsop{j}{k} =
   \bsop{i}{j}\tjop{(\tau)j}{k}(\bfx)
 - \tjop{(\tau)i}{j}(\bfx)\bsop{j}{k} = 0
\]
\label{t:ertwo}
\end{thm}

\noindent {\bf Proof:} \,\,
The condition that $\Sigma$ is a symmetry of the lattice gas with the given
product structure asserts that $\sop{a}{b} = \bsop{i(a)}{i(b)}
\dsop{\bfx(a)}{\bfx(b)}$ is a permutation matrix on all the bits of the
system which can be extended to a permutation matrix
$\sop{\alpha}{\beta}$ on subsets of $B$, satisfying
\[
   \sinvop{\alpha}{\beta} \advop{\beta}{\gamma} \sop{\gamma}{\delta} =
   \advop{\alpha}{\delta}
\]
and
\[
   \sinvop{\alpha}{\beta} \kop{\beta}{\gamma} \sop{\gamma}{\delta} =
   \kop{\alpha}{\delta}.
\]
Since by their definitions, the functions $f$ and $g$ connecting means
and CCF's are invariant under permutations, it follows from
Eq.~(\ref{eq:dexp}) and the invariance of the local Boltzmann
equilibrium at $\bfx$ that
\[
   \sinvop{\alpha}{\beta} \ckop{\beta}{\gamma} \sop{\gamma}{\delta} =
   \ckop{\alpha}{\delta},
\]
where we assume that all zero-order single particle means are evaluated
at $\bfx$, as in the diagrammatic expansion.  It then follows from
Eqs.~(\ref{eq:fubar}), (\ref{eq:rj}), and the fact that
$\mixten{\delta}{\alpha}{\beta}$ commutes with $\Sigma$ that
\[
   \sinvop{ a}{b} \cjop{b}{c} \sop{c}{d} = \cjop{a}{d}.
\]
and
\[
\bsinvop{i}{j}\jop{j}{k}(\bfx)\bsop{k}{l}=\jop{i}{l}(\bfx),
\]
proving the first part of the theorem.  From the constraint that
$\dot{\Sigma}$ leave the point $\bfx$ fixed, we have
\bgeas
   \tjop{i}{l} (\bfx)
   & = & \sum_\bfw \cjop{a(i,\bfx)}{a(l,\bfw)} \\
   & = & \sum_\bfw \dsinvop{ \bfx}{\bfy} \bsinvop{ i}{j}
                   \cjop{a(j,\bfy)}{a(k,\bfz)}
                   \bsop{k}{l} \dsop{\bfz}{\bfw} \\
   & = & \sum_\bfz \bsinvop{ i}{j}
                   \cjop{a(j,\bfx)}{a(k,\bfz)}
                   \bsop{k}{l} \\
   & = & \bsinvop{ i}{j}
         \tjop{j}{k} (\bfx)
         \bsop{k}{l},
\eeas
so the second assertion of the theorem is proven.$\Box$

\begin{corr}
If the sequence $\{\tjop{(\tau)i}{j}\}$ is convergent as
$\tau\rightarrow\infty$, then Theorems~\ref{t:erone} and \ref{t:ertwo}
hold in this limit.  That is,
\[
\qsub{i}^\nu (\tjij - \jij) = (\tjij - \jij) \qsup{j}_\nu = 0,
 \; \; \; {\rm for} \; \nu \in H,
\]
and
\[
   \bsop{i}{j}\jop{j}{k} -  \jop{i}{j}\bsop{j}{k} =
   \bsop{i}{j}\tjop{j}{k} - \tjop{i}{j}\bsop{j}{k} = 0,
\]
for $\bsop{i}{j}$ as in Theorem \ref{t:ertwo}.
\end{corr}

For a wide class of lattice gases, including all the lattice gases
described in Section~\ref{ssec:odtp}, these two theorems suffice to
demonstrate that the eigenvectors of the $\tilde{J}$ matrix coincide
with those of $J$, so that the only effect of correlations is to
renormalize the eigenvalues of $J$.  The essential point is that we
can classify the eigenvectors of $J$ by their transformation
properties under the group of symmetries of the lattice gas.
Theorem \ref{t:ertwo} asserts that $\tilde{J}$ can only mix eigenvectors
with identical symmetry properties.  Thus, if no two kinetic
eigenvectors of $J$ share the same symmetry properties, then
$\tilde{J}$ must be diagonal with respect to the basis of eigenvectors
of $J$.  More generally, if the representation of the group of
symmetries on the space of kinetic eigenvectors of $J$ breaks up into
irreducible representations in such a way that no irreducible
representation appears more than once, then $\tilde{J}$ is diagonal
with respect to the eigenvectors of $J$.

As a simple example, consider the 1D3P diffusive lattice gas.  The
right kinetic eigenvectors of $J$ are
\[
   \qsubp{2}
   = \frac{1}{2} \left(\begin{array}{r} -1 \\  0 \\ +1 \end{array}\right)
\]
and
\[
   \qsubp{3}
   = \frac{1}{6}
\left(\begin{array}{r} -1 \\ +2 \\ -1 \end{array}\right).
\]
Under the symmetry transformation $\bfx\rightarrow -\bfx$,  $+
\leftrightarrow -$, these
eigenvectors transform with eigenvalues $\sop{2}{2}=-1$,
$\sop{3}{3}=+1$.  Thus, these eigenvectors cannot be mixed by
$\tilde{J}$, and so their eigenvalues are separately renormalized.
Explicitly, in matrix notation, with respect to the basis $\qsubp{i}$,
we have
\[
   \begin{array}{cc}
      \tilde{J} = \left(
                  \begin{array}{ccc}
                  0 & 0 & 0 \\
                  0 & \mixten{j}{2}{2} & \mixten{j}{2}{3} \\
                  0 & \mixten{j}{3}{2} & \mixten{j}{3}{3}
                  \end{array}
                  \right), &
        \bar{\Sigma} = \left(
                  \begin{array}{ccc}
                  +1 & 0 & 0 \\
                  0 & -1 & 0 \\
                  0 & 0 & +1
                  \end{array}
                  \right).
   \end{array}
\]
By Theorem~\ref{t:ertwo}, $\tilde{J}$ commutes with $\bar{\Sigma}$, so
$\mixten{j}{2}{3}=\mixten{j}{3}{2}=0$.

For most standard lattice gases, a similar analysis of the symmetry
properties of the eigenvectors of $J$ shows that no irreducible
representation of the symmetry group appears more than once, so that the
eigenvalues of $J$ are renormalized in a straightforward fashion.  For
those lattice gases where this cannot be shown, it is necessary to
repeat the entire Chapman-Enskog analysis using the renormalized
$\tilde{J}$ matrix.  Note that Theorem~\ref{t:ertwo} implies that any
eigenvectors of $J$ which lie in the same irreducible representation of
the symmetry group must have identical eigenvalues in the matrices $J$
and also in $\tilde{J}$.

\subsection{Renormalization Effects and Higher-Order Collision Operators}
\label{ssec:correlationsandcollision}

We conclude this section with a discussion of the effects of the
higher-order collision operators $\epsilon C_1$ and $\epsilon^2 C_2$ in
the full kinetic theory.  We show that the second-order term only
appears in the source term for the hydrodynamic equation as in
Eq.~(\ref{eq:hydsour}), and does not generate extra renormalization
effects.  On the other hand, we find that the first-order collision
operator not only appears in the advection coefficient as in
Eq.~(\ref{eq:hydadv}), but generates a set of additional correlations
which modify the advection coefficient by effectively renormalizing the
components of the first-order collision operator itself.

We begin by discussing the second-order term $\epsilon^2 C_2$.  This
part of the collision operator only appears in the first-order
conservation equation (\ref{eq:hyd}).  The corrections to this term due
to correlations are of one higher order in $\epsilon$, and can clearly
be neglected in the entire analysis.  Thus, inclusion of this term only
generates the source term in Eq.~(\ref{eq:hyd}) in the manner described
in Subsection~\ref{ssec:foce}.

Now, we consider the effects of including a first-order term $\epsilon
C_1$ in the collision operator.  Recall that this term is restricted
to obey the conservation laws, but is not required to satisfy
semi-detailed balance.  Because this part of the collision operator
obeys the conservation laws, it does not appear directly in the
first-order conservation equation.  It does appear in the linearized
Boltzmann equation (\ref{eq:eqfornone}).  However, corrections to the
term in this equation due to correlations are again of higher order in
$\epsilon$, so that no change is necessary to this equation due to
renormalization effects.

At this point, one might imagine that inclusion of the first-order
term in the collision operator does not necessitate any further
modification to the exact hydrodynamic equation in the scaling limit
other than the effects described in Section 3.  However, this is not
the case.  In fact, the inclusion of this term in the collision
operator has a nontrivial effect on Eq.~(\ref{eq:cdlin}), which
describes the propagation of correlated quantities in the system.

Eq.~(\ref{eq:cdlin}) gives an expression to order $\epsilon$ for
$\Gamma^{\widehat{\alpha}}$ at time $t + \Delta t$ in terms of
quantities at time $t$.  In the derivation of Eq.~(\ref{eq:cdlin}), we
used the fact that the zero-order means $N^i_0$ describe a local
Boltzmann equilibrium which does not generate correlations through the
collision operator $C_0$.  When we include the first-order collision
operator $C_1$, we must include the fact that $\epsilon C_1^i (N^*_0)$
need not vanish.  Thus, the correct form of Eq.~(\ref{eq:cdlin}) in this
case is
\bge
   \ggamsup{\widehat{\alpha}}(\tpdt) =
   \advop{\widehat{\alpha}}{\widehat{\beta}} \left(
   \epsilon \ckop{\widehat{\beta}}{a}
   \nnsup{a}_1 (t) +
   \ckop{\widehat{\beta}}{\widehat{\gamma}} \ggamsup{\widehat{\gamma}}(t)
   + \epsilon I^{\widehat{\beta}}
\right), \label{eq:correctcdl}
\ee
where we have defined $I^{\widehat{\beta}} = 0$ whenever $ |L_{\widehat{\beta}}
| > 1$, and
\bge
   I^{\widehat{\beta}} = I^{\mu} \equiv
   \sum_{\nu \subseteq \mu}
   \left(\prod_{i \in \mu\setminus \nu} -N^i_0 \right)
   \sum_{\xi}
   \left(\subscrpt{\left.
   \frac{\pdv }{\pdv \epsilon}
   \right|}{\epsilon =0}\vop{\nu}{\xi} \right)
   \prod_{j \in \xi} N^j_0, \label{eq:extrasource}
\ee
whenever $ L_{\widehat{\beta}}= \{\bfy\}$ and $ \widehat{\beta}_\bfy =
\mu(\widehat{\beta}) = \mu$
for some $\bfy \in L$; all means $N_0$ in this equation are evaluated at
the point $\bfy$.  In Eq.~(\ref{eq:extrasource}), we have used the
collision operator on means $\vop{\nu}{\gamma}$ defined by using the
collision operator $C^i = C_0^i + \epsilon C_1^i$.  However, the CVC's
used in Eq.~(\ref{eq:correctcdl}) should still be evaluated with respect
to the zero-order collision operator $C_0$.

Combining Eqs.~(\ref{eq:correctcdl}) and (\ref{eq:exact1}), and
including the first order collision operator as in
Eq.~(\ref{eq:eqfornone}), we get
\[
\nnsup{a}(\tpdt)  =
   \advop{a}{ b} \left(N^b (t)+
	\epsilon \cjop{b}{c}
	\nnsup{c}_1 (t)+
	\epsilon {\cal I}^b \right),
\]
where
\[
   {\cal I}^b =
   C^{i (b)}_1 (N^*_0 (\bfx (b))) +
   \ckop{b}{\widehat{\alpha}}
   (\advop{\widehat{\alpha}}{\widehat{\beta}}  I^{\widehat{\beta}} +
   \advop{\widehat{\alpha}}{\widehat{\beta}}
   \ckop{\widehat{\beta}}{\widehat{\gamma}}
   (\advop{\widehat{\gamma}}{\widehat{\delta}}I^{\widehat{\delta}} +
   \ldots)).
\]
By applying the same analysis used in Section 3, we find that the
renormalized hydrodynamic equations for the theory are of the same form
as Eq.~(\ref{eq:hyd}); however, the inclusion of the first-order
collision term changes the result for the renormalized advection
coefficient to be
\[
   \advco{\mu}(\qqsupp{\ast}) =
   \frac{c}{\dt}\sum_{\nu\in K}
   \frac{\mixten{\tilde{\bfdelta} (1)}{\mu}{\nu}
         \supsub{\tilde{\cal C}}{\nu}{1}
         \left(\qqsupp{\ast}\right)}
        {(- \tilde{\lambda}^\nu)},
\]
where the renormalized generalized Kronecker delta function and
eigenvalues are defined with respect to the renormalized $J$ matrix,
and where the renormalized collision operator $\tilde{{\cal C}}_1^\nu$
is given by
\[
\tilde{{\cal C}}_1^\nu (\bfx)= q^\nu_i {\cal I}^{a (i,\bfx)}.
\]
Just as for the renormalized $J$ matrix, we evaluate all means $N_0$
appearing in CVC's and $I^\mu$ at the point $\bfx$.  This
simplification depends again upon the convergence of the infinite
series of terms in $\tilde{{\cal C}}_1^\nu$.

We can express the renormalized collision operator in terms of an
infinite diagrammatic sum, analogous to the sum (\ref{eq:jsum}) for
the renormalized matrix $\tilde{J}$.
Specifically, we have
\[
{\cal I}^a  = C^{i (a)}_1  +
\sum_{k = 1}^{ \infty}
\sum_{T \in \ctop{i (a)}{\mu}(\bfx (a), k)}
 W (T) I^{\mu}.
\]

\newpage
\section{Examples of Vertices and Renormalization}
\setcounter{equation}{0}
\label{sec:examples}

We will now apply the methods of the previous section to compute the
vertex factors for the example lattice gases described in section
4.  We also derive expressions for the renormalized
transport coefficients for all these lattice gases.

\subsection{1D3P Lattice Gas}
\label{ssec:odtpex}
\subsubsection{Vertices}

Beginning with either the collision operator (\ref{eq:collision1}) or
the state transition table (see Subsection~\ref{ssec:odtp}) for the 1D3P
lattice gas, we can calculate the mean vertex coefficients
$\mixten{V}{\mu}{\nu}$ using Eqs.~(\ref{eq:vertexmean1}) and
(\ref{eq:vertexmean2}).  The nonzero mean vertex coefficients are given
by
\begin{eqnarray*}
\mixten{V}{B}{B} & = & \mixten{V}{\emptyset}{\emptyset} = 1\\
\mixten{V}{i}{j} & = & \mixten{\delta}{i}{j}\\
\mixten{V}{i}{\widehat{j}} & = &  p (3\mixten{\delta}{i}{j}- 1)\\
\mixten{V}{\widehat{i}}{ \widehat{j}} & = &  p +\mixten{\delta}{i}{j} (1 - 3p).
\end{eqnarray*}
Note that since the ensemble-averaged collision operator is invariant
under permutations (relabeling) on the bits, the mean vertex
coefficients also have this symmetry.

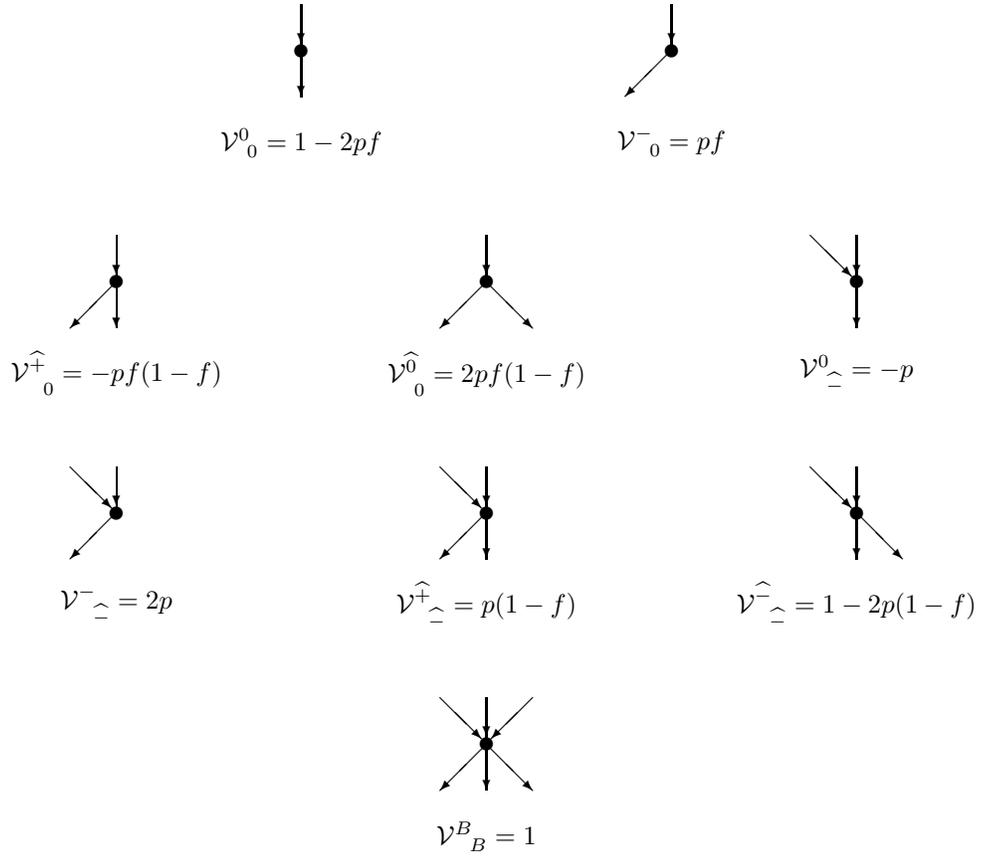
\begin{figure}
\centering
\begin{picture}(350,350)(0,0)
\put(70.,306.25){
\begin{picture}(70,70)(0,0)
\put(35.,52.5){\vector(0,-1){15.}}
\put(35.,35.){\vector(0,-1){17.5}}
\put(35.,35.){\circle*{5.}}
\put(35.,0.){\makebox(0,0){\footnotesize $\cvop{0}{0} = 1-2pf$}}
\end{picture}}
\put(210.,306.25){
\begin{picture}(70,70)(0,0)
\put(35.,52.5){\vector(0,-1){15.}}
\put(35.,35.){\vector(-1,-1){17.5}}
\put(35.,35.){\circle*{5.}}
\put(35.,0.){\makebox(0,0){\footnotesize $\cvop{-}{0} = pf$}}
\end{picture}}
\put(0.,218.75){
\begin{picture}(70,70)(0,0)
\put(35.,52.5){\vector(0,-1){15.}}
\put(35.,35.){\vector(-1,-1){17.5}}
\put(35.,35.){\vector(0,-1){17.5}}
\put(35.,35.){\circle*{5.}}
\put(35.,0.){\makebox(0,0){\footnotesize $\cvop{\widehat{+}}{0} = -pf(1-f)$}}
\end{picture}}
\put(140.,218.75){
\begin{picture}(70,70)(0,0)
\put(35.,52.5){\vector(0,-1){15.}}
\put(35.,35.){\vector(-1,-1){17.5}}
\put(35.,35.){\vector(1,-1){17.5}}
\put(35.,35.){\circle*{5.}}
\put(35.,0.){\makebox(0,0){\footnotesize $\cvop{\widehat{0}}{0} = 2pf(1-f)$}}
\end{picture}}
\put(280.,218.75){
\begin{picture}(70,70)(0,0)
\put(17.5,52.5){\vector(1,-1){15.7322}}
\put(35.,52.5){\vector(0,-1){15.}}
\put(35.,35.){\vector(0,-1){17.5}}
\put(35.,35.){\circle*{5.}}
\put(35.,0.){\makebox(0,0){\footnotesize $\cvop{0}{\widehat{-}} = -p$}}
\end{picture}}
\put(0.,131.25){
\begin{picture}(70,70)(0,0)
\put(17.5,52.5){\vector(1,-1){15.7322}}
\put(35.,52.5){\vector(0,-1){15.}}
\put(35.,35.){\vector(-1,-1){17.5}}
\put(35.,35.){\circle*{5.}}
\put(35.,0.){\makebox(0,0){\footnotesize $\cvop{-}{\widehat{-}} = 2p$}}
\end{picture}}
\put(140.,131.25){
\begin{picture}(70,70)(0,0)
\put(17.5,52.5){\vector(1,-1){15.7322}}
\put(35.,52.5){\vector(0,-1){15.}}
\put(35.,35.){\vector(-1,-1){17.5}}
\put(35.,35.){\vector(0,-1){17.5}}
\put(35.,35.){\circle*{5.}}
\put(35.,0.){\makebox(0,0){\footnotesize
   $\cvop{\widehat{+}}{\widehat{-}} = p(1-f)$}}
\end{picture}}
\put(280.,131.25){
\begin{picture}(70,70)(0,0)
\put(17.5,52.5){\vector(1,-1){15.7322}}
\put(35.,52.5){\vector(0,-1){15.}}
\put(35.,35.){\vector(0,-1){17.5}}
\put(35.,35.){\vector(1,-1){17.5}}
\put(35.,35.){\circle*{5.}}
\put(35.,0.){\makebox(0,0){\footnotesize
   $\cvop{\widehat{-}}{\widehat{-}} = 1-2p(1-f)$}}
\end{picture}}
\put(140.,43.75){
\begin{picture}(70,70)(0,0)
\put(17.5,52.5){\vector(1,-1){15.7322}}
\put(35.,52.5){\vector(0,-1){15.}}
\put(52.5,52.5){\vector(-1,-1){15.7322}}
\put(35.,35.){\vector(-1,-1){17.5}}
\put(35.,35.){\vector(0,-1){17.5}}
\put(35.,35.){\vector(1,-1){17.5}}
\put(35.,35.){\circle*{5.}}
\put(35.,0.){\makebox(0,0){\footnotesize $\cvop{B}{B} = 1$}}
\end{picture}}
\end{picture}
\caption{Vertex Factors for 1D3P Lattice Gas}
\label{fig:1d3pvertices}
\end{figure}

Using the equilibrium value $f$ for the mean occupation numbers
$\nnsupo{i}{0}$ (see Eq.~(\ref{eq:equilibrium1})), the expression for
the CVC's, Eq.~(\ref{eq:vdef}), reads
\[
   \cvop{\alpha}{\beta} =
   \sum_{\mu  \subseteq\alpha, \nu \supseteq \beta}
   (-1)^{|\alpha | - | \mu|}
   f^{| \alpha | + | \nu | - | \mu | - | \beta |}
	 \vop{\mu}{\nu}.
\]
For instance, we have
\begin{eqnarray*}
   \cvop{\widehat{+}}{0}  &=& f (\mixten{V}{\widehat{+}}{\widehat{+}}
	+\mixten{V}{\widehat{+}}{\widehat{-}}
	-\mixten{V}{0}{0}-\mixten{V}{-}{0})
	- f^2 (\mixten{V}{0}{\widehat{+}} +\mixten{V}{-}{\widehat{+}}
	+\mixten{V}{0}{\widehat{-}}+\mixten{V}{-}{\widehat{-}}) \\
	& =& -pf (1 - f)
\end{eqnarray*}
The remaining nonzero CVC's are given by the
equations
\begin{eqnarray}
\mixten{{\cal V}}{B}{B} & = &
   \mixten{{\cal V}}{\emptyset}{\emptyset} = 1\nonumber\\
\mixten{{\cal V}}{i}{j}  & = &  pf +\mixten{\delta}{i}{j}(1 - 3pf)\nonumber\\
\mixten{{\cal V}}{i}{\widehat{j}}  & = & \mixten{\delta}{i}{j}3p - p\nonumber\\
\mixten{{\cal V}}{\widehat{i}}{j}
  & = & \mixten{\delta}{i}{j}3pf (1 - f)- pf (1 - f)\nonumber\\
\mixten{{\cal V}}{ \widehat{i}}{ \widehat{j}}
  & = & p (1 - f)+\mixten{\delta}{i}{j}(1 - 3p (1 - f)).
\label{eq:1to2}
\end{eqnarray}
Note that the CVC's are also symmetric under an arbitrary permutation on
the particle labels.  The nonvanishing correlation vertex factors are
depicted graphically in Fig.~\ref{fig:1d3pvertices}; only a single
vertex is shown in each equivalence class under the permutation
symmetry.

\subsubsection{Renormalization of Diffusivity}

As was shown in Section~\ref{ssec:eigrnrm}, the symmetry of the $1D3P$
lattice gas under spatial inversion is sufficient to ensure that the
effect of correlations is simply to renormalize the eigenvalues
$\lamsupp{2}$ and $\lamsupp{3}$.  From the Chapman-Enskog analysis, we
know that the diffusivity is given in the hydrodynamic limit by
\bge
\tilde{D} = \frac{c^2}{3\dt}
   \left(\frac{2}{-\supscrpt{\tilde{\lambda}}{2}} -1 \right),
\label{eq:edoodtplga}
\ee
where $\supscrpt{\tilde{\lambda}}{2}$ is the eigenvalue of the
vector $(-1, 0,
+1)$ in the matrix $\tilde{J}$.  In this subsection we will find the set of
diagrams which contribute to the renormalization of $\lamsupp{2}$.

Factoring out the initial and closure vertices, we can write the
renormalized matrix $\tilde{J}$ as
\bge
\tjij(\bfx) = \mixten{J}{i}{j}
+ \cvop{i}{ \widehat{\nu}}
\cvop{ \widehat{\mu}}{j}
\yop{\widehat{\nu}}{\widehat{\mu}}(\bfx),
\label{eq:rjodtp}
\ee
where
\[
   \yop{\nu}{\mu}(\bfx) =
   \sum_{k=1}^\infty \sum_{T \in \ctop{\nu}{\mu}(\bfx,k)} W(T).
\]
{}From the fact that $\cvop{i}{B}= \cvop{B}{i}= 0 $ for all $i \in B$, we
see that the only nonzero contributions to Eq.~(\ref{eq:rjodtp}) can come
{}from terms of the form $\cvop{ i}{\widehat{l}}
\yop{\widehat{l}}{\widehat{k}} \cvop{\widehat{k}}{j}$.  From the
inversion symmetry, we observe that
\[
   \yop{\widehat{-l}}{\widehat{-k}} = \yop{\widehat{l}}{\widehat{k}}.
\]
Thus, Eq.~(\ref{eq:rjodtp}) can be explicitly rewritten as
\[
   \tjij = \mixten{J}{i}{j}+2\left(\xop{i \widehat{+}}{\widehat{+}j}
\yop{\widehat{+}}{\widehat{+}} +
           \xop{i \widehat{-}}{\widehat{+}j} \yop{\widehat{+}}{\widehat{-}} +
           \xop{i \widehat{0}}{\widehat{+}j} \yop{\widehat{+}}{\widehat{0}} +
           \xop{i \widehat{+}}{\widehat{0}j}
\yop{\widehat{0}}{\widehat{+}}\right) +
           \xop{i \widehat{0}}{\widehat{0}j} \yop{\widehat{0}}{\widehat{0}},
\]
where
\[
   \xop{i\widehat{l}}{\widehat{k}j} =\cvop{i}{\widehat{k}}
\cvop{\widehat{l}}{j}.
\]
To evaluate $\supscrpt{\tilde{\lambda}}{2}$, we need now only calculate
the eigenvalues of $\qrow{2}{i}$ and $\qcol{2}{j}$ with respect to the
matrices $\xop{i\widehat{l}}{\widehat{k}j}$, for each pair of values for
$l$ and $k$.  Evaluating
\begin{eqnarray*}
\qrow{2}{i} \cvop{i}{\widehat{+}} & = & - \qrow{2}{i} \cvop{i}{\widehat{-}}
=3 p,\\
\cvop{\widehat{+}}{i} \qcol{2}{i} & = & - \cvop{\widehat{-}}{i} \qcol{2}{i}
= \frac{3}{2} p f (1- f),\\
\qrow{2}{i} \cvop{i}{\widehat{0}} & = & - \cvop{\widehat{0}}{i} \qcol{2}{i}
= 0,
\end{eqnarray*}
we have
\begin{eqnarray}
\supscrpt{\tilde{\lambda}}{2} & = &  \lamsupp{2} + 2\qrow{2}{i}
\xop{i\widehat{+}}{\widehat{+}j} \qcol{2}{j} \yop{\widehat{+}}{\widehat{+}}
+ 2\qrow{2}{i} \xop{i\widehat{+}}{\widehat{-}j}
\qcol{2}{j}\yop{\widehat{+}}{\widehat{-}}
\label{eq:1D3Pcorrectiona} \\
& = &  - 3p f + 9 p^2 f (1- f) (\yop{\widehat{+}}{\widehat{+}}
- \yop{\widehat{+}}{\widehat{-}}) \nonumber.
\end{eqnarray}Diagrammatically, this equation can be expressed as
\bge
   \supscrpt{\tilde{\lambda}}{2} = - 3p f + 9 p^2 f (1- f)
   \left(\sum_{\cdots}
\begin{picture}(50,50)(0,22)
\put(35.,50.){\vector(0,-1){17.5}}
\put(35.,50.){\vector(-1,-1){18.2322}}
\put(25.,25.){\makebox(0,0){$\cdots$}}
\put(15.,20.){\vector(0,-1){17.5}}
\put(35.,20.){\vector(-1,-1){18.2322}}
\put(35.,50.){\circle*{5.}}
\put(35.,30.){\circle*{5.}}
\put(15.,30.){\circle*{5.}}
\put(15.,20.){\circle*{5.}}
\put(35.,20.){\circle*{5.}}
\put(15.,0.){\circle*{5.}}
\end{picture}
-\sum_{\cdots}
\begin{picture}(50,50)(0,22)
\put(35.,50.){\vector(0,-1){17.5}}
\put(35.,50.){\vector(-1,-1){18.2322}}
\put(25.,25.){\makebox(0,0){$\cdots$}}
\put(35.,20.){\vector(0,-1){17.5}}
\put(15.,20.){\vector(1,-1){18.2322}}
\put(35.,50.){\circle*{5.}}
\put(35.,30.){\circle*{5.}}
\put(15.,30.){\circle*{5.}}
\put(15.,20.){\circle*{5.}}
\put(35.,20.){\circle*{5.}}
\put(35.,0.){\circle*{5.}}
\end{picture}
   \right),
   \label{eq:1D3Pcorrection}
\ee
where the notation in brackets indicates summation of the products of
all internal vertex factors over all diagrams with the depicted initial
and final configurations.  Together, Eqs.~(\ref{eq:edoodtplga}) and
(\ref{eq:1D3Pcorrection}), with vertices given in
Fig.~\ref{fig:1d3pvertices}, constitute an {\sl exact} expression for
the diffusivity of the 1D3P lattice gas.

\subsection{Burgers' Equation Lattice Gas}
\subsubsection{Vertices}

{}From either the collision operator or
the state transition table for this lattice gas
(see Subsection~\ref{ssec:belg}) we can
calculate the nonzero mean vertex coefficients, which are given by
\begin{eqnarray*}
\mixten{V}{B}{B} & = & \mixten{V}{\emptyset}{\emptyset} = 1\\
\mixten{V}{+}{B} & = & - a\\
\mixten{V}{-}{B} & = & a\\
\mixten{V}{+}{\pm} & = & \frac{1 + a}{2}  \\
\mixten{V}{-}{\pm} & = & \frac{1 - a}{2}.
\end{eqnarray*}

Using the equilibrium value $f$ for the mean occupation numbers
$\nnsupo{\pm}{0}$, we can calculate the correlation vertex
coefficients.  Recall that we calculate the CVC's using only the
zero-order mean occupation numbers.  The  CVC's are given by
\begin{eqnarray*}
\mixten{{\cal V}}{B}{B} & = & \mixten{{\cal V}}{\emptyset}{\emptyset} = 1\\
\mixten{{\cal V}}{i}{j} & = & \frac{1}{2} \\
\mixten{{\cal V}}{i}{B} & = &
\mixten{{\cal V}}{B}{j} = 0 \\
\mixten{{\cal V}}{\emptyset}{B} & = &
\mixten{{\cal V}}{B}{\emptyset} = 0.
\end{eqnarray*}
In particular, note that all the CVC's which modify the number of
correlated quantities are zero.  Thus, in this lattice gas, no ${\cal
O}(\epsilon)$ correlations are generated by gradients in $N_1$, and
correlations cannot affect the hydrodynamic equation by influencing the
single-particle means.

Because of the first order collision operator $C_1$ which does not
satisfy semi-detailed balance, correlations might also be generated by
the quantities $I^{\widehat{\mu}}$.  From Eq.~(\ref{eq:extrasource}), we
have
\[
I^B = 0.
\]
Thus, we find that for this lattice gas, no correlations are generated
to ${\cal O}(\epsilon)$.  Furthermore, even if correlations existed,
they would not couple back to the hydrodynamic equations, since
$\mixten{{\cal V}}{\pm}{B} = 0$.  It follows that the standard
Chapman-Enskog analysis gives the correct results for the transport
coefficients.  In fact, this result was proven using other methods in
\cite{lebopres}.

\subsection{2D4P Lattice Gas}
\subsubsection{Vertices}

\begin{figure}
\centering
\begin{picture}(470,450)(0,0)
\put(117.5,427.5){
\begin{picture}(40,40)(0,0)
\put(-20.,0.){\vector(1,0){17.5}}
\put(0.,0.){\vector(1,0){20.}}
\put(0.,0.){\circle*{5.}}
\put(-6.,-24.){\makebox(0,0){\scriptsize $1-2\nu(1-\nu)$}}
\end{picture}}
\put(235.,427.5){
\begin{picture}(40,40)(0,0)
\put(-20.,0.){\vector(1,0){17.5}}
\put(0.,0.){\vector(-1,0){20.}}
\put(0.,0.){\circle*{5.}}
\put(-6.,-24.){\makebox(0,0){\scriptsize $2\nu(1-\nu)$}}
\end{picture}}
\put(352.5,427.5){
\begin{picture}(40,40)(0,0)
\put(-20.,0.){\vector(1,0){17.5}}
\put(0.,0.){\vector(1,0){20.}}
\put(0.,0.){\vector(0,1){20.}}
\put(0.,0.){\circle*{5.}}
\put(-6.,-24.){\makebox(0,0){\scriptsize $-\nu(1-\nu)(1-2\nu)$}}
\end{picture}}
\put(117.5,360.){
\begin{picture}(40,40)(0,0)
\put(20.,0.){\vector(-1,0){17.5}}
\put(0.,0.){\vector(1,0){20.}}
\put(0.,0.){\vector(0,1){20.}}
\put(0.,0.){\circle*{5.}}
\put(-6.,-24.){\makebox(0,0){\scriptsize $\nu(1-\nu)(1-2\nu)$}}
\end{picture}}
\put(235.,360.){
\begin{picture}(40,40)(0,0)
\put(0.,20.){\vector(0,-1){17.5}}
\put(0.,0.){\vector(1,0){20.}}
\put(0.,0.){\vector(0,1){20.}}
\put(0.,0.){\vector(-1,0){20.}}
\put(0.,0.){\circle*{5.}}
\put(-6.,-24.){\makebox(0,0){\scriptsize $2\mu^2(1-\mu)^2$}}
\end{picture}}
\put(352.5,360.){
\begin{picture}(40,40)(0,0)
\put(0.,-20.){\vector(0,1){17.5}}
\put(0.,0.){\vector(1,0){20.}}
\put(0.,0.){\vector(0,1){20.}}
\put(0.,0.){\vector(-1,0){20.}}
\put(0.,0.){\circle*{5.}}
\put(-6.,-24.){\makebox(0,0){\scriptsize $-2\mu^2(1-\mu)^2$}}
\end{picture}}
\put(117.5,292.5){
\begin{picture}(40,40)(0,0)
\put(20.,0.){\vector(-1,0){17.5}}
\put(0.,20.){\vector(0,-1){17.5}}
\put(0.,0.){\vector(1,0){20.}}
\put(0.,0.){\circle*{5.}}
\put(-6.,-24.){\makebox(0,0){\scriptsize $1-2\nu$}}
\end{picture}}
\put(235.,292.5){
\begin{picture}(40,40)(0,0)
\put(-20.,0.){\vector(1,0){17.5}}
\put(0.,20.){\vector(0,-1){17.5}}
\put(0.,0.){\vector(1,0){20.}}
\put(0.,0.){\circle*{5.}}
\put(-6.,-24.){\makebox(0,0){\scriptsize $1-2\nu$}}
\end{picture}}
\put(352.5,292.5){
\begin{picture}(40,40)(0,0)
\put(0.,-20.){\vector(0,1){17.5}}
\put(20.,0.){\vector(-1,0){17.5}}
\put(0.,0.){\vector(1,0){20.}}
\put(0.,0.){\vector(0,1){20.}}
\put(0.,0.){\circle*{5.}}
\put(-6.,-24.){\makebox(0,0){\scriptsize $2\mu(1-\mu)-2\nu(1-\nu)$}}
\end{picture}}
\put(117.5,225.){
\begin{picture}(40,40)(0,0)
\put(-20.,0.){\vector(1,0){17.5}}
\put(0.,-20.){\vector(0,1){17.5}}
\put(0.,0.){\vector(1,0){20.}}
\put(0.,0.){\vector(0,1){20.}}
\put(0.,0.){\circle*{5.}}
\put(-6.,-24.){\makebox(0,0){\scriptsize $2\nu(1-\nu)+2\mu(1-\mu)$}}
\end{picture}}
\put(235.,225.){
\begin{picture}(40,40)(0,0)
\put(20.,0.){\vector(-1,0){17.5}}
\put(0.,20.){\vector(0,-1){17.5}}
\put(0.,0.){\vector(1,0){20.}}
\put(0.,0.){\vector(0,1){20.}}
\put(0.,0.){\circle*{5.}}
\put(-6.,-24.){\makebox(0,0){\scriptsize $1-2\nu(1-\nu)-2\mu(1-\mu)$}}
\end{picture}}
\put(352.5,225.){
\begin{picture}(40,40)(0,0)
\put(-20.,0.){\vector(1,0){17.5}}
\put(20.,0.){\vector(-1,0){17.5}}
\put(0.,0.){\vector(1,0){20.}}
\put(0.,0.){\vector(-1,0){20.}}
\put(0.,0.){\circle*{5.}}
\put(-6.,-24.){\makebox(0,0){\scriptsize $1$}}
\end{picture}}
\put(117.5,157.5){
\begin{picture}(40,40)(0,0)
\put(20.,0.){\vector(-1,0){17.5}}
\put(0.,20.){\vector(0,-1){17.5}}
\put(0.,0.){\vector(1,0){20.}}
\put(0.,0.){\vector(0,1){20.}}
\put(0.,0.){\vector(-1,0){20.}}
\put(0.,0.){\circle*{5.}}
\put(-6.,-24.){\makebox(0,0){\scriptsize $-\mu(1-\mu)(1-2\mu)$}}
\end{picture}}
\put(235.,157.5){
\begin{picture}(40,40)(0,0)
\put(0.,-20.){\vector(0,1){17.5}}
\put(20.,0.){\vector(-1,0){17.5}}
\put(0.,0.){\vector(1,0){20.}}
\put(0.,0.){\vector(0,1){20.}}
\put(0.,0.){\vector(-1,0){20.}}
\put(0.,0.){\circle*{5.}}
\put(-6.,-24.){\makebox(0,0){\scriptsize $\mu(1-\mu)(1-2\mu)$}}
\end{picture}}
\put(352.5,157.5){
\begin{picture}(40,40)(0,0)
\put(0.,-20.){\vector(0,1){17.5}}
\put(20.,0.){\vector(-1,0){17.5}}
\put(0.,20.){\vector(0,-1){17.5}}
\put(0.,0.){\vector(1,0){20.}}
\put(0.,0.){\circle*{5.}}
\put(-6.,-24.){\makebox(0,0){\scriptsize $-2$}}
\end{picture}}
\put(117.5,90.){
\begin{picture}(40,40)(0,0)
\put(-20.,0.){\vector(1,0){17.5}}
\put(0.,-20.){\vector(0,1){17.5}}
\put(0.,20.){\vector(0,-1){17.5}}
\put(0.,0.){\vector(1,0){20.}}
\put(0.,0.){\circle*{5.}}
\put(-6.,-24.){\makebox(0,0){\scriptsize $+2$}}
\end{picture}}
\put(235.,90.){
\begin{picture}(40,40)(0,0)
\put(-20.,0.){\vector(1,0){17.5}}
\put(0.,-20.){\vector(0,1){17.5}}
\put(20.,0.){\vector(-1,0){17.5}}
\put(0.,0.){\vector(1,0){20.}}
\put(0.,0.){\vector(0,1){20.}}
\put(0.,0.){\circle*{5.}}
\put(-6.,-24.){\makebox(0,0){\scriptsize $1-2\nu$}}
\end{picture}}
\put(352.5,90.){
\begin{picture}(40,40)(0,0)
\put(-20.,0.){\vector(1,0){17.5}}
\put(20.,0.){\vector(-1,0){17.5}}
\put(0.,20.){\vector(0,-1){17.5}}
\put(0.,0.){\vector(1,0){20.}}
\put(0.,0.){\vector(0,1){20.}}
\put(0.,0.){\circle*{5.}}
\put(-6.,-24.){\makebox(0,0){\scriptsize $-(1-2\mu)$}}
\end{picture}}
\put(117.5,22.5){
\begin{picture}(40,40)(0,0)
\put(-20.,0.){\vector(1,0){17.5}}
\put(20.,0.){\vector(-1,0){17.5}}
\put(0.,20.){\vector(0,-1){17.5}}
\put(0.,0.){\vector(1,0){20.}}
\put(0.,0.){\vector(0,1){20.}}
\put(0.,0.){\vector(-1,0){20.}}
\put(0.,0.){\circle*{5.}}
\put(-6.,-24.){\makebox(0,0){\scriptsize $2\mu(1-\mu)$}}
\end{picture}}
\put(235.,22.5){
\begin{picture}(40,40)(0,0)
\put(-20.,0.){\vector(1,0){17.5}}
\put(0.,-20.){\vector(0,1){17.5}}
\put(20.,0.){\vector(-1,0){17.5}}
\put(0.,0.){\vector(1,0){20.}}
\put(0.,0.){\vector(0,1){20.}}
\put(0.,0.){\vector(-1,0){20.}}
\put(0.,0.){\circle*{5.}}
\put(-6.,-24.){\makebox(0,0){\scriptsize $1-2\mu(1-\mu)$}}
\end{picture}}
\put(352.5,22.5){
\begin{picture}(40,40)(0,0)
\put(-20.,0.){\vector(1,0){17.5}}
\put(0.,-20.){\vector(0,1){17.5}}
\put(20.,0.){\vector(-1,0){17.5}}
\put(0.,20.){\vector(0,-1){17.5}}
\put(0.,0.){\vector(1,0){20.}}
\put(0.,0.){\vector(0,1){20.}}
\put(0.,0.){\vector(-1,0){20.}}
\put(0.,0.){\vector(0,-1){20.}}
\put(0.,0.){\circle*{5.}}
\put(-6.,-24.){\makebox(0,0){\scriptsize $1$}}
\end{picture}}
\end{picture}
\caption{Vertex Factors for 2D4P Lattice Gas}
\label{fig:tdfpvertices}
\end{figure}
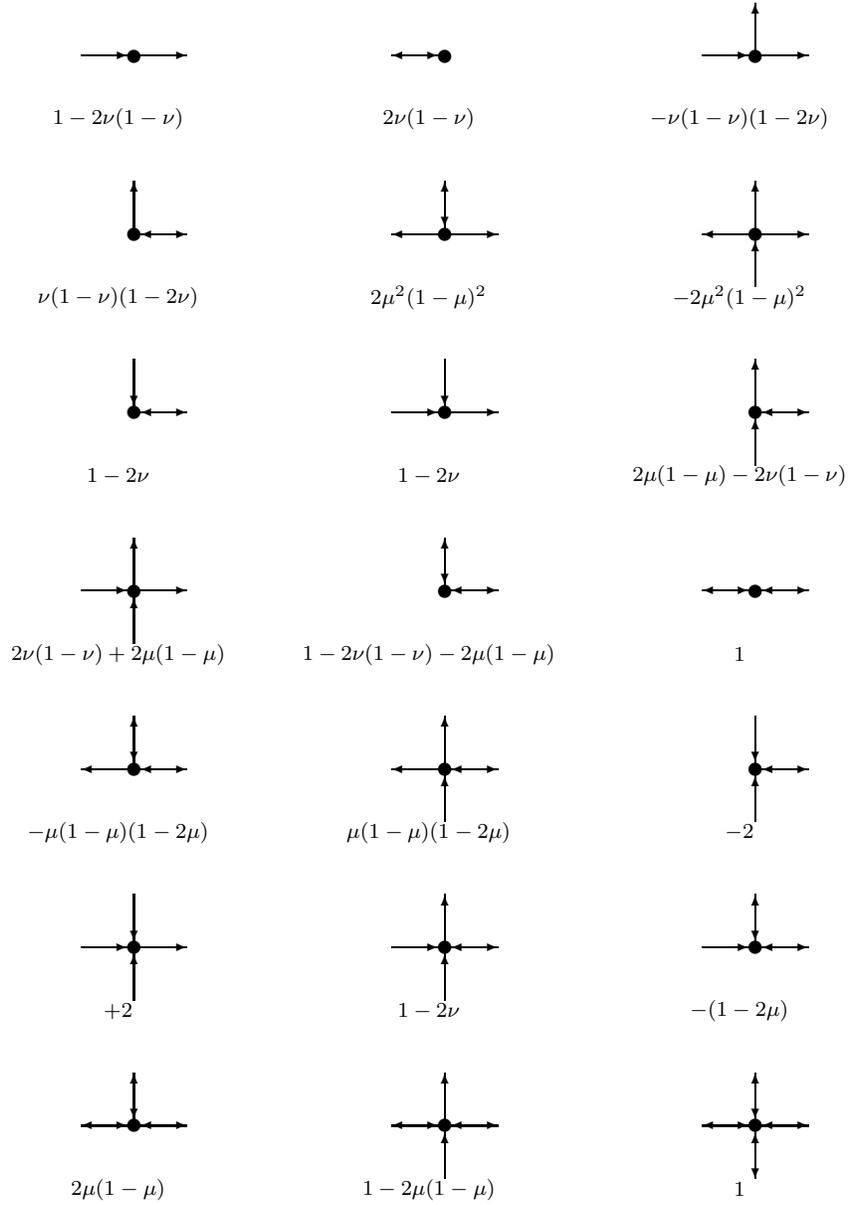

Using the general formulae, Eqs.~(\ref{eq:vertexmean2}) and
(\ref{eq:vdef}), to calculate the correlation vertex coefficients
$\cvop{\alpha}{\beta}$ for the 2D4P lattice gas, we arrive at the values
for the CVC's which are depicted in Fig.~\ref{fig:tdfpvertices}.
In this figure we have only included a single example of each
equivalence class of vertices under the symmetry group generated by
reflections across either axis.  Furthermore, we have only included a
single example of each pair of vertices which are related by a
$90^\circ$ rotation.  The CVC's for such vertices are related by
exchanging the two particle densities $\mu \leftrightarrow \nu$.  For
example, we have
\begin{eqnarray}
\cvop{\vertex1000}{\vertex1000} & = & 1 - 2 \nu + 2 \nu^2\\
\cvop{\vertex0100}{\vertex0100} & = & 1 - 2 \mu + 2 \mu^2.
\end{eqnarray}

\subsubsection{Renormalization of Eigenvalues}

Applying the theorems of Subsection~\ref{ssec:eigrnrm} to the 2D4P
lattice gas, we find that the renormalized $J$-matrix must be of the
form
\[
\tilde{J} = \left(\begin{array}{cccc}
           \tilde{\lambda}^4/2 & 0 & - \tilde{\lambda}^4/2 & 0 \\
          0 &  \tilde{\lambda}^3/2 & 0 & - \tilde{\lambda}^3/2 \\
          - \tilde{\lambda}^4/2 & 0 &  \tilde{\lambda}^4/2 & 0 \\
          0 & - \tilde{\lambda}^3/2 & 0 &  \tilde{\lambda}^3/2
          \end{array}
    \right).
\]
The renormalized hydrodynamic equations are
\bgeas
  \frac{\pdv \mu}{\pdv t} & = &
  \frac{\pdv}{\pdv x}
    \left(
       \tilde{{\cal D}}(\nu, \mu) \frac{\pdv \mu}{\pdv x}
    \right) \\
  \frac{\pdv \nu}{\pdv t} & = &
  \frac{\pdv}{\pdv y}
    \left(
       \tilde{{\cal D}}(\mu, \nu) \frac{\pdv \nu}{\pdv y}
    \right),
\eeas
with
\[
  \tilde{{\cal D}}(\mu,\nu) =
		\frac{c^2}{2\dt}\left(\frac{2}{(-\tilde{\lambda}^3)}-1\right)
\]
and
\[
  \tilde{{\cal D}}(\nu, \mu) =
		\frac{c^2}{2\dt}\left(\frac{2}{(-\tilde{\lambda}^4)}-1\right).
\]
In order to calculate the renormalized eigenvalue
$\tilde{\lambda}^4$ it is only necessary to calculate the component
$\mixten{\tilde{J}}{1}{1}$ of the renormalized $J$-matrix.  Because
the lattice gas is invariant under the combination of a $90^\circ$
rotation and the exchange $\mu \leftrightarrow \nu$, the eigenvalue
$\tilde{\lambda}^3$ can be calculated from $\tilde{\lambda}^4$ by the
exchange of particle densities.

Using a similar notation to that developed for the 1D3P lattice gas in
Section \ref{sec:examples}.1.2, we can write the renormalized matrix
$\tilde{J}$ in the form of Eq.~(\ref{eq:rjodtp}).
Using the symmetry properties of the lattice gas to prove the equality
of quantities $\mixten{Y}{\alpha}{\beta}$ which are related by
reflections across each axis, we can simplify the expression for the
shift in $\tilde{\lambda}^4$ to
\begin{eqnarray*}
\delta \lambda^4 & = &    8 \nu (\nu - 1) (2 \nu - 1)^2
	\left[-\mixten{Y}{\vertex1100}{\vertex1100}
	+\mixten{Y}{\vertex0110}{\vertex1100}
	+\mixten{Y}{\vertex0011}{\vertex1100}
	-\mixten{Y}{\vertex1001}{\vertex1100}\right]
+  16 \nu^2 (\nu - 1)^2
	\left[\mixten{Y}{\vertex1101}{\vertex1101}
	-\mixten{Y}{\vertex0111}{\vertex1101}\right]\\
& &+  16 \nu (\nu - 1) (2\nu - 1)
	\left[-\mixten{Y}{\vertex1101}{\vertex1100}
	+\mixten{Y}{\vertex0111}{\vertex1100}\right]
+  16 \nu^2 (\nu - 1)^2 (2 \nu - 1)
	\left[\mixten{Y}{\vertex1100}{\vertex1101}
	-\mixten{Y}{\vertex0110}{\vertex1101}\right].
\end{eqnarray*}
Similarly, we have
\begin{eqnarray*}
\delta \lambda^3 & = &    8 \mu (\mu - 1) (2 \mu - 1)^2
	\left[-\mixten{Y}{\vertex1001}{\vertex1001}
	+\mixten{Y}{\vertex1100}{\vertex1001}
	+\mixten{Y}{\vertex0110}{\vertex1001}
	-\mixten{Y}{\vertex0011}{\vertex1001}\right]
+  16 \mu^2 (\mu - 1)^2
	\left[\mixten{Y}{\vertex1011}{\vertex1011}
	-\mixten{Y}{\vertex1110}{\vertex1011}\right]\\
& &+  16 \mu (\mu - 1) (2\mu - 1)
	\left[-\mixten{Y}{\vertex1011}{\vertex1001}
	+\mixten{Y}{\vertex1110}{\vertex1001}\right]
+  16 \mu^2 (\mu - 1)^2 (2 \mu - 1)
	\left[\mixten{Y}{\vertex1001}{\vertex1011}
	-\mixten{Y}{\vertex1100}{\vertex1011}\right].
\end{eqnarray*}

These equations describe completely the renormalization of the
hydrodynamic equations due to correlations.  Each term
$\mixten{Y}{\alpha}{\beta}$ corresponds to a set of diagrams with a
specific set of outgoing and incoming virtual particles at the initial
and final vertices of the diagram.

A rather dramatic simplification of the eigenvalue renormalization
equations occurs when the equilibrium particle densities $\mu$ and $\nu$
are equal to $1/2$.  When $\nu = 1/2$, we have
\bge
\delta \lambda^4 =
	\left[\mixten{Y}{\vertex1101}{\vertex1101}
	-\mixten{Y}{\vertex0111}{\vertex1101}\right],
\label{eq:tdfpexa}
\ee
and when $\mu = 1/2$ similarly
\bge
\delta \lambda^3=
	\left[\mixten{Y}{\vertex1011}{\vertex1011}
	-\mixten{Y}{\vertex1110}{\vertex1011}\right].
\label{eq:tdfpexb}
\ee
It follows that for these particular values of $\mu$ and $\nu$, the set
of diagrams which give a nonzero contribution to the renormalization of
the eigenvalues is reduced to only those diagrams which have 3 outgoing
virtual particles at the initial vertex, and 3 incoming virtual
particles at the final vertex.  Thus, for example, the ring and
2-particle BBGKY approximations for this lattice gas vanish at the
equilibrium described by $\mu = \nu = 1/2$.

In this particular case, the $\tau=3$ short-$\tau$ approximation is
given by Eqs.~(\ref{eq:tdfpexa}) and (\ref{eq:tdfpexb}), with the first
$Y$ term in each expression vanishing (because it is impossible to
connect the outgoing and incoming particles in two time steps), and with
the second arising from a single diagram of weight $1/8$.  It follows
that the corrected eigenvalues are equal to $-9/8$ in this simple
approximation.  This leads to $\tilde{\cal D}/{\cal D}=7/9=0.777\ldots$,
which may be compared with the experimental value of
${\cal D}_{\mbox{expt}}/{\cal D}\sim 0.71$~\cite{bmbcdltdfp}.

In a future paper~\cite{bt2}, we will discuss in more detail the results
of summing various subsets of diagrams for this lattice gas and compare
the results to experimental data.  The ring kinetic theory for this
lattice gas has been worked out in~\cite{vanroij}; in this reference, a
ring-like approximation is also used to treat noninteracting
three-particle correlations, giving nonzero correction in the
$\mu=\nu=1/2$ case.

\subsection{FHP-I Lattice Gas}
\subsubsection{Vertices}

Using the general formulae (\ref{eq:vertexmean2}) and (\ref{eq:vdef}) to
calculate the correlation vertex coefficients $\cvop{\alpha}{\beta}$ for
the FHP-I lattice gas results in nearly 300 nonvanishing CVC's that are
independent in the sense that they are not related by symmetries.  While
it is a straightforward task for a symbolic algebra computer program to
compute and work with these quantities, it would not be useful to
present all the results in this paper.  Instead, we present only the
initial (one-to-many) vertices, the propagator (one-to-one) vertices,
and the closure (many-to-one) vertices.  These vertices are sufficient
to compute all the diagrams in the kinetic ring approximation, and in
the $\tau = 3$ short-$\tau$ approximation.
\begin{figure}
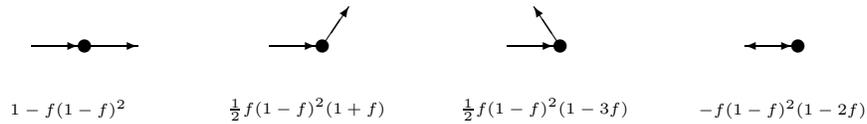

\centering

\caption{Propagator Vertex Factors for FHP-I Lattice Gas}
\label{fig:fhpverticesb}
\end{figure}
\begin{figure}
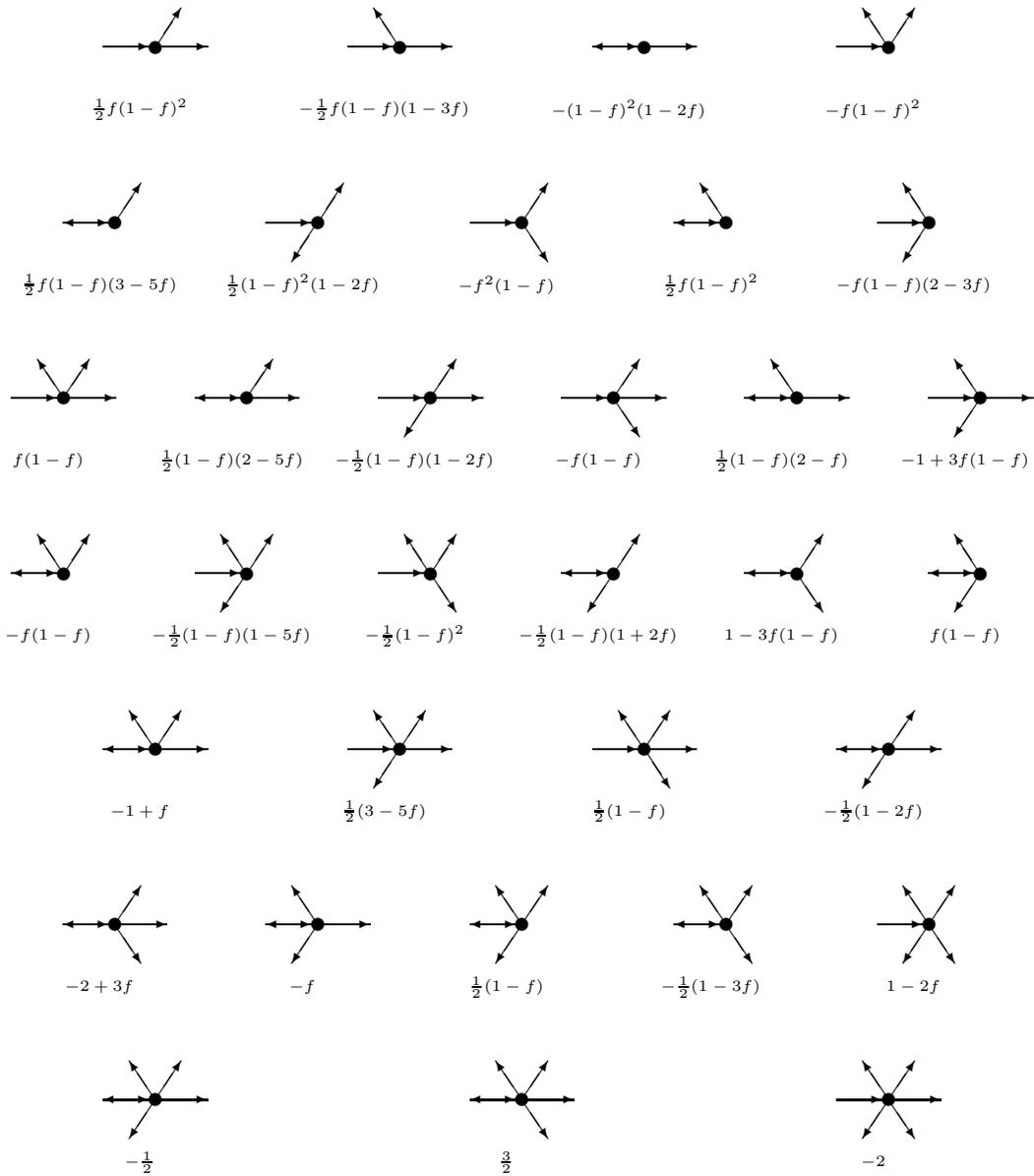

\centering
%
\caption{Initial Vertex Factors for FHP-I Lattice Gas}
\label{fig:fhpverticesa}
\end{figure}
\begin{figure}
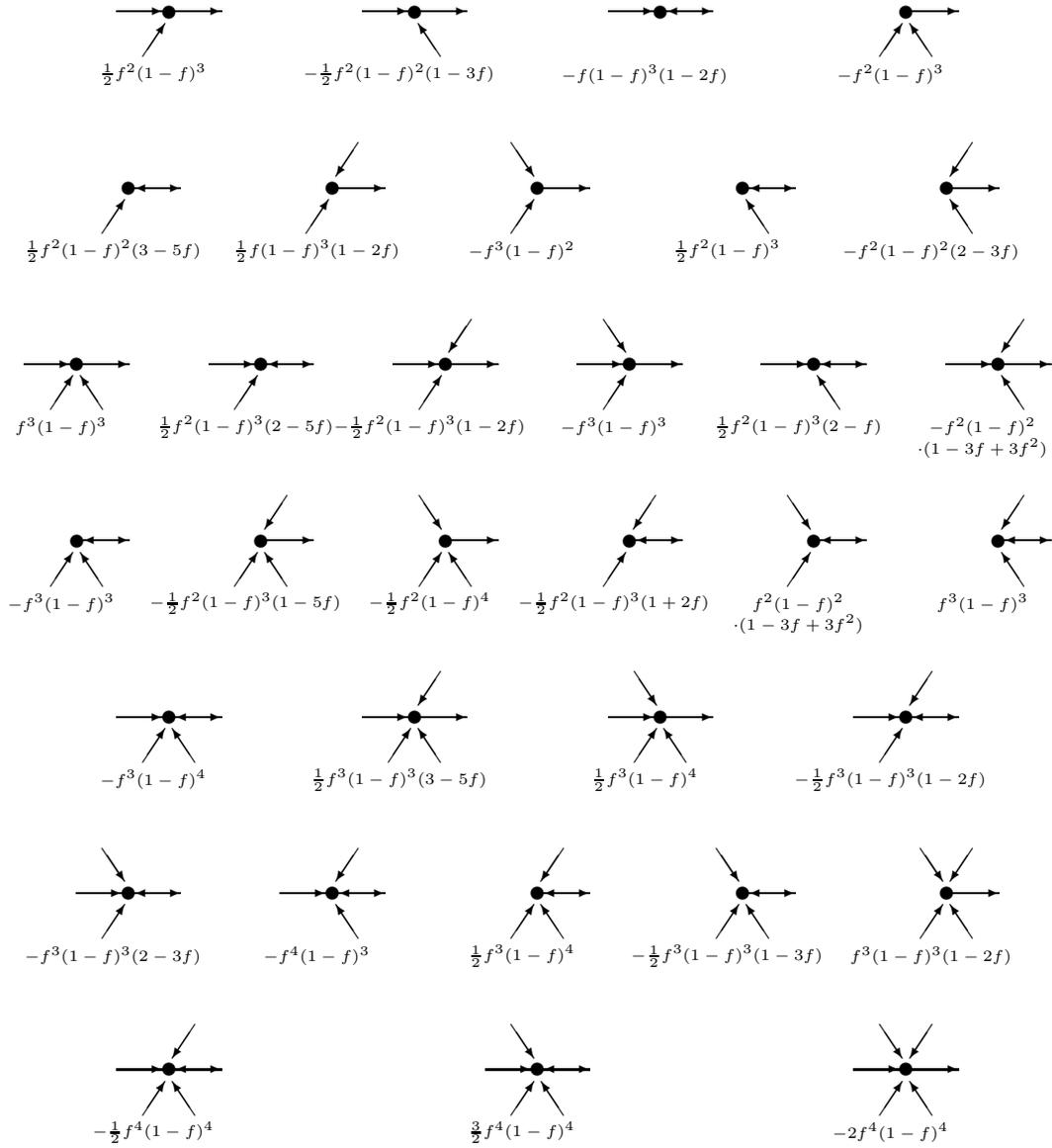

\centering
%
}
\end{picture}
\caption{Closure Vertex Factors for FHP-I Lattice Gas}
\label{fig:fhpverticesc}
\end{figure}

\subsubsection{Renormalization of Eigenvalues}

The only transport coefficient of the FHP-I lattice gas which undergoes
renormalization is the shear viscosity $\nu$.  From the general
arguments in Subsection~\ref{ssec:eigrnrm}, we see that the renormalized
eigenvalues $\supscrpt{\tilde{\lambda}}{4}$ and
$\supscrpt{\tilde{\lambda}}{6}$ must be equal, as they lie in a single
irreducible representation of the symmetry group generated by a
$60^\circ$ rotation.  In terms of these renormalized eigenvalues of the
$\tilde{J}$ matrix, the renormalized shear viscosity $\tilde{\nu}$ is
given by
\[
\tilde{\nu} = \frac{c^2}{8 \Delta t}
\left( \frac{2}{ (-\tilde{\lambda}^4)}  -1 \right).
\]
The renormalized eigenvalue $\tilde{\lambda}^4$ can be extracted from
the $\tilde{J}$ matrix by taking the components
\[
\tilde{\lambda}^4 = 6 (\tjop{1}{1}+\tjop{1}{2}).
\]
As in the previous examples, we can  express the eigenvalue shift in
the form
\[
\delta \lambda^4 =
6\left(\xop{1 \widehat{\mu}}{\widehat{\nu}1}
\yop{\widehat{\nu}}{\widehat{\mu}} +
\xop{1 \widehat{\mu}}{\widehat{\nu}2}
\yop{\widehat{\nu}}{\widehat{\mu}} \right).
\]
By collecting coefficients of terms $\yop{\widehat{\mu}}{\widehat{\nu}}$
which are related by symmetry, the expression for the eigenvalue shift
reduces to a sum over more than a hundred terms.

We will not explore the diagrammatic expansion of the FHP-I lattice gas
further in this paper.  However, we have used the general expression for
the renormalized shear viscosity to calculate several simple
renormalization effects.  In particular, one finds that in the ring
approximation, the sum over diagrams diverges logarithmically; this
result is well-known for incompressible fluids in 2-dimensions, and has
been verified using other methods for the FHP-I lattice gas~\cite{Ern}.

\newpage
\section{Approximations and Numerical Results \newline
         for the 1D3P Lattice Gas}
\label{sec:approximations}

In this section we discuss in detail the sets of diagrams for the 1D3P
lattice gas which correspond to the various approximation methods
described in Section~\ref{sec:exact}.7, and perform the associated
partial diagrammatic sums.  The goal of this detailed analysis is to
use the simple 1D3P lattice gas as a test case to study the relative
accuracy of the different approximations and the relative difficulty
of computing these approximations.  The results for this lattice gas
hopefully give a good indication of what approaches will lead to
useful results for more complicated lattice gases.

Correlations cause corrections to the Boltzmann diffusivity of the 1D3P
lattice gas which are as large as 5\% for certain values of the particle
density $f$ and bounce probability $p$.  We describe in this section the
results of numerical calculations of the partial diagrammatic sums
corresponding to the various approximation methods, and compare to
empirical results from computer simulations of this lattice gas.  In
Subsection~\ref{sec:approximations}.1 we calculate corrections to the
diffusivity in the short-$\tau$ approximation.  By graphing these
corrections and comparing to experiment, we see that as $\tau$ increases
these approximations give corrections to the Boltzmann approximation
which seem to converge to the correct experimental values.  The
convergence is slow, however, and since the calculation of these
corrections is computationally quite expensive, it is difficult to
estimate the asymptotic value of the diffusivity to a high degree of
accuracy using these approximations.  In
Subsection~\ref{sec:approximationsbbgky} we apply the partial BBGKY
summation prescription and again graph the results compared to
experimental values.  For small values of $k$ the $k$-particle BBGKY
diagrammatic summation converges rapidly, and in most ranges for which
we have calculated the results, these approximations approach
monotonically the experimental values as $k$ increases.  In this
subsection we also include a proof that the 2-particle BBGKY
approximation converges for arbitrary values of $p,f > 0$.  In
Subsection~\ref{sec:approximationsexpand} we observe that in the
vicinity of the density value $f = 1$, the corrections to the Boltzmann
approximation can be expanded in a power series in the variable
$\varepsilon =1 - f$.  The diagrams which contribute corrections of
order $\varepsilon^n$ are the diagrams which have $n$ or fewer vertices
where a single virtual particle branches out to two virtual particles
(``1-2'' vertices).  This expansion in $\varepsilon$ is roughly
equivalent to the familiar expansion in the continuum theory or in
quantum field theory in terms of the number of loops.  Using this
expansion, we can numerically evaluate the successive derivatives of the
correction term at $f = 1$, and compare these results to experiment.
The corrections thus calculated agree exactly with our experimental
results to within the statistical accuracy of the experimental results.
This calculation has the additional feature that it is possible to prove
that the sum over all diagrams which contribute to a given order in
$\varepsilon$ converges.  This convergence follows from the fact that
the sum over all diagrams which contain a fixed number of 1-2 vertices
is convergent; we prove this for the case of a single 1-2 vertex using
an argument which can be generalized in a straightforward fashion.  In
Subsection~\ref{sec:approximationsring} we consider the ring
approximation from an analytic perspective, and compare this
approximation to the closely related 2-particle BBGKY approximation.
Finally, in Subsection~\ref{sec:approximations}.5 we briefly compare the
results of the varied approximation methods used in this section.

\subsection{The Short-$\tau$ Approximation}

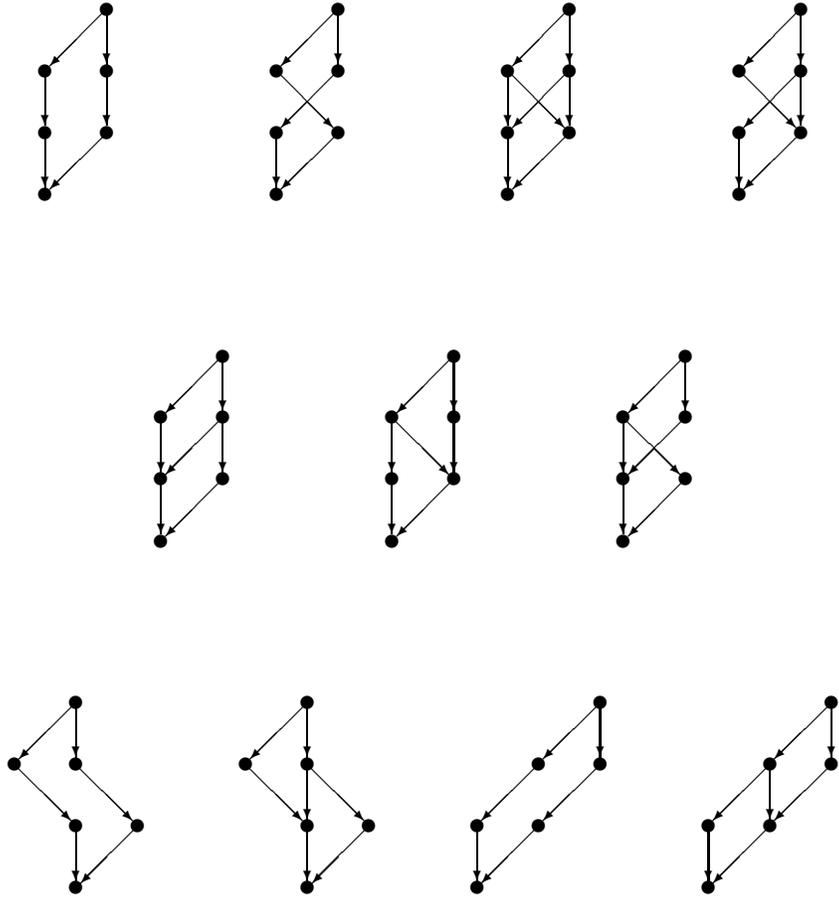
\begin{figure}
\centering
\begin{picture}(350,350)(0,0)
\put(43.75,306.25){
\begin{picture}(70,70)(0,0)
\put(23.31,46.69){\vector(0,-1){20.81}}
\put(46.69,46.69){\vector(0,-1){20.81}}
\put(46.69,70.){\vector(-1,-1){21.5422}}
\put(46.69,70.){\vector(0,-1){20.81}}
\put(23.31,23.31){\vector(0,-1){20.81}}
\put(46.69,23.31){\vector(-1,-1){21.5422}}
\put(46.69,70.){\circle*{5.}}
\put(23.31,46.69){\circle*{5.}}
\put(46.69,46.69){\circle*{5.}}
\put(23.31,23.31){\circle*{5.}}
\put(46.69,23.31){\circle*{5.}}
\put(23.31,0.){\circle*{5.}}
\end{picture}}
\put(131.25,306.25){
\begin{picture}(70,70)(0,0)
\put(23.31,46.69){\vector(1,-1){21.5422}}
\put(46.69,46.69){\vector(-1,-1){21.5422}}
\put(46.69,70.){\vector(-1,-1){21.5422}}
\put(46.69,70.){\vector(0,-1){20.81}}
\put(23.31,23.31){\vector(0,-1){20.81}}
\put(46.69,23.31){\vector(-1,-1){21.5422}}
\put(46.69,70.){\circle*{5.}}
\put(23.31,46.69){\circle*{5.}}
\put(46.69,46.69){\circle*{5.}}
\put(23.31,23.31){\circle*{5.}}
\put(46.69,23.31){\circle*{5.}}
\put(23.31,0.){\circle*{5.}}
\end{picture}}
\put(218.75,306.25){
\begin{picture}(70,70)(0,0)
\put(23.31,46.69){\vector(0,-1){20.81}}
\put(23.31,46.69){\vector(1,-1){21.5422}}
\put(46.69,46.69){\vector(-1,-1){21.5422}}
\put(46.69,46.69){\vector(0,-1){20.81}}
\put(46.69,70.){\vector(-1,-1){21.5422}}
\put(46.69,70.){\vector(0,-1){20.81}}
\put(23.31,23.31){\vector(0,-1){20.81}}
\put(46.69,23.31){\vector(-1,-1){21.5422}}
\put(46.69,70.){\circle*{5.}}
\put(23.31,46.69){\circle*{5.}}
\put(46.69,46.69){\circle*{5.}}
\put(23.31,23.31){\circle*{5.}}
\put(46.69,23.31){\circle*{5.}}
\put(23.31,0.){\circle*{5.}}
\end{picture}}
\put(306.25,306.25){
\begin{picture}(70,70)(0,0)
\put(23.31,46.69){\vector(1,-1){21.5422}}
\put(46.69,46.69){\vector(-1,-1){21.5422}}
\put(46.69,46.69){\vector(0,-1){20.81}}
\put(46.69,70.){\vector(-1,-1){21.5422}}
\put(46.69,70.){\vector(0,-1){20.81}}
\put(23.31,23.31){\vector(0,-1){20.81}}
\put(46.69,23.31){\vector(-1,-1){21.5422}}
\put(46.69,70.){\circle*{5.}}
\put(23.31,46.69){\circle*{5.}}
\put(46.69,46.69){\circle*{5.}}
\put(23.31,23.31){\circle*{5.}}
\put(46.69,23.31){\circle*{5.}}
\put(23.31,0.){\circle*{5.}}
\end{picture}}
\put(87.5,175.){
\begin{picture}(70,70)(0,0)
\put(23.31,46.69){\vector(0,-1){20.81}}
\put(46.69,46.69){\vector(-1,-1){21.5422}}
\put(46.69,46.69){\vector(0,-1){20.81}}
\put(46.69,70.){\vector(-1,-1){21.5422}}
\put(46.69,70.){\vector(0,-1){20.81}}
\put(23.31,23.31){\vector(0,-1){20.81}}
\put(46.69,23.31){\vector(-1,-1){21.5422}}
\put(46.69,70.){\circle*{5.}}
\put(23.31,46.69){\circle*{5.}}
\put(46.69,46.69){\circle*{5.}}
\put(23.31,23.31){\circle*{5.}}
\put(46.69,23.31){\circle*{5.}}
\put(23.31,0.){\circle*{5.}}
\end{picture}}
\put(175.,175.){
\begin{picture}(70,70)(0,0)
\put(23.31,46.69){\vector(0,-1){20.81}}
\put(23.31,46.69){\vector(1,-1){21.5422}}
\put(46.69,46.69){\vector(0,-1){20.81}}
\put(46.69,70.){\vector(-1,-1){21.5422}}
\put(46.69,70.){\vector(0,-1){20.81}}
\put(23.31,23.31){\vector(0,-1){20.81}}
\put(46.69,23.31){\vector(-1,-1){21.5422}}
\put(46.69,70.){\circle*{5.}}
\put(23.31,46.69){\circle*{5.}}
\put(46.69,46.69){\circle*{5.}}
\put(23.31,23.31){\circle*{5.}}
\put(46.69,23.31){\circle*{5.}}
\put(23.31,0.){\circle*{5.}}
\end{picture}}
\put(262.5,175.){
\begin{picture}(70,70)(0,0)
\put(23.31,46.69){\vector(0,-1){20.81}}
\put(23.31,46.69){\vector(1,-1){21.5422}}
\put(46.69,46.69){\vector(-1,-1){21.5422}}
\put(46.69,70.){\vector(-1,-1){21.5422}}
\put(46.69,70.){\vector(0,-1){20.81}}
\put(23.31,23.31){\vector(0,-1){20.81}}
\put(46.69,23.31){\vector(-1,-1){21.5422}}
\put(46.69,70.){\circle*{5.}}
\put(23.31,46.69){\circle*{5.}}
\put(46.69,46.69){\circle*{5.}}
\put(23.31,23.31){\circle*{5.}}
\put(46.69,23.31){\circle*{5.}}
\put(23.31,0.){\circle*{5.}}
\end{picture}}
\put(43.75,43.75){
\begin{picture}(70,70)(0,0)
\put(35.,70.){\vector(-1,-1){21.5422}}
\put(35.,70.){\vector(0,-1){20.81}}
\put(35.,23.31){\vector(0,-1){20.81}}
\put(58.31,23.31){\vector(-1,-1){21.5422}}
\put(11.69,46.69){\vector(1,-1){21.5422}}
\put(35.,46.69){\vector(1,-1){21.5422}}
\put(35.,70.){\circle*{5.}}
\put(11.69,46.69){\circle*{5.}}
\put(35.,46.69){\circle*{5.}}
\put(35.,23.31){\circle*{5.}}
\put(58.31,23.31){\circle*{5.}}
\put(35.,0.){\circle*{5.}}
\end{picture}}
\put(131.25,43.75){
\begin{picture}(70,70)(0,0)
\put(35.,46.69){\vector(0,-1){20.81}}
\put(35.,70.){\vector(-1,-1){21.5422}}
\put(35.,70.){\vector(0,-1){20.81}}
\put(35.,23.31){\vector(0,-1){20.81}}
\put(58.31,23.31){\vector(-1,-1){21.5422}}
\put(11.69,46.69){\vector(1,-1){21.5422}}
\put(35.,46.69){\vector(1,-1){21.5422}}
\put(35.,70.){\circle*{5.}}
\put(11.69,46.69){\circle*{5.}}
\put(35.,46.69){\circle*{5.}}
\put(35.,23.31){\circle*{5.}}
\put(58.31,23.31){\circle*{5.}}
\put(35.,0.){\circle*{5.}}
\end{picture}}
\put(218.75,43.75){
\begin{picture}(70,70)(0,0)
\put(58.31,70.){\vector(-1,-1){21.5422}}
\put(58.31,70.){\vector(0,-1){20.81}}
\put(11.69,23.31){\vector(0,-1){20.81}}
\put(35.,23.31){\vector(-1,-1){21.5422}}
\put(35.,46.69){\vector(-1,-1){21.5422}}
\put(58.31,46.69){\vector(-1,-1){21.5422}}
\put(58.31,70.){\circle*{5.}}
\put(35.,46.69){\circle*{5.}}
\put(58.31,46.69){\circle*{5.}}
\put(11.69,23.31){\circle*{5.}}
\put(35.,23.31){\circle*{5.}}
\put(11.69,0.){\circle*{5.}}
\end{picture}}
\put(306.25,43.75){
\begin{picture}(70,70)(0,0)
\put(35.,46.69){\vector(0,-1){20.81}}
\put(58.31,70.){\vector(-1,-1){21.5422}}
\put(58.31,70.){\vector(0,-1){20.81}}
\put(11.69,23.31){\vector(0,-1){20.81}}
\put(35.,23.31){\vector(-1,-1){21.5422}}
\put(35.,46.69){\vector(-1,-1){21.5422}}
\put(58.31,46.69){\vector(-1,-1){21.5422}}
\put(58.31,70.){\circle*{5.}}
\put(35.,46.69){\circle*{5.}}
\put(58.31,46.69){\circle*{5.}}
\put(11.69,23.31){\circle*{5.}}
\put(35.,23.31){\circle*{5.}}
\put(11.69,0.){\circle*{5.}}
\end{picture}}
\end{picture}
\caption{Diagrams Contributing to the $\tau=4$ Correction for the
1D3P Lattice Gas}
\label{fig:1d3pdiagrams}
\end{figure}

Consider the corrections to the eigenvalue $\supscrpt{\lambda}{2}$ in
the Chapman-Enskog analysis of the 1D3P lattice gas arising from
diagrams of fixed length $\tau$.  The first few such corrections are
easy to evaluate by hand.  For $\tau = 3$, the only diagrams
contributing to Eq.~(\ref{eq:1D3Pcorrectiona}) are the two diagrams
$T_1$ and $T_2$ shown in Fig.~\ref{fig:diags}.  These two diagrams
shift the eigenvalue $\supscrpt{\lambda}{2}$ by
\begin{eqnarray}
\delta \lamsupp{2}  =   \tilde{\lambda}^{2 (3)}
- \lamsupp{2}&= &
9p^2 f (1- f) [(fp)^2 - fp (1 - 2fp)] \nonumber \\
 & = & - 9p^3 f^2 (1 - f) (1- 3fp),\label{eq:t3}
\end{eqnarray}
where we denote by $\tilde{\lambda}^{2 (3)} $ the eigenvalue of $q^2_i$
in the $\tau = 3$ matrix $ \tjop{(3)i}{j}$.  For $\tau=4$, there are 22
diagrams which contribute to Eq.~(\ref{eq:1D3Pcorrectiona}).  The 11
diagrams contributing to $\yop{\widehat{+}}{\widehat{+}}$ are shown in
Fig.~\ref{fig:1d3pdiagrams}; the diagrams contributing to
$\yop{\widehat{+}}{\widehat{-}}$ can be generated from these by simply
changing the directions of the final pair of virtual particles.
The correction to $\supscrpt{\lambda}{2}$ for $\tau =4$ is given by
\bge
\delta \lamsupp{2} = \tilde{\lambda}^{2 (4)}-
\tilde{\lambda}^{2 (3)}=
9p^4 f^2 (1-f) (1-3fp)
\left( 4-3f-3f^2 p \right).
\label{eq:t4}
\ee
The complete correction in the $\tau = 4$ short-$\tau$ approximation is
given by summing the shifts in Eqs.~(\ref{eq:t3}) and (\ref{eq:t4}).  As
$\tau$ increases, the number of diagrams contributing to
$\delta\lamsupp{2}$ increases exponentially, and it rapidly becomes
impractical to compute the exact correction, even using numerical
computing techniques, without some means of simplifying or approximating
the calculation.  We have calculated the corrections to $\lamsupp{2}$
including diagrams up to $\tau =5$.  The resulting short-$\tau$
approximations $\tilde{\lambda}^{2 (\tau)}$ are plotted against $f$ and
compared to experiment for $p=1/4$ and $1/2$ in Figs.~\ref{fig:finite1}
and \ref{fig:finite2}, respectively.  Although the results of this
calculation are a great improvement over the Boltzmann approximation,
and clearly appear to be converging to the experimental values as $\tau$
increases, the oscillatory nature of these approximations (in $\tau$) is
an undesirable feature which makes it difficult to use partial results
to put bounds on the actual diffusivity.  Nonetheless, it is clear from
the graphs in Figs.~\ref{fig:finite1} and \ref{fig:finite2} that this
formalism includes empirically measurable effects which are completely
dropped in the Boltzmann approximation.
\begin{figure}
\centering
\setlength{\unitlength}{0.240900pt}
\ifx\plotpoint\undefined\newsavebox{\plotpoint}\fi
\sbox{\plotpoint}{\rule[-0.500pt]{1.000pt}{1.000pt}}%

\caption{Short-$\tau$ Approximations for $p=1/4$}
\label{fig:finite1}
\end{figure}

\subsection{BBGKY Approximations}
\label{sec:approximationsbbgky}

We now consider the partial BBGKY approximations for the 1D3P lattice
gas.  As described in Section~\ref{sec:exact}, the $k$-particle BBGKY
approximation is given by summing over all diagrams which have at most
$k$ simultaneously correlated virtual particles.  For each $k$, the
approximation thus consists of an infinite number of diagrams.  This
approximation can be reduced to a finite sum by also limiting the
lengths of the diagrams to some maximum size $\tau$ as in the
short-$\tau$ approximation.  Because for fixed $k$, the computational
complexity of the summation of graphs of length $\leq\tau$ grows
polynomially in $\tau$ rather than exponentially as in the short-$\tau$
approximation, it is easier to compute the limit of the set of
$k$-particle diagrams as $\tau\rightarrow\infty$ than the complete set
of diagrams in this limit.  For $k = 2$, it is possible to prove that
this infinite sum of diagrams must in fact converge; we derive this
result later in this subsection.  For $k > 2$, we do not have a complete
proof of convergence; however, numerical evidence indicates that for
each $k$, the infinite $k$-particle BBGKY sum of diagrams is convergent.
By using methods like those used in the following subsection, it may be
possible to prove that for each $k$ the BBGKY approximation converges.

\begin{figure}
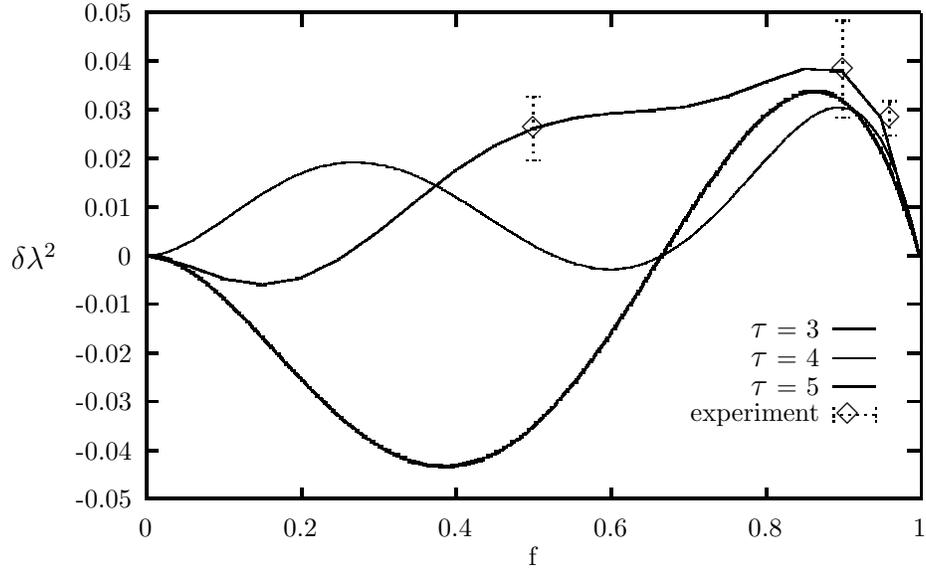

\centering
\setlength{\unitlength}{0.240900pt}
\ifx\plotpoint\undefined\newsavebox{\plotpoint}\fi
\sbox{\plotpoint}{\rule[-0.500pt]{1.000pt}{1.000pt}}%

\caption{Short-$\tau$ Approximations for $p=1/2$}
\label{fig:finite2}
\end{figure}

\begin{figure}
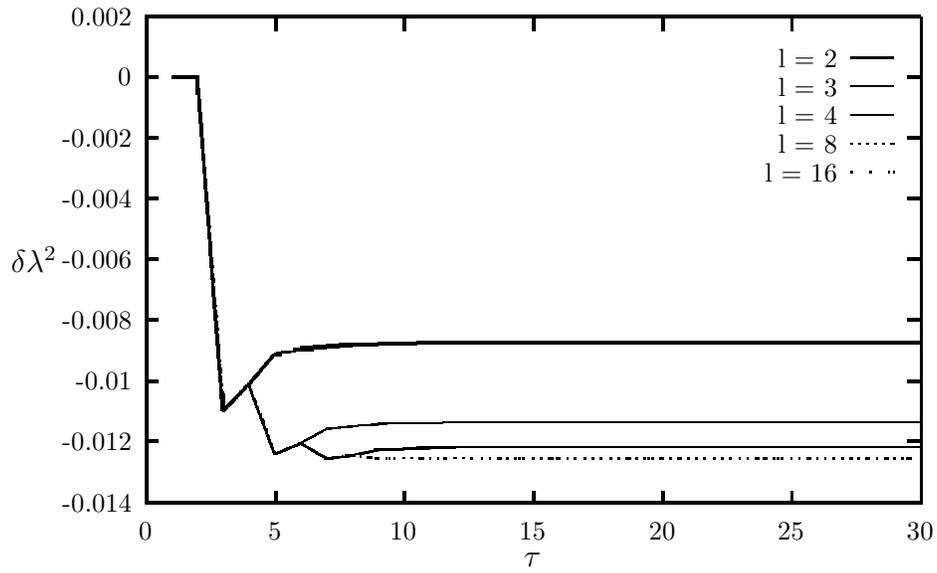

\centering
\setlength{\unitlength}{0.240900pt}
\ifx\plotpoint\undefined\newsavebox{\plotpoint}\fi
\sbox{\plotpoint}{\rule[-0.500pt]{1.000pt}{1.000pt}}%
%
\caption{Two-Particle BBGKY Corrections for $(p,f)=(0.25,0.5)$}
\label{fig:examplebbgky}
\end{figure}

\begin{figure}
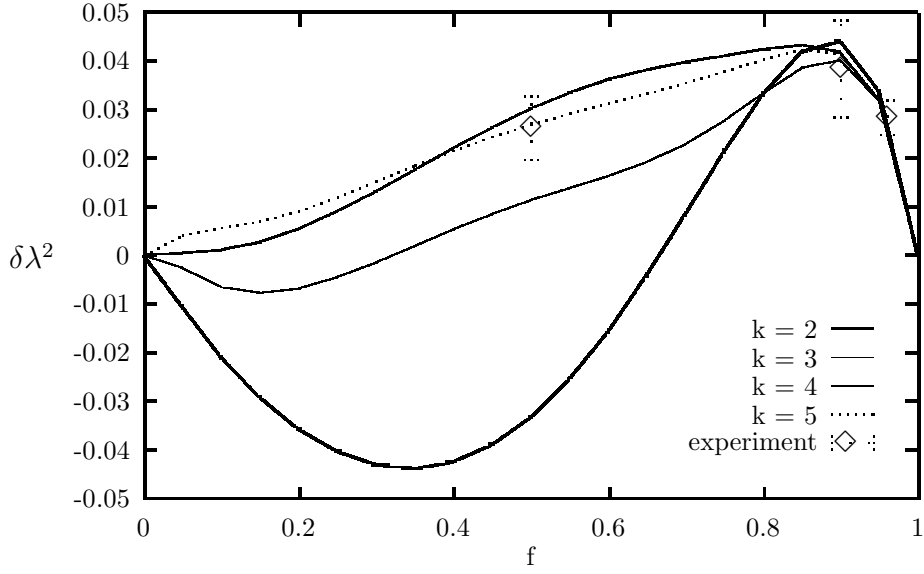

\centering
\setlength{\unitlength}{0.240900pt}
\ifx\plotpoint\undefined\newsavebox{\plotpoint}\fi
\sbox{\plotpoint}{\rule[-0.500pt]{1.000pt}{1.000pt}}%
%
\caption{Partial BBGKY Approximations for $p=1/2$}
\label{fig:bbgky2}
\end{figure}

We have used a computer to numerically calculate the limit of the full
$k$-particle BBGKY approximation for certain values of $f$ and $p$.
The algorithm we used was to  sum all diagrams of length $\leq \tau$
on a lattice of width $l$, then to take the limits as $\tau, l
\rightarrow \infty$.  As an example, for a characteristic pair of
values $(p,f)= (.25,.5)$, we have graphed in Fig.~\ref{fig:examplebbgky}
the corrections due to 2-particle BBGKY diagrams of length $ \leq \tau$
on a lattice of size $l$ for all $\tau<30$ and for lattice sizes 2, 3,
4, 8, and 16.  The curve for $l = 8$ is indistinguishable from, and
hidden by, the curve for $ l = 16$.  Note that all diagrams of length
$\tau$ are correctly summed as long as $l\geq\tau$ for all $k$; for $k =
2$, however, many diagrams cancel so that larger values of $\tau$ give
exact results, such as for $l = 2$, $\tau = 4$.  It is clear from this
graph not only that the sum over diagrams converges rapidly, but also
that the major part of the sum arises from the contributions of diagrams
which are of limited width.  We have numerically approximated the limits
of the $k$-particle BBGKY sums for $k \leq 5$.  The results of this
calculation are graphed in Figs.~\ref{fig:bbgky1} and \ref{fig:bbgky2},
for the same ranges of values for $p,f$ which were used for the
short-$\tau$ approximations in Figs.~\ref{fig:finite1} and
\ref{fig:finite2}.
\clearpage
\clearpage
\begin{figure}
\centering
\setlength{\unitlength}{0.240900pt}
\ifx\plotpoint\undefined\newsavebox{\plotpoint}\fi
\sbox{\plotpoint}{\rule[-0.500pt]{1.000pt}{1.000pt}}%
%
\caption{Partial BBGKY Approximations for $p=1/4$}
\label{fig:bbgky1}
\end{figure}
\clearpage

We conclude this subsection with a proof that the 2-particle BBGKY
approximation is convergent.  Although this proof is carried out for the
particular 1D3P lattice gas, it generalizes easily to any other system.
We begin by considering a matrix $M$ on the space $B_2$ of 2-particle
subsets of $B$.  The matrix element $\mixten{M}{\widehat{i}}{\widehat{j}}$ is
defined by summing over all 2-particle diagrams which begin with
outgoing VP's $\widehat{j}$ at an initial vertex and conclude with incoming
VP's $\widehat{i}$ at a final vertex, and which have the property that all
internal vertices have a single incoming and outgoing VP.  In
particular, this means that not only are we restricting to 2-particle
diagrams, but we are also selecting that subset of diagrams which have
no interaction vertices other than the initial and final points.
Because the internal vertices $\cvop{i}{j}$ describing propagation of a
single correlated quantity are symmetric in $i$ and $j$, and satisfy the
conditions
\[
0 \leq \cvop{i}{j} \leq 1
\]
and
\bge
\sum_{i} \cvop{i}{j} = \sum_{i} \cvop{j}{i} = 1,
\label{eq:condition2}
\ee
it is fairly straightforward to see that the matrix elements of $M$ must
satisfy the same constraints.  Intuitively, the above conditions on
$\cvop{i}{j}$ can be interpreted as arising from a description of the
propagation of a single correlated quantity as a discrete random walk,
where the direction of the walk at a given time is dependent on the
direction of the walk at the previous time step, according to the rule
that the (virtual) particle will change direction with probability $g =
fp$ and will continue moving in the same direction with probability $1 -
2g$.  The possible directions of the walk correspond to the lattice
vectors ${\bf e}_0, {\bf e}_\pm$.  With this interpretation, the matrix
element $\mixten{M}{\widehat{i}}{\widehat{j}}$ gives the probability
that a pair of virtual particles which begin in state $\widehat{j}$ will
collide for the first time in state $\widehat{i}$.  Because for $g > 0$
the probability is 1 that two particles beginning at the same lattice
site will eventually collide, it follows that condition
(\ref{eq:condition2}) must hold for the matrix $M$.  The remaining
conditions on $M$ of symmetry and positivity follow immediately from the
random walk interpretation.

It is now possible to give a complete description of the 2-particle
BBGKY approximation in terms of the matrix $M$.  The 2-particle BBGKY
correction to the transport coefficients of any lattice gas can be
described in terms of the correction to the $J$-matrix
\[
   \delta \jop{i}{j}
   =  \cvop{i}{lm}\cvop{np}{j}
	\mixten{\bar{M}}{lm}{np},
\]
where $\bar{M}$ is a matrix in the space $B_2$ defined by summing over
all 2-particle diagrams.  This matrix is defined in a similar fashion to
$M$; however, now we are including diagrams which have an arbitrary
number of 2-particle interaction vertices
$\cvop{\widehat{i}}{\widehat{j}}$.  By considering these 2-particle
interactions as another matrix ${\cal V}$ in the space $B_2$, we can
write a matrix equation for $\bar{M}$ in terms of $M$ and ${\cal V}$,
\[
\bar{M} = M + M {\cal V} \bar{M}.
\]
Formally, this equation has the solution
\bge
\bar{M} = M (1 - {\cal V} M)^{-1}.
\label{eq:bbgkym}
\ee
The matrix ${\cal V}$ restricted to the space $B_2$ is easily seen to
be another positive definite symmetric matrix which satisfies
Eq.~(\ref{eq:condition2}).  It follows that the matrix ${\cal V}M$ has the
same properties.  Unfortunately, this appears to lead to a difficulty;
namely, it is a consequence of Eq.~(\ref{eq:condition2}) that all these
matrices have eigenvalues of 1, with associated eigenvectors
$(1,1,1)$.  This means that the matrix $1 - {\cal V}M$ has a 0
eigenvalue, and thus has no inverse.  However, we have an additional
factor working in our favor.  That is, we are not interested in
computing the complete matrix $\bar{M}$.  Rather, we are interested in
computing the part of that matrix which contributes to the
renormalization of the eigenvalue $\supscrpt{\lambda}{2}$.  The
correction to this eigenvalue due to the 2-particle BBGKY
approximation is given by
\bge
\delta \lamsupp{2} = 9 p^2 f (1 - f) \left[
\mixten{\bar{M}}{\widehat{+}}{\widehat{+}}-
\mixten{\bar{M}}{\widehat{-}}{\widehat{+}} \right].
\label{eq:b2}
\ee
The quantity in  brackets can be rewritten in matrix form as
\[
q^2 \bar{M}  q_2 = q^2 \left(M + M {\cal V}M + M {\cal V}M {\cal V}M+
\cdots\right) q_2,
\]
where in the basis $(\widehat{+}, \widehat{0}, \widehat{-})$, $q^2 =
(-1,0,+1)$ is the usual left eigenvector of the $J$-matrix, and $q_2$ is
the corresponding right eigenvector.  By the same symmetry arguments
which we used in Subsection~\ref{ssec:eigrnrm} to prove that the vectors
$q^i$ are eigenvectors of the renormalized $J$-matrix, it follows that
$q^2$ must be an eigenvector of $M$ with some eigenvalue $l$.
Similarly, $q^2$ is an eigenvector of ${\cal V}$ (considered as a matrix
in the space $B_2$).  It is straightforward to verify {}from the vertex
rules that the eigenvalue of $q^2$ in this matrix is $1 - 3p + 3 pf$.
Thus, due to the fact that $q^2$ is orthogonal to the eigenvector
$(1,1,1)$ we avoid the divergence associated with the unit eigenvalue of
this vector.  In terms of the eigenvalue $l$ it is possible to rewrite
Eq.~(\ref{eq:b2}) in the form
\bge
\delta \lamsupp{2} = 9 \frac{p^2 f (1 - f) l}{1 - l (1 - 3 p + 3 pf)},
\label{eq:rewritten}
\ee
which is manifestly a finite renormalization.  We have thus shown that
for the 1D3P lattice gas, the 2-particle BBGKY correction is finite for
any $p,f$.  To check this result, we have numerically estimated the
matrix $M$ for certain values of $p,f$ and verified that
Eq.~(\ref{eq:rewritten}) gives a correction which agrees with the
numerical results obtained from the general BBGKY computer code
described above.  As an example, for $(p,f)= (.25,.5)$ we get a matrix
$M$ which is approximately
\[
M =\left( \begin{array}{lll}
0.305894 & 0.293698 & 0.400407 \\ 0.297612 & 0.404777 &
0.297612 \\ 0.400407 & 0.293698 & 0.305894
\end{array} \right).
\]
The eigenvalue of $q^2$ for this matrix is $l = -0.094513$.  The estimated
2-particle BBGKY correction to the eigenvalue
$\supscrpt{\lambda}{2}$ is thus
\[
\delta \lamsupp{2} = \frac{9l}{64-40 l} \sim -0.012550,
\]
in exact agreement with the results computed numerically and graphed
in Fig.~\ref{fig:examplebbgky}.

A particularly simple example of this formalism arises in the case $g
= fp = 1/ 3$.  In this case, the random walk which $M$ is described in
terms of is a true random walk, with the probability at each time step
of each of the 3 possible directions being exactly 1/3, independent of
the direction of the previous step.  It follows immediately from a
consideration of the matrix $M$ that the eigenvalue $l$ of $q^2$ is in
this case 0, which implies that the shift to the diffusivity arising
{}from the 2-particle BBGKY approximation is 0 whenever $g = 1/3$.
Note, however, that the higher $k$ BBGKY approximations do {\sl not}
generally vanish in this case.  For example, in
Fig.~\ref{fig:bbgky2} the 2-particle BBGKY approximation
vanishes at the point $(p,f) = (1/2,2/3)$; however, the higher $k$
approximations do not vanish and are closer to the experimental
results.

\subsection{Expansion Around $f = 1$}
\label{sec:approximationsexpand}

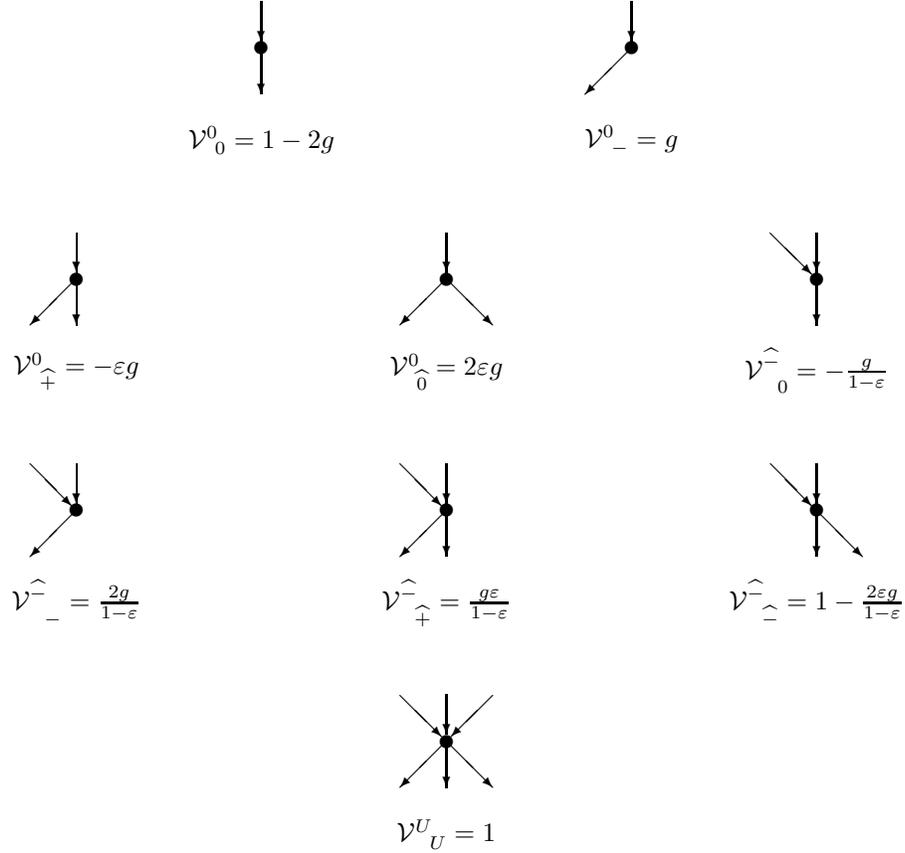
\begin{figure}
\centering
\begin{picture}(350,350)(0,0)
\put(70.,306.25){
\begin{picture}(70,70)(0,0)
\put(35.,52.5){\vector(0,-1){15.}}
\put(35.,35.){\vector(0,-1){17.5}}
\put(35.,35.){\circle*{5.}}
\put(35.,0.){\makebox(0,0){\footnotesize $\cvop{0}{0} = 1-2g$}}
\end{picture}}
\put(210.,306.25){
\begin{picture}(70,70)(0,0)
\put(35.,52.5){\vector(0,-1){15.}}
\put(35.,35.){\vector(-1,-1){17.5}}
\put(35.,35.){\circle*{5.}}
\put(35.,0.){\makebox(0,0){\footnotesize $\cvop{0}{-} = g$}}
\end{picture}}
\put(0.,218.75){
\begin{picture}(70,70)(0,0)
\put(35.,52.5){\vector(0,-1){15.}}
\put(35.,35.){\vector(-1,-1){17.5}}
\put(35.,35.){\vector(0,-1){17.5}}
\put(35.,35.){\circle*{5.}}
\put(35.,0.){\makebox(0,0){\footnotesize
   $\cvop{0}{\widehat{+}} = -\varepsilon g$}}
\end{picture}}
\put(140.,218.75){
\begin{picture}(70,70)(0,0)
\put(35.,52.5){\vector(0,-1){15.}}
\put(35.,35.){\vector(-1,-1){17.5}}
\put(35.,35.){\vector(1,-1){17.5}}
\put(35.,35.){\circle*{5.}}
\put(35.,0.){\makebox(0,0){\footnotesize
   $\cvop{0}{\widehat{0}} = 2\varepsilon g$}}
\end{picture}}
\put(280.,218.75){
\begin{picture}(70,70)(0,0)
\put(17.5,52.5){\vector(1,-1){15.7322}}
\put(35.,52.5){\vector(0,-1){15.}}
\put(35.,35.){\vector(0,-1){17.5}}
\put(35.,35.){\circle*{5.}}
\put(35.,0.){\makebox(0,0){\footnotesize
   $\cvop{\widehat{-}}{0} = -\frac{g}{1-\varepsilon}$}}
\end{picture}}
\put(0.,131.25){
\begin{picture}(70,70)(0,0)
\put(17.5,52.5){\vector(1,-1){15.7322}}
\put(35.,52.5){\vector(0,-1){15.}}
\put(35.,35.){\vector(-1,-1){17.5}}
\put(35.,35.){\circle*{5.}}
\put(35.,0.){\makebox(0,0){\footnotesize
   $\cvop{\widehat{-}}{-} = \frac{2g}{1-\varepsilon}$}}
\end{picture}}
\put(140.,131.25){
\begin{picture}(70,70)(0,0)
\put(17.5,52.5){\vector(1,-1){15.7322}}
\put(35.,52.5){\vector(0,-1){15.}}
\put(35.,35.){\vector(-1,-1){17.5}}
\put(35.,35.){\vector(0,-1){17.5}}
\put(35.,35.){\circle*{5.}}
\put(35.,0.){\makebox(0,0){\footnotesize
   $\cvop{\widehat{-}}{\widehat{+}} = \frac{g\varepsilon}{1-\varepsilon}$}}
\end{picture}}
\put(280.,131.25){
\begin{picture}(70,70)(0,0)
\put(17.5,52.5){\vector(1,-1){15.7322}}
\put(35.,52.5){\vector(0,-1){15.}}
\put(35.,35.){\vector(0,-1){17.5}}
\put(35.,35.){\vector(1,-1){17.5}}
\put(35.,35.){\circle*{5.}}
\put(35.,0.){\makebox(0,0){\footnotesize
   $\cvop{\widehat{-}}{\widehat{-}} = 1-\frac{2\varepsilon g}{1-\varepsilon}$}}
\end{picture}}
\put(140.,43.75){
\begin{picture}(70,70)(0,0)
\put(17.5,52.5){\vector(1,-1){15.7322}}
\put(35.,52.5){\vector(0,-1){15.}}
\put(52.5,52.5){\vector(-1,-1){15.7322}}
\put(35.,35.){\vector(-1,-1){17.5}}
\put(35.,35.){\vector(0,-1){17.5}}
\put(35.,35.){\vector(1,-1){17.5}}
\put(35.,35.){\circle*{5.}}
\put(35.,0.){\makebox(0,0){\footnotesize
   $\cvop{U}{U} = 1$}}
\end{picture}}
\end{picture}
\caption{Correlation Vertex Coefficients in Terms of $g, \varepsilon$}
\label{fig:epsiloncvc}
\end{figure}

In this subsection we consider the renormalization of
$\supscrpt{\lambda}{2}$ when the particle density $f$ approaches 1.  In
this regime, it is possible to expand the eigenvalue
$\supscrpt{\tilde{\lambda}}{2}$, and thus the diffusivity, in the
quantity $\varepsilon\equiv 1-f$.  \label{pg:zw} By rewriting the
correlation vertex coefficients in terms of $\varepsilon$, we can
ascertain which diagrams contribute to each order in $\varepsilon$.  The
CVC's are shown in terms of $\varepsilon$ and $g = fp$ in
Fig.~\ref{fig:epsiloncvc}.  Note that both vertices which take a single
incoming virtual particle to two virtual particles (1-2 vertices) are
proportional to $\varepsilon$.  Since these vertices are the only
nonzero vertices which increase the number of virtual particles in a
diagram, it follows that the set of diagrams which contribute to order
$\varepsilon^n$ must be a subset of the diagrams in the $(n +
1)$-particle BBGKY approximation.  In fact, aside from the interactions
described by the vertices with 2 and 3 virtual particles in both the
incoming and outgoing states, the ordering of diagrams in $\varepsilon$
is equivalent to the loop ordering of diagrams which is commonly used in
continuum kinetic theory and quantum field theory.  Because the sets of
diagrams which contribute to the corrections for low orders in
$\varepsilon$ are fairly simple (but infinite), we can numerically
evaluate these partial diagrammatic sums to get a prediction for the
low-order derivatives of $D$ around $f = 1$.  Because the factor $g$
appears in the 1-1 vertices $\mixten{{\cal V}}{i}{j}$, it is convenient
to fix this quantity while evaluating the derivatives of $D$ with
respect to $\varepsilon$.  The numerical calculation of the low-order
coefficients as a sum over diagrams converges quite rapidly.  In fact,
it can be shown that for each $n$, the coefficient of $\varepsilon^n$
gives a convergent sum; we outline a proof of this fact at the end of
this subsection.

As the simplest example of this type of calculation, consider the set of
diagrams which contribute to order $\varepsilon$ in
$\supscrpt{\tilde{\lambda}}{2}$.  Because the correction factor in
Eq.~(\ref{eq:1D3Pcorrection}) is itself proportional to $\varepsilon$, the
diagrams which contribute linearly in $\varepsilon$ cannot contain any
internal 1-2 vertices; the contribution to the eigenvalue from such
diagrams is equivalent to the 2-particle BBGKY contribution using the
limits of the 2-2 vertices as $\varepsilon \rightarrow 0$.  Using the
results of the previous subsection, this contribution is finite, and can
be computed just as the 2-particle BBGKY correction is calculated above.
As an example, consider again the case where $g = 1/3$.  In this case,
we expect the derivative of $\tilde{\lambda}^2$ at $f = 1$ to be 0; this
result seems to be in agreement with experiment.

\begin{figure}
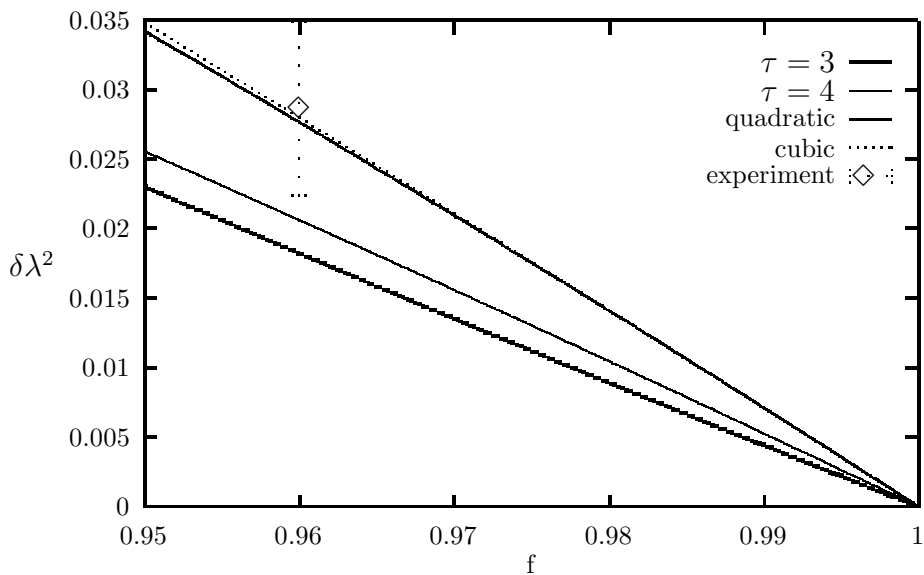

\centering
\setlength{\unitlength}{0.240900pt}
\ifx\plotpoint\undefined\newsavebox{\plotpoint}\fi
\sbox{\plotpoint}{\rule[-0.500pt]{1.000pt}{1.000pt}}%

\caption{Quadratic and Cubic Approximations to $\tilde{\lambda}^2$
Around $f = 1$ for $g = fp = 0.48$}
\label{fig:cubic}
\end{figure}

We have numerically calculated the first $3$ derivatives of
$\tilde{\lambda}^2$ at the point $f = 1$ for several values of the
parameter $g$.  The results are completely in agreement with
experimental results when $1 - f\ll 1$.  As an example, we have graphed
the quadratic and cubic approximations to $\delta \lambda^2
=\tilde{\lambda}^2-\lambda^2$ for $g = 0.48$ in Fig.~\ref{fig:cubic},
and compared to experimental data at the point $ f = 0.96$.  For
comparison, the curves describing the $\tau = 3$ and $\tau = 4$
short-$\tau$ approximations are also graphed in this region.  Note that
the region of the graph with $f < 0.96$ corresponds to $p > 0.5$ and is
unphysical.

\begin{figure}
\centering
\begin{picture}(200,100)(- 100,- 50)
\thicklines
\put(- 90, 0){\vector(1,0){8}}
\put(- 90, 0){\line(1,0){10}}
\put(- 80, 0){\circle*{4}}
\put(- 80, 0){\vector(1,2){6}}
\put(- 80, 0){\line(1,2){7}}
\put(- 80, 0){\vector(1,-2){6}}
\put(- 80, 0){\line(1,-2){7}}
\put(- 73, 14){\vector(1, -2){11}}
\put(- 73, -14){\vector(1, 2){11}}
\put(-  52, 0){\makebox(0,0){$\cdots$}}
\put(- 42, - 8){\vector(1,2){10}}
\put(- 42, - 8){\line(1,2){11}}
\put(- 42, 8){\vector(1, -2){10}}
\put(- 42, 8){\line(1,-2){11}}
\put(- 31, 14){\circle*{4}}
\put(- 31, -14){\circle*{4}}
\put(- 31, 14){\line(1, -2){14}}
\put(- 17, - 14){\line(1,0){1}}
\put(- 16, - 14){\vector(1,2){9}}
\put(- 31, 14){\line(1,0){10}}
\put(- 21, 14){\vector(1, -2){14}}
\put(- 31, - 14){\line(1,0){5}}
\put(- 26, - 14){\line(1,2){14}}
\put(- 12, 14){\line(1,0){1}}
\put(- 11, 14){\vector(1, -2){4}}
\put(0,0){\makebox(0,0){$\cdots$}}
\put(80, 0){\vector(1,0){10}}
\put(80, 0){\circle*{4}}
\put( 74, - 12){\vector(1, 2){6}}
\put(80, 0){\line(-1, -2){7}}
\put( 74,  12){\vector(1,-2){6}}
\put(80, 0){\line(-1,2){7}}
\put( 62, 8){\vector(1,-2){11}}
\put(62, - 8){\vector(1, 2){11}}
\put( 52, 0){\makebox(0,0){$\cdots$}}
\put(31,  -14){\vector(1, 2){11}}
\put(31,  14){\vector(1, -2){11}}
\put(31, -14){\circle*{4}}
\put(31, 14){\circle*{4}}
\put(17, 14){\vector(1,-2){14}}
\put(17, 14){\line(-1, 0){1}}
\put(16, 14){\line(-1, -2){9}}
\put(21, -14){\vector(1, 0){10}}
\put(21, -14){\line(-1,2){14}}
\put(26, 14){\vector(1, 0){5}}
\put(26, 14){\line(-1, -2){14}}
\put(12, -14){\line(-1, 0){1}}
\put(11, -14){\line(-1,2){4}}
\thinlines
\put(-  75, - 30){\line( 0,1){60}}
\put(- 36, - 30){\line( 0,1){60}}
\put(- 26, - 30){\line( 0,1){60}}
\put(-  70, - 30){\makebox(0,0){$\widehat{j}$}}
\put( - 41, - 30){\makebox(0,0){$w$}}
\put( - 21, - 30){\makebox(0,0){$v$}}
\put(  75, - 30){\line( 0,1){60}}
\put( 36, - 30){\line( 0,1){60}}
\put( 26, - 30){\line( 0,1){60}}
\put(  70, - 30){\makebox(0,0){$\widehat{i}$}}
\put( 41, - 30){\makebox(0,0){$x$}}
\put( 21, - 30){\makebox(0,0){$y$}}
\put(- 55, - 50){\makebox(0,0){\tiny $ (1 - M {\cal V})^{-1}M$}}
\put(0, - 50){\makebox(0,0){\tiny $M (1 - {\cal V}M)^{-1}$}}
\put( 55, - 50){\makebox(0,0){\tiny $M (1 -  {\cal V}M)^{-1}$}}
\put(- 31, 40){\makebox(0,0){\scriptsize ${\cal V}$}}
\put(31, 40){\makebox(0,0){\scriptsize ${\cal V}$}}
\end{picture}
\caption{Diagrammatic Sketch}
\label{fig:sketch}
\end{figure}
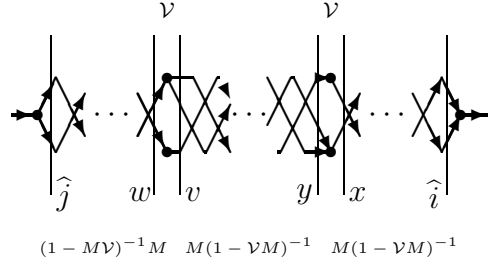

We conclude this subsection with an outline of the proof that the sum
over diagrams contributing to the $\varepsilon^n$ term in
$\tilde{\lambda}^2$ is convergent for any fixed value of $n$.  This
proof is similar in nature to the proof of convergence for the
2-particle BBGKY approximation in the previous subsection, but is
slightly more subtle.  As mentioned above, to prove the desired result
it will suffice to show that the contribution from all diagrams with
$n$ vertices of the 1-2 type gives a convergent sum for all values of
$n$.  As the simplest example beyond the 2-particle BBGKY
approximation, we consider the contribution to the order $\varepsilon^2$
term arising from diagrams with a single internal 1-2 vertex.  Using
an analogous notation to that used in the previous subsection for the
analysis of the 2-particle BBGKY approximation, the contribution from
the diagrams with a single internal 1-2 vertex can be written as
\[
\delta \lamsupp{2} = 9 p^2 f (1 - f) \left(
\mixten{\bar{M}}{\widehat{+}}{\widehat{+}}-
\mixten{\bar{M}}{\widehat{-}}{\widehat{+}}
\right),
\]
where $\bar{M}$ is given by
\begin{eqnarray}
\mixten{\bar{M}}{\widehat{i}}{\widehat{j}}  & = &
\mixten{[(1 - M{\cal V} )^{-1}]}{\widehat{i}}{\widehat{k}}
\mixten{M}{\widehat{k}}{x}
\mixten{{\cal V}}{x}{y}
\mixten{M}{y}{z}\nonumber \\ & & \; \; \; \; \times
\mixten{[(1 - {\cal V} M)^{-1}]}{z}{v}
\mixten{{\cal V}}{v}{w}
\mixten{M}{w}{\widehat{m}}
\mixten{[(1 -  {\cal V}M)^{-1}]}{\widehat{m}}{\widehat{j}}.
\label{eq:complicated}
\end{eqnarray}
In this equation, the variables $x$ and $w$ are summed over all distinct
elements of the set of pairs of bits at distinct lattice points, modded
out by equivalence under translation.  The elements of this set
$B_{1,1}$ are in 1-1 correspondence with triplets $(n,i,j)$, with $n >
0$ giving the distance between the lattice points of the two bits, and
$i,j$ denoting the elements of the set $B$ corresponding to the two
bits.  The variables $y, z,$ and $v$ are similarly summed over all
elements of the set $B_{1,2}$ of triplets $(n,i, \widehat{j})$
representing a single bit $i$ at a distance $n$ from a pair of bits
$\widehat{j}\in B_2$.  This set contains elements with $n$ both positive
and negative, since the lattice point with a single VP can be on either
side of the lattice point with two VP's; there is a single element of
the set $B_{1,2}$ with $n = 0$ corresponding to the situation where all
3 particles are at the same lattice site.  With this constraint on the
summation variables, the matrices $M$ and ${\cal V}$ are defined in an
analogous fashion to the 2-particle BBGKY case.  For example, the matrix
$\mixten{M}{y}{z}$ is the matrix on the space $B_{1,2}$ whose entries
correspond to the probability that a set of 3 particles beginning in
state $z$ will first have a collision when they are in the state $y$.
Similarly, the matrix $\mixten{M}{\widehat{k}}{x}$ gives the sum over
all 2-particle diagrams with no (internal) collisions which begin in the
state described by $x$ and end with the collision described by the state
$\widehat{k}$.  The matrix $\mixten{{\cal V}}{x}{y}$ always describes
the product of correlation vertex coefficients involved in a collision
from a state $y$ to a state $x$; when both $y$ and $x$ are in $B_{1,2}$,
we insist that the two particles remain at the same lattice site.  Note
that if a collision involves virtual particles at more than one lattice
point, the distances between the vertices must be commensurate in both
states for this matrix element to be nonzero.  In
Eq.~(\ref{eq:complicated}), the matrix inverses of the form $(1 -
x)^{-1}$ should be taken to be shorthand for the formal expansion
$\sum_{i}x^i$, as these matrices generally have a unit eigenvalue.  Just
as in the BBGKY case, it is straightforward to verify that all matrices
considered here which are square (have both indices taking values in the
same space) are symmetric, nonnegative, and satisfy condition
(\ref{eq:condition2}).  Eq.~(\ref{eq:complicated}) can be sketched
diagrammatically as in Fig.~\ref{fig:sketch}.

By noting that $\bar{M}$  and its expression (\ref{eq:complicated})
are symmetric,  an identical argument to that in the previous subsection
tells us that the contribution to the eigenvalue
$\supscrpt{\lambda}{2}$ can be described by the matrix formula
\begin{eqnarray}
\delta \lamsupp{2}& = &
9 \frac{p^2 f (1 - f) l^2}{[1 - l (1 - 3 p + 3 pf)]^2} \nonumber \\
& & \; \; \;\times q^2_{k} \left(
\mixten{M}{\widehat{k}}{x}
\mixten{{\cal V}}{x}{y}
\mixten{M}{y}{z}
\mixten{[(1 - {\cal V} M)^{-1}]}{z}{v}
\mixten{{\cal V}}{v}{w}
\mixten{M}{w}{\widehat{m}}
\right)  q_2^{m}.
\label{eq:simpler}
\end{eqnarray}
We have thus reduced the problem of proving that
Eq.~(\ref{eq:complicated}) is convergent to the problem of proving
convergence for Eq.~(\ref{eq:simpler}).  There are several key arguments
necessary to proving the convergence of this remaining sum.  The first
step is to prove that any matrix element of the form
\[
\mixten{\widehat{M}}{x}{w} =
\mixten{{\cal V}}{x}{y}
\mixten{M}{y}{z}
\mixten{[(1 - {\cal V} M)^{-1}]}{z}{v}
\mixten{{\cal V}}{v}{w}
\]
is convergent, and has an absolute value bounded above by some number
$\Xi$.  The second step is to argue that the infinite sum
\bge
\sum_{w \in B_{1,1}} \mixten{M}{w}{\widehat{ m}}
\label{eq:expectedwalktime}
\ee
converges for all $\widehat{m}$ and thus can be bounded above by another
number $\Delta$.  Once these two facts are shown, it follows
immediately that total eigenvalue shift is bounded above by
\[
|\delta \lamsupp{2} | < 9 \frac{p^2 f (1 - f) l^2}{[1 - l (1 - 3 p + 3
pf)]^2} \Xi \Delta^2.
\]

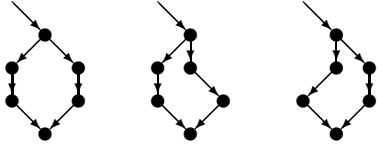
\begin{figure}
\centering
\begin{picture}(165,50)(0,0)
\put(0.,0.){
\begin{picture}(50,50)(0,0)
\put(12.5,50.){\vector(1,-1){10.7322}}
\put(25.,37.5){\circle*{5.}}
\put(25.,37.5){\vector(-1,-1){10.7322}}
\put(25.,37.5){\vector(1,-1){10.7322}}
\put(12.5,25.){\circle*{5.}}
\put(37.5,25.){\circle*{5.}}
\put(12.5,25.){\vector(0,-1){10.}}
\put(37.5,25.){\vector(0,-1){10.}}
\put(12.5,12.5){\circle*{5.}}
\put(37.5,12.5){\circle*{5.}}
\put(12.5,12.5){\vector(1,-1){10.7322}}
\put(37.5,12.5){\vector(-1,-1){10.7322}}
\put(25.,0.){\circle*{5.}}
\end{picture}}
\put(54.945,0.){
\begin{picture}(50,50)(0,0)
\put(12.5,50.){\vector(1,-1){10.7322}}
\put(25.,37.5){\circle*{5.}}
\put(25.,37.5){\vector(-1,-1){10.7322}}
\put(25.,37.5){\vector(0,-1){10.}}
\put(12.5,25.){\circle*{5.}}
\put(25.,25.){\circle*{5.}}
\put(12.5,25.){\vector(0,-1){10.}}
\put(25.,25.){\vector(1,-1){10.7322}}
\put(12.5,12.5){\circle*{5.}}
\put(37.5,12.5){\circle*{5.}}
\put(12.5,12.5){\vector(1,-1){10.7322}}
\put(37.5,12.5){\vector(-1,-1){10.7322}}
\put(25.,0.){\circle*{5.}}
\end{picture}}
\put(110.055,0.){
\begin{picture}(50,50)(0,0)
\put(12.5,50.){\vector(1,-1){10.7322}}
\put(25.,37.5){\circle*{5.}}
\put(25.,37.5){\vector(1,-1){10.7322}}
\put(25.,37.5){\vector(0,-1){10.}}
\put(37.5,25.){\circle*{5.}}
\put(25.,25.){\circle*{5.}}
\put(37.5,25.){\vector(0,-1){10.}}
\put(25.,25.){\vector(-1,-1){10.7322}}
\put(12.5,12.5){\circle*{5.}}
\put(37.5,12.5){\circle*{5.}}
\put(12.5,12.5){\vector(1,-1){10.7322}}
\put(37.5,12.5){\vector(-1,-1){10.7322}}
\put(25.,0.){\circle*{5.}}
\end{picture}}
\end{picture}
\caption{Cancelling Diagrams}
\label{fig:examplecancellation}
\end{figure}

We will now proceed to prove these two necessary convergence results.
We first show that there is an upper bound $\Xi$ on the matrix
elements $\mixten{\widehat{M}}{x}{w}$.  For a fixed value of $x$, the
matrix elements $\mixten{{\cal V}}{x}{y}$ form a vector in $B_{1,2}$
which we will refer to as $t_{y}$.  An examination of the collision
rules (\ref{eq:1to2}) tells us that $t$ has only 18 nonzero
components; 8 of magnitude $- g^2 \varepsilon$, 4 of magnitude $2g^2
\varepsilon $, 4 of magnitude $- g(1-2g) \varepsilon$ and 2 of magnitude
$2g(1-2g) \varepsilon$.  We will now consider the result $s$ of
multiplying this vector by the matrix $M$
\[
s_{z} = t_{y} \mixten{M}{y}{z}.
\]
Because the sum of the components of $t$ is 0, the same must be true
of the vector $s$.  Furthermore, if we define a norm on $t$ by
\[
| t | = \sum_{y} | t_{y} |,
\]
then  we can proceed to show that
\bge
| s | \leq | t | (1 - g^6 (1-2g)^2).
\label{eq:relativebound}
\ee
This result follows because there is always at least one set of
diagrams of order $g^6 (1-2g)^2$ which cancel between the positive and
negative elements of $t$ (in practice, much cancellation occurs;
however, we are interested here only in the convergence.)  An example
of three diagrams giving such a cancellation is given in
Fig.~\ref{fig:examplecancellation}.  The vertex factors from these
diagrams are identical; however, the diagrams connect to components of
$t_y$ with opposite sign.  We do not depict the motion of the extra
particle, which can be assumed to be constant for the three diagrams.
{}From Eq.~(\ref{eq:relativebound}),
it follows that every matrix element of $\widehat{M}$ is bounded above
by
\[
|\mixten{\widehat{M}}{x}{w}| \leq \Xi =
\frac{24 pg \varepsilon}{g^6 (1-2g)^2}.
\]
Thus, we have an upper bound of
the desired form for the matrix elements of $\widehat{M}$.

Finally, it remains to demonstrate that the infinite sum
(\ref{eq:expectedwalktime}) is bounded above for all $\widehat{m}$.  This
sum, however, is simply equal to the expected value of the number of
time steps necessary for a pair of random walks beginning in state
$\widehat{ m}$ to collide.  To see this, observe that every diagram with
weight $W (T)$ containing two noncolliding random walks which
contributes to $\mixten{M}{\widehat{i}}{\widehat{ m}}$ will contribute to the
sum (\ref{eq:expectedwalktime}) a total of $\tau W (T)$ where $\tau$
is the length of the diagram; the factor $\tau$ appears because the
diagram can be chopped in half at any point $x$ and will contribute a
factor of $W (T)$ for each such division.  Given this interpretation,
however, it is clear that the sum is convergent due to the standard
result that in one dimension a random walk will return to any point on
the lattice in a finite expected time (this result is usually stated
for random walks without memory; however, a generalization to random
walks with memory and nonzero bounce probability is straightforward.)
In conclusion, we have proven both convergence bounds which were
needed to demonstrate conclusively the convergence of the sum over
diagrams which contain a single 1-2 vertex.  This implies the
convergence of the order $\varepsilon^2$ term in the diffusivity around
$f = 1$.  It is fairly straightforward to generalize these arguments
to the coefficient of an arbitrary order $\varepsilon^n$ by showing that
the sum over diagrams with any fixed number of 1-2 vertices is
convergent; however, the details become correspondingly more complex
and are left as an exercise to the reader.  Note also, that we do not
have any reason other than empirical results to believe that the
expansion of $\tilde{\lambda}^2$ in $\varepsilon$ has a nonzero radius of
convergence, even though the coefficients themselves are proven to be
finite.

\subsection{Ring Approximation}
\label{sec:approximationsring}

We now consider the ring approximation for the 1D3P lattice gas.  In
this subsection we study the ring approximation from two perspectives.
First, we show that the ring approximation can be formulated in terms of
the $M$ matrix in the same form as the 2-particle BBGKY approximation,
where the collision matrix ${\cal V}$ is replaced by a new effective
collision matrix.  This formalism gives an analytic relationship between
the corrections to $\lambda^2$ from the ring approximation and the
2-particle BBGKY approximation.  Second, we perform an explicit analysis
of the ring approximation in Fourier space, and describe the complete
contribution to the eigenvalue renormalization from this approximation.

\begin{figure}
\centering
\begin{picture}(100,100)(0,0)
\put(50.,100.){\circle*{5.}}
\put(50.,100.){\vector(-1,-1){18.2322}}
\put(50.,100.){\vector(0,-1){17.5}}
\put(30.,80.){\circle*{5.}}
\put(30.,80.){\vector(1,-1){18.2322}}
\put(50.,80.){\circle*{5.}}
\put(50.,80.){\vector(0,-1){17.5}}
\put(50.,60.){\circle*{5.}}
\put(50.5,60.5){\vector(1,-1){18.2322}}
\put(49.5,59.5){\vector(1,-1){18.2322}}
\put(70.,40.){\circle*{5.}}
\put(70.,40.){\vector(-1,-1){18.2322}}
\put(70.,40.){\vector(0,-1){17.5}}
\put(50.,20.){\circle*{5.}}
\put(50.,20.){\vector(1,-1){18.2322}}
\put(70.,20.){\circle*{5.}}
\put(70.,20.){\vector(0,-1){17.5}}
\put(70.,0.){\circle*{5.}}
\end{picture}
\caption{Unphysical diagram
present in the ring approximation}
\label{fig:movingtogether}
\end{figure}
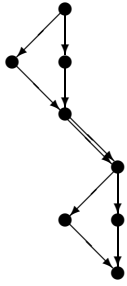

The ring approximation is taken by summing over all independent paths
for two separate virtual particles to propagate from one point to
another, using for each virtual particle the 1-1 CVC's as weights on the
vertices of the independent paths.  The set of diagrams associated with
this approximation is closely related to the 2-particle BBGKY set of
diagrams; however, there are two important differences.  The first
essential difference is that because the two VP's are moving
independently, there is no constraint dictating that the two VP's cannot
move along the same lattice vector at some time step $\tau$.  Thus,
diagrams such as Fig.~\ref{fig:movingtogether} must be included in
this approximation.  The second essential difference is that even when
the two VP's enter a vertex from different directions described by the
state $\widehat{i}$ and leave in different directions $\widehat{j}$, the
amplitude for such a transition is no longer given by ${\cal V}$ but
rather by the collision matrix $U$ with elements
\[
\mixten{U}{\widehat{j}}{\widehat{i}} =\mixten{\delta}{j}{i} [(1 - 2g) (1 - 3g)]
 + g (1 - g).
\]
Despite these differences, it is still possible to formulate an
expression for the ring approximation which is identical in form to
Eq.~(\ref{eq:b2}), where the matrix $\bar{M}$ is no longer defined by
Eq.~(\ref{eq:bbgkym}) but rather by
\[
\bar{M} = M (1 - {\cal U} M)^{-1},
\]
where the matrix ${\cal U}$ has elements $\mixten{{\cal
U}}{\widehat{j}}{\widehat{i}}$ giving the total amplitude for all processes
where a pair of particles $\widehat{i}$ come together at some vertex,
perhaps travel together for several steps, and then separate in
directions $\widehat{j}$ at the same or a later vertex.  Algebraically,
${\cal U}$ is given by
\[
{\cal U} = U + S \cdot T,
\]
where the matrix $\mixten{T}{k}{\widehat{i}}$ gives the amplitude for an
incoming pair of particles $\widehat{i}$ to both move in direction $k$
according to the separate ${\cal V}$ 1-1 vertices, and the matrix
$\mixten{S}{\widehat{j}}{k}$ gives the total amplitude for a pair of
particles both moving in direction $k$ to eventually separate in
directions $\widehat{j}$.  The matrix $T$ can easily be computed from
${\cal V}$, and  has matrix elements
\[
\mixten{T}{k}{\widehat{i}} =\mixten{\delta}{k}{i}[g (3g - 1)]
+ g (1 - 2g).
\]
The matrix $S$ is slightly more complicated.  By symmetry, the
elements of this matrix are given by
\[
\mixten{S}{\widehat{j}}{k} =\mixten{\delta}{j}{k} [s - d] + d,
\]
for some  values of  the functions $s (g),d (g)$.  From the
definitions of these functions, one finds that they must satisfy the
recursion relations
\begin{eqnarray}
s  & = & 2g^2 + (1 - 2g)^2 s + 2 g^2 d ,  \\
d  & = & 2g (1 - 2g) + (1 - 2g)^2 d + g^2 s + g^2 d.
\end{eqnarray}
These equations have the solution
\begin{eqnarray}
s  & = & \frac{- 3g}{3g - 4}, \\
d  & = & \frac{3g - 2}{3g - 4}.
\end{eqnarray}
Plugging these values for $s$ and $d$ into  $S$, and computing the
eigenvalue $w$ of $q^2$ with respect to the resulting matrix ${\cal
U}$, we find
\[
\mixten{{\cal U}}{\widehat{j}}{\widehat{i}} =\mixten{\delta}{j}{i} [1 - 9g
\left(\frac{3g - 2}{3g - 4}\right)] + 3g \left(\frac{3g - 2}{3g - 4}\right),
\]
and thus
\bge
w = 1 - 9g \frac{3g - 2}{3g - 4}.
\label{eq:wequation}
\ee

As a result of this analysis, we have an expression similar to
Eq.~(\ref{eq:rewritten}) for the eigenvalue shift,
\bge
\delta \lamsupp{2} = 9 \frac{p^2 f (1 - f) l}{1 - l w},
\label{eq:rewrittenr}
\ee
where $w$ is given by Eq.~(\ref{eq:wequation}).
We thus have not only managed to prove that the ring approximation for
the 1D3P lattice gas is convergent, but we  can also derive from
Eqs.~(\ref{eq:rewritten}) and (\ref{eq:rewrittenr}) an explicit algebraic
relationship between the eigenvalue renormalization from the ring
approximation and that from the 2-particle BBGKY approximation.  If we
write the shift in $\lambda^2$ due to the ring approximation as
$\delta$, and the shift from the 2-particle BBGKY approximation as
$\delta'$, we have the equation
\bgea
\lefteqn{\frac{\delta (4 - 3p f)}{\delta (4 - 21 pf  + 27
p^2 f^2)+ 9 p^2 f (1 - f)(4 - 3pf)}}  \nonumber\\
& = &\frac{\delta'}{\delta' (1 - 3p + 3p f) +9p^2 f (1 - f)}.
\label{eq:ecwnc}
\eea
As a specific example, we can plug in the explicit value of
$\delta' = -0.012550$ for   $(p,f)= (.25,.5)$, to find that
\[
\delta = -0.012696.
\]
This result is in exact correspondence with numerical calculations.

To derive the correction due to the ring approximation from first
principles, note that the weight of any ring diagram is the product of
contributions from each of the two virtual particles.  Denote by
$\mixten{y}{i}{k}\bfxt$ the factor contributed by a virtual particle
beginning at the origin at $\tau=0$ in direction $k$, and terminating at
$\bfx$ at $\tau=t$ in direction $i$.  In terms of these factors, the
ring approximation can be written
\bge
   \mixten{Y}{\{i,j\}}{\{k,l\}} = \sum_{\bfx}\sum_{t=2}^\infty
   (\mixten{y}{i}{k}\bfxt \mixten{y}{j}{l}\bfxt +
    \mixten{y}{i}{l}\bfxt \mixten{y}{j}{k}\bfxt).
\label{eq:ringeo}
\ee
Furthermore, because the ring approximation considers only one-point to
one-point vertices, the $\mixten{y}{i}{k}\bfxt$ factors obey the linear,
homogeneous dynamical equation,
\[
\mixten{y}{i}{k}(\bfxci,\tpdt) = \cvop{i}{j}\mixten{y}{j}{k}\bfxt.
\]
Because this equation is linear and homogeneous, it can be solved
exactly in Fourier space.  The result can be expressed as a Fourier
series in $\bfx$ and $t$.  When these results for
$\mixten{y}{i}{a}\bfxt$ and $\mixten{y}{j}{b}\bfxt$, with Fourier
summation variables $\bfk$ and $\bfk'$ respectively, are inserted into
Eq.~(\ref{eq:ringeo}), the summation over time is nothing more than a
geometric series, and one of the Fourier summations yields a Kronecker
delta in $\bfk$ and $\bfk'$.  The result for
$\mixten{Y}{\{i,j\}}{\{a,b\}}$ is then reduced to one summation over
$\bfk$.  Passing to the hydrodynamic limit, this becomes a quadrature.
For a $D$-dimensional lattice gas, one obtains at most a $D$-dimensional
quadrature.

Note that this program can be carried out to compute the ring
approximation for any lattice gas.  Since the relationship
(\ref{eq:ecwnc}) between the ring approximation and the $k=2$ BBGKY
truncation is also generalizable to any lattice gas, it follows that the
latter approximation is also reduced to quadrature.

\subsection{Comparison of Approximations}

We conclude this section with a brief discussion of the relative
effectiveness of the various methods used to compute partial
diagrammatic sums.  We have calculated explicitly in this section the
corrections to the diffusivity in the 1D3P lattice gas which arise from
various methods of truncating the complete diagrammatic summation.  The
short-$\tau$ approximations seem to converge, however require
exponential time to compute, so that achieving a high degree of accuracy
with this method is difficult.  The $k$-particle BBGKY approximations
can be calculated in polynomial time, and seem to converge rapidly for
each value of $k$.  It seems likely that for most other lattice gases of
interest, these approximations will behave similarly, and thus that in
general the BBGKY approximations will be the more efficient of these two
methods to achieve a maximal degree of accuracy with a minimum of
computation.

We have also discussed the ring approximation, and shown that the result
of this approximation is closely related to the 2-particle BBGKY
approximation through an algebraic relation.  Since the ring
approximation can be performed analytically for most lattice gases, at
least in an asymptotic sense, this approximation is generally useful for
indicating the convergence properties of the diagrammatic summation for
any given lattice gas.  In general, for lattice gases in 2 dimensions
which preserve momentum as well as particle number, the ring
approximation diverges logarithmically in the lattice size~\cite{Kad}.
This divergence can be reproduced from the diagrammatic formalism
directly; this issue will be discussed in more detail in a future
paper~\cite{bt2}.

For the 1D3P lattice gas we have studied in this section, we found that
in the vicinity of the parameter value $f = 1$, it is possible to expand
the sum over diagrams in a power series in $1 - f$, giving a
perturbation series analogous to the loop expansion in field theory or
continuum kinetic theory.  We showed that each term in this expansion
corresponds to a convergent sum.  In other lattice gases, similar
expansions may be useful in calculating the effect of renormalization in
the vicinity of certain parameter values.  When expansions of this type
are possible, they are generally more accurate than any of the other
approximation methods.

Finally, in Fig.~\ref{f:comparison}, we compare numerically the
results of the different approximation methods described above to the
Boltzmann and experimental calculations of the transport coefficients.

\begin{figure}
\begin{center}
\makebox[2.5in]{
\begin{tabular}{ | l | r |}
\hline
\hline
Approximation
 &
$\delta \lambda$ \\
\hline
\hline
{\bf Experiment} & {\bf -0.081 $ \pm $ .0045} \\
\hline
$k = 5$ BBGKY & -0.0094\\
\hline
$k = 4$ BBGKY & -0.0095\\
\hline
$k = 3$ BBGKY & -0.0101\\
\hline
$k = 2$ BBGKY & -0.0125\\
\hline
 $\tau =  5$ & -0.0079\\
\hline
 $\tau = 4$ & -0.0046\\
\hline
 $\tau = 3$ & -0.0110\\
\hline
Ring & -0.0127\\
\hline
Boltzmann & 0\\
\hline
\end{tabular}
}
\hspace*{1in}
\makebox[2.5in]{
\begin{tabular}{ | l | r |}
\hline
\hline
Approximation
 &
$\delta \lambda$ \\
\hline
\hline
{\bf Experiment} & {\bf -0.0111 $ \pm $ .0022} \\
\hline
$k = 5$ BBGKY & -0.0100\\
\hline
$k = 4$ BBGKY &  -0.0106\\
\hline
$k = 3$ BBGKY & -0.0134\\
\hline
$k = 2$ BBGKY & -0.0207\\
\hline
 $\tau =  5$ & -0.0086\\
\hline
 $\tau = 4$ & -0.0006\\
\hline
 $\tau = 3$ & -0.0164\\
\hline
Ring & -0.0216\\
\hline
Boltzmann & 0\\
\hline
\end{tabular}
} \\
\vspace*{.1in}
\makebox[2.5in]{$(p,f) = (0.25, 0.5) $}
\hspace*{1in}
\makebox[2.5in]{$(p,f) = (0.33333,0.33333) $} \\
\vspace*{.5in}
\begin{tabular}{ | l | r |}
\hline
\hline
Approximation
 &
$\delta \lambda$ \\
\hline
\hline
{\bf Experiment} & {\bf 0.0384 $ \pm $ .0099} \\
\hline
$k = 5$ BBGKY & 0.0414\\
\hline
$k = 4$ BBGKY & 0.0417\\
\hline
$k = 3$ BBGKY & 0.0402\\
\hline
$k = 2$ BBGKY & 0.0441\\
\hline
 $\tau =  5$ & 0.0380\\
\hline
 $\tau = 4$ & 0.0305\\
\hline
 $\tau = 3$ & 0.0319\\
\hline
Ring & 0.0373\\
\hline
Boltzmann & 0\\
\hline
\end{tabular} \\
\vspace*{.1in}
$(p,f)= (0.5,0.9)$ \\
\vspace*{.2in}
\end{center}
\caption[x]{\footnotesize  Comparison of approximation methods for 1D3P}
\label{f:comparison}
\end{figure}

\newpage
\section{Conclusions}
In this paper, we have presented a complete kinetic theory of lattice
gases, applied it to four model lattice gases, and compared the
predictions of the theory to experiment for one of these models.

The approach presented in this paper opens up a wide range of possible
new work on discrete kinetic theory.  By applying these techniques to
compute deviations from the Boltzmann predictions for commonly used
lattice gases, the results of simulations can be more accurately
interpreted.  Lattice gases are currently being used, both in industrial
and academic settings, for computational fluid dynamics
calculations~\cite{kim}; to ensure the accuracy of these calculations,
it is essential to account for the renormalization effects that we have
studied here.

In addition to quantitative refinement of lattice gas calculations, the
theory presented here provides a tool with which to investigate
fundamental physical phenomena in nonequilibrium statistical systems.
In recent years, for example, lattice gases have been used to model many
different hydrodynamic systems, including reaction-diffusion equations,
and other systems capable of spontaneous self-organization.  It is
known~\cite{lw} that the Boltzmann approximation does not yield accurate
results for the transport coefficients of such systems, unless the
reactants are allowed to diffuse for several steps between reactions in
order to artificially supress the correlations that develop~\cite{dab}.
Thus, it is likely that the methods described in this paper will be
directly applicable to these systems, providing an important correction
to their theory.  More interestingly, these methods will also provide
insight into the extremely subtle flow and agglomeration of
interparticle correlations in pattern-forming lattice gases, and hence
into the dynamical basis of self-organization.

\section*{Acknowledgements}
One of us (BMB) would like to acknowldege helpful conversations with and
encouragement from Professors M.H. Ernst and E.G.D. Cohen, and from Dr.
B. Hasslacher.  In addition, he would like to acknowledge the
hospitality of the Information Mechanics Group at the M.I.T. Laboratory
for Computer Science where he was a visiting scientist during a portion
of this work.  This work was supported in part by Thinking Machines
Corporation, and in part by the divisions of Applied Mathematics of the
U.S. Department of Energy under contracts DE-FG02-88ER25065 and
DE-FG02-88ER25066.

\appendix

\newpage
\section{Glossary of Notation}

In this appendix, we list all the important symbols used in this paper,
giving the page number where they were first used (if appropriate) and a
brief description (if appropriate).
\begin{tabbing}
\hspace{1.0truein} \= \hspace{1.2truein} \= \hspace{3.5truein} \kill
{\bf SYMBOL} \> {\bf PAGE} \> {\bf DESCRIPTION} \\
\vspace{0.5truein}
\glossitem{\ast}!{pg:aa}!{Used in place of an index to indicate possible
          functional dependence on all values of an index}
\glossitem{\emptyset}!{}!{Empty set}
\glossitem{\widehat{\phantom{\alpha}}}!{pg:ab}!{Placed over a set to indicate
          that it contains strictly more than one element}
\glossitem{\sim}!{pg:bm}!{Equivalence relation between two states}
\glossitem{\alpha, \beta, \ldots}!{pg:ab}!{Used to indicate subsets of the
          set ${\cal B}$ of bits in the entire lattice}
\glossitem{\alpsupp{\mu}}!{pg:ac}!{Parameters of Fermi-Dirac equilibrium}
\glossitem{\beta_\bfx}!{pg:ad}!{Set of bit numbers for all particles in
          $\beta$ at site $\bfx$}
\glossitem{\ggamsup{\alpha}}!{pg:ae}!{Connected correlation function}
\glossitem{\mixten{\Gamma}{\eta}{\mu\nu}}!{pg:af}!{Fermi connection}
\glossitem{\mixten{\bfgamma (k)}{\eta}{\mu\nu}}!{pg:ag}!{Generalized Fermi
          connection}
\glossitem{\delta (x,y)}!{pg:ah}!{Kronecker delta of two bits}
\glossitem{\epsilon}!{pg:ai}!{Perturbation expansion parameter}
\glossitem{\varepsilon}!{pg:zw}!{$1-f$ for the 1D3P lattice gas}
\glossitem{\lamsupp{\mu}}!{pg:aj}!{Eigenvalue of $\jij$}
\glossitem{\mu (\{a_1, \ldots, a_j\})}!{pg:zx}!{Notation for
          $\{i (a_1),  \ldots, i (a_j)\}$}
\glossitem{\mu, \nu, \ldots}!{pg:ab}!{Used to indicate subsets of the set $B$
          of bits at a single lattice site}
\glossitem{\mu_T(\tau)}!{pg:ak}!{Set of outgoing virtual particles in diagram
          $T$ at timestep $\tau$}
\glossitem{\nu_T(\tau)}!{pg:ak}!{Set of incoming virtual particles in diagram
          $T$ at timestep $\tau$}
\glossitem{\pi (\alpha)}!{pg:ae}!{Set of all partitions of set $\alpha$}
\glossitem{\sigma_T(\tau)}!{pg:ak}!{Number of outgoing virtual particles in
          diagram $T$ at timestep $\tau$}
\glossitem{a,b,\ldots}!{pg:al}!{Used to enumerate all the bits on the
          lattice}
\glossitem{a(i,\bfx)}!{pg:al}!{Bit number on the lattice of the $i$th bit at
          site $\bfx$}
\glossitem{a(s'\rightarrow s)}!{pg:am}!{Microscopic transition matrix element}
\glossitem{A(s'\rightarrow s)}!{pg:an}!{Ensemble-averaged transition matrix
          element}
\glossitem{\advco{\mu}(\qqsupp{\ast})}!{pg:ao}!{Advection
          coefficients in hydrodynamic equation}
\glossitem{\advop{b}{c}}!{pg:ap}!{Advection operator}
\glossitem{B}!{pg:al}!{The set of $n$ bits at a site}
\glossitem{{\cal B}}!{pg:al}!{The set of $N$ bits on the lattice}
\glossitem{c}!{pg:zz}!{Characteristic lattice spacing}
\glossitem{\csup{i}}!{pg:aa}!{Microscopic collision operator}
\glossitem{\bfci}!{pg:aq}!{Lattice vector along which particles can move}
\glossitem{\ccsup{i}}!{pg:ar}!{Ensemble averaged collision operator}
\glossitem{\ccsupo{i}{\ell}}!{pg:as}!{$\ell$th order, ensemble-averaged
          collision operator}
\glossitem{D}!{pg:zy}!{Spatial dimension of lattice}
\glossitem{\cdop{\mu}{\xi} (\qqsupp{\ast})}!{pg:ao}!{Diffusion coefficients
          in hydrodynamic equation}
\glossitem{\bfesup{i}}!{pg:zz}!{Dimensionless lattice vectors}
\glossitem{\fsup{\alpha}}!{pg:au}!{Functional dependence of multipoint means
          on connected correlation functions}
\glossitem{\gsup{\alpha}}!{pg:av}!{Functional dependence of connected
          correlation functions on multipoint means}
\glossitem{g_{\mu\xi}}!{pg:aw}!{Fermi metric}
\glossitem{\subscrpt{\bf g (k)}{\mu\xi}}!{pg:ag}!{Generalized Fermi Metric}
\glossitem{H}!{pg:ax}!{Set of hydrodynamic modes, $\{1,\ldots,n_c\}$}
\glossitem{\mixten{\bfdelta (k)}{\mu}{\nu}}!{pg:ay}!{Generalized
          Kronecker delta}
\glossitem{i,j,\ldots}!{pg:al}!{Used to enumerate all the bits at a
          given site}
\glossitem{i(a)}!{pg:al}!{The bit number (at its site) of the $a$th bit on
          the lattice}
\glossitem{\jij}!{pg:az}!{Jacobian matrix of collision operator at
          equilibrium}
\glossitem{\tjij}!{pg:ba}!{Renormalized Jacobian of collision operator}
\glossitem{\cjop{b}{c}}!{pg:bb}!{Exact propagator for one-point means}
\glossitem{\mixten{k}{i}{\nu}}!{pg:bc}!{Coefficient of dependence of
          $\csup{i}$ on $\prod_{j\in\nu}\nsup{j}$}
\glossitem{k(T)}!{pg:bd}!{Length of diagram $T$}
\glossitem{K}!{pg:ax}!{Set of kinetic modes, $\{n_c+1,\ldots,n\}$}
\glossitem{\kop{\beta}{\gamma}}!{pg:be}!{Collision operator for multipoint
          means}
\glossitem{\ckop{\beta}{\gamma}}!{pg:bf}!{Collision operator for connected
          correlation functions}
\glossitem{L}!{pg:al}!{Lattice}
\glossitem{L_\beta}!{pg:ad}!{Subset of points in $L$ which contain at least
          one particle in set $\beta$}
\glossitem{n}!{pg:al}!{Number of bits per site}
\glossitem{\nixt}!{pg:al}!{$i$th bit at site $\bfx$ at timestep $t$}
\glossitem{\nsup{a}(t)}!{pg:al}!{The value of the $a$th bit on the lattice at
          timestep $t$}
\glossitem{\supscrpt{n}{p}, \supscrpt{n}{q}, \supscrpt{n}{r}}!{pg:bg}!{
          Sometimes
          used to denote stochastic bits, when no ambiguity will arise.}
\glossitem{n_c}!{pg:bh}!{Number of conserved quantities}
\glossitem{N}!{pg:al}!{Total number of bits on the lattice}
\glossitem{\nnsup{\alpha}}!{pg:bi}!{Multipoint mean}
\glossitem{\nnixt}!{pg:bj}!{Ensemble average of $\nixt$}
\glossitem{\nnsupo{i}{\ell}}!{pg:bk}!{$\ell$th order, ensemble-averaged
          distribution function}
\glossitem{\ppsup{\alpha}}!{pg:bl}!{Multipoint probability distribution}
\glossitem{\qsup{\mu}\bfxt}!{pg:bm}!{Value of $\mu$th conserved quantity at
          position $\bfx$ at timestep $t$}
\glossitem{\qrowd}!{pg:bm}!{Coefficient of dependence of $\mu$th conserved
          quantity on bit $i$, for $\mu=1,\ldots,n_c$; also left
          eigenvector of $\jij$ for $\mu=1,\ldots,n$}
\glossitem{\qcold}!{pg:bn}!{Right eigenvector of $\jij$ for $\mu=1,\ldots,n$}
\glossitem{\qqsupp{\mu}\bfxt}!{pg:bo}!{Ensemble average of $\qsup{\mu}\bfxt$}
\glossitem{s}!{pg:al}!{A state of a site}
\glossitem{s^i}!{pg:al}!{The value of the $i$th bit at a site in state $s$}
\glossitem{S}!{pg:al}!{The set of $2^n$ states of a site}
\glossitem{\supscrpt{\cal S}{\mu} (\qqsupp{\ast})}!{pg:ao}!{Source term in
          hydrodynamic equation}
\glossitem{t}!{pg:al}!{Discrete time}
\glossitem{\dt}!{pg:bp}!{Timestep}
\glossitem{T}!{pg:bd}!{A diagram}
\glossitem{\ctop{\mu}{\mu}(k)}!{pg:bd}!{Set of diagrams of length $k$ with at
          least two virtual particles at each timestep}
\glossitem{\vop{\mu}{\nu}}!{pg:bq}!{Coefficient of dependence of
          $\prod_{i\in\mu}[\nsup{i}+\csup{i}(\nsup{\ast})]$ on
          $\prod_{j\in\nu}\nsup{j}$}
\glossitem{\cvop{\alpha_\bfx}{\beta_\bfx}}!{pg:br}!{Factor contributed by one
          vertex, at site $\bfx$, to ${\cal K}$ coefficients}
\glossitem{W(T)}!{pg:bd}!{Weight of diagram $T$}
\glossitem{\bfx}!{pg:al}!{Position of a site on lattice $L$}
\glossitem{\bfx (a)}!{pg:al}!{The site of the $a$th bit on the lattice}
\end{tabbing}

\end{document}